\documentclass{utarticle}

\usepackage{graphicx,amsmath,amssymb,amsthm,amsxtra,bm,url}
\usepackage[pdfborder={0 0 0}]{hyperref}

\newcommand{\G}[1][]{\ensuremath{\mathcal{G}_{#1}}}
\newcommand{\bl}[1]{\ensuremath{\bm{#1}}}
\newcommand{\I}{\bl{I}}
\newcommand{\Id}{\operatorname{Id}}

\DeclareMathOperator{\inp}{\cdot}
\DeclareMathOperator{\lin}{\rfloor}
\DeclareMathOperator{\rin}{\lfloor}
\DeclareMathOperator{\out}{\wedge}

\newcommand{\grinv}[2][]{\ensuremath{#2^{*#1}}}
\newcommand{\doublegrinv}[1]{\ensuremath{#1^{**}}}
\newcommand{\rev}[1]{\ensuremath{#1^\dagger}}
\newcommand{\doublerev}[1]{\ensuremath{#1^{\dagger\dagger}}}
\newcommand{\grinvrev}[1]{\ensuremath{#1^{*\dagger}}}
\newcommand{\revgrinv}[1]{\ensuremath{#1^{\dagger *}}}
\newcommand{\clifconj}[1]{\ensuremath{#1^\ddagger}}
\newcommand{\doubleclifconj}[1]{\ensuremath{#1^{\ddagger\ddagger}}}

\newcommand{\scprod}[2]{\ensuremath{#1 * #2}}
\newcommand{\commute}[2]{\ensuremath{#1 \times #2}}

\newcommand{\grade}[2][]{\ensuremath{\left\langle #2 \right\rangle_{#1}}}
\newcommand{\weight}{\operatorname{weight}}
\newcommand{\dual}[1]{\ensuremath{#1^\perp}}
\newcommand{\invdual}[1]{\ensuremath{#1^{-\perp}}}

\newcommand{\Ker}{\operatorname{Ker}}
\newcommand{\Range}{\operatorname{Range}}
\newcommand{\rank}{\operatorname{rank}}
\newcommand{\nullity}{\operatorname{null}}
\newcommand{\adj}[1]{\ensuremath{\overline{#1}}}
\newcommand{\doubleadj}[1]{\ensuremath{\overline{\overline{#1}}}}
\newcommand{\Out}[1]{\ensuremath{\left[#1\right]}}

\newcommand{\half}{\ensuremath{\frac{1}{2}}}
\newtheorem{axiom}{Axiom}
\newtheorem{thm}{Theorem}
\newcommand{\bp}{\begin{quotation} \begin{proof}}
\newcommand{\ep}{\end{proof} \end{quotation}}
\DeclareMathOperator{\sgn}{sgn}
\allowdisplaybreaks[4]

\title{Geometric Algebra}

\author{Eric Chisolm}

\begin{document}

\Abstract{
This is an introduction to \textit{geometric algebra}, an alternative to traditional vector
algebra that expands on it in two ways:
\begin{enumerate}
\item In addition to scalars and vectors, it defines new objects representing subspaces 
          of any dimension. 
\item It defines a product that's strongly motivated by geometry and can be taken 
          between any two objects.  For example, the product of two vectors taken in a 
          certain way represents their common plane.
\end{enumerate}
This system was invented by William Clifford and is more commonly known as Clifford 
algebra.  It's actually older than the vector algebra that we use today (due to Gibbs) and 
includes it as a subset.  Over the years, various parts of Clifford algebra have been 
reinvented independently by many people who found they needed it, often not realizing 
that all those parts belonged in one system.  This suggests that Clifford had the right idea, 
and that geometric algebra, not the reduced version we use today, deserves to be 
the standard ``vector algebra."  My goal in these notes is to describe geometric algebra 
from that standpoint and illustrate its usefulness.  The notes are work in progress; I'll keep 
adding new topics as I learn them myself.
}

\maketitle

\tableofcontents


\section{Introduction}
\label{intro}

\subsection{Motivation}
\label{motive}

I'd say the best intuitive definition of a vector is ``anything that can be represented 
by arrows that add head-to-tail."  Such objects have \emph{magnitude} (how long 
is the arrow) and \emph{direction} (which way does it point).  Real numbers have 
two analogous properties: a \emph{magnitude} (absolute value) and a \emph{sign} 
(plus or minus).  Higher-dimensional objects in real vector spaces
also have these properties: for example, a surface element is a plane with a 
\emph{magnitude} (area) and an \emph{orientation} (clockwise or counterclockwise).  
If we associate real scalars with zero-dimensional spaces, then we can say that scalars, 
vectors, planes, etc.\ have three features in common:
\begin{enumerate}
\item An \emph{attitude}: exactly which subspace is represented.
\item A \emph{weight}: an amount, or a length, area, volume, etc. 
\item An \emph{orientation}: positive or negative, forward or backward, clockwise or 
          counterclockwise.  No matter what the dimension of the space, there are always 
          only two orientations.
\end{enumerate}
If spaces of any dimension have these features, and we have algebraic objects representing 
the zero- and one-dimensional cases, then maybe we could make objects representing 
the other cases too.  This is exactly what geometric algebra gives us; in fact, it goes farther 
by including all of these objects on equal footing in a single system, in which anything can be added 
to or multiplied by anything else.  I'll illustrate by starting in three-dimensional Euclidean space.

My goal is to create a product of vectors, called the \emph{geometric product}, which 
will allow me to build up objects that represent all the higher-dimensional subspaces.
Given two vectors $u$ and $v$, traditional vector algebra lets us perform two operations 
on them: the dot product (or inner product) and the cross product.  The dot 
product is used to project one vector along another; the projection of $v$ along $u$ is
\begin{equation} 
P_u(v) = \frac{u \inp v}{\ |u|^2}u
\label{vecproj}
\end{equation}
where $u \inp v$ is the inner product and $|u|^2 = u \inp u$ is the
square of the length of $u$. The cross product represents the oriented plane defined by $u$
and $v$; it points along the normal to the plane and its direction indicates orientation.  This
has two limitations:
\begin{enumerate}
\item It works only in three dimensions, because only there does every plane have a unique 
          normal.  
\item Even where it works, it depends on an arbitrarily chosen convention: whether to use
          the right or left hand to convert orientations to directions.  So the resulting vector does
          not simply represent the plane itself.
\end{enumerate}
Because of this, I'll replace the cross product with a new object that 
represents the plane directly, and it will generalize beyond three dimensions as
easily as vectors themselves do.

I begin with a formal product of vectors $uv$ that obeys the usual rules for multiplication;
for example, it's associative and distributive over addition.  Given these rules I can write
\begin{equation} uv = \half(uv+vu) + \half(uv-vu).   \label{decomp} \end{equation}
The first term is symmetric and bilinear, just like a generic inner 
product; therefore I set it equal to the Euclidean inner product, or
\begin{equation} \half(uv+vu) := u \inp v.   \end{equation}
I can immediately do something interesting with this: notice that $u^2 = u \inp u = |u|^2$, 
so the square of any vector is just its squared length.  Therefore, the vector
\begin{equation} u^{-1} := \frac{u}{\ u^2} \end{equation}
is the multiplicative inverse of $u$, since obviously $uu^{-1} = u^2 / u^2 = 1$.
So in a certain sense we can divide by vectors.  That's neat. By the way, the 
projection of $v$ along $u$ from Eq.~\eqref{vecproj} can now be written 
\begin{equation} 
P_u(v) = (v \inp u) u^{-1}.  
 \label{bettervecproj}
 \end{equation}
In non-Euclidean spaces, some vectors are null, so they aren't invertible.  That means 
that this projection operator won't be defined.  As it turns out, projection along noninvertible 
vectors doesn't make sense geometrically; I'll explain why in Section \ref{proj}.  Thus we 
come for the first time to a consistent theme in geometric algebra:  algebraic properties of 
objects frequently have direct geometric meaning.

What about the second term in Eq.~\eqref{decomp}?  I call it the \emph{outer product} or
\emph{wedge product} and represent it with the symbol $\out$, so now the geometric
product can be decomposed as
\begin{equation} uv = u \inp v + u \out v. \label{def} \end{equation} 
To get some idea of what $u \out v$ is, I'll use the fact that it's antisymmetric in 
$u$ and $v$, while $u \inp v$ is symmetric, to modify Eq.~\eqref{def} and get
\begin{equation} vu = u \inp v - u \out v. \end{equation}
Multiplying these equations together I find
\begin{equation}
uvvu = (u \inp v)^2 - (u \out v)^2.
\label{prodsq}
\end{equation}
Now $vv = |v|^2,$ and the same is true for $u$, while $u \inp v =
|u|\,|v|\cos\theta$, so
\begin{equation}
(u \out v)^2 = -|u|^2|v|^2\sin^2\theta.
\end{equation}
So whatever $u \out v$ is, its square has two properties:
\begin{enumerate}
\item It's a negative scalar.  (Just like an imaginary number, without my having
          to introduce them separately.  Hmm.)
\item Aside from the minus sign, it is the square of the magnitude of the 
          cross product.
\end{enumerate}
The first property means that $u \out v$ is neither scalar nor vector, while the second 
property makes it look like a good candidate for the plane spanned by the vectors.  
$u \out v$ will turn out to be something called a \emph{simple bivector} or 
\emph{$2$-blade}, so $2$-blades represent planes with an area and 
an orientation (interchange $u$ and $v$ and you change the sign of $u 
\out v$).  There's no unique parallelogram associated with $u \out 
v$ because for any $\lambda$,
\begin{equation}
u \out (v + \lambda u) = u \out v.
\end{equation}
So sliding the tip of one side along the direction of the other side 
changes the parallelogram but not the outer product.  
It is the plane (attitude), area (weight), and orientation that the outer 
product defines uniquely.  With these definitions, the product of two 
vectors turns out to be the sum of two very different objects: a scalar 
and a bivector.  For the moment think of such a sum as purely formal, 
like the sum of a real and an imaginary number.

Later I'll define the outer product of any number of vectors, and 
this product will be associative:
\begin{equation} 
(u \out v) \out w = u \out (v \out w) = u \out v \out w.
\end{equation}
This guy is called a \emph{simple trivector} or \emph{$3$-blade}, and it represents 
the three-dimensional space spanned by its factors, again with a weight (volume) 
and orientation.  We can also form $4$-blades, $5$-blades, and so on up to the 
dimension of whatever vector space we're in.  Each of these represents a subspace 
with the three attributes of attitude, weight, and orientation.  These $r$-blades and 
their sums, called \emph{multivectors}, make up the entire geometric algebra.  
(Even scalars are included as $0$-vectors.)  The geometric product 
of vectors can be extended to the whole algebra; you can multiply any two objects 
together, which lets you do all sorts of useful things.  Just multiplying vectors already 
lets us do a lot, as I'll show now.

\subsection{Simple applications}
\label{simpleapps}

I'll start by solving two standard linear algebra problems.  Let's suppose a plane is 
spanned by vectors $a$ and $b$, and you have a known vector $x$ in the plane that 
you want to expand in terms of $a$ and $b$.  Therefore you want scalars $\alpha$ and 
$\beta$ such that
\begin{equation} x = \alpha a + \beta b. \end{equation}
To solve this, take the outer product of both sides with $a$; since $a \out a = 0$,
you get
\begin{equation} a \out x = \beta a \out b. \end{equation}
It will turn out in Euclidean space that every nonzero vector, $2$-blade, and so on is
invertible, so this can be solved to get
\begin{equation} \beta = (a \out x) (a \out b)^{-1}. \end{equation}
This makes sense geometrically: both $a \out x$ and $a \out b$ are
bivectors in the same plane, so one should be a scalar multiple of the other.  Since 
$\beta$ is effectively a ratio of areas, I'm going to write instead
\begin{equation} \beta = \frac{a \out x}{a \out b}. \end{equation}
The problem with this is that it could mean either $(a \out x) (a \out b)^{-1}$ 
or $(a \out b)^{-1}(a \out x)$; but in this case they're the same, so there's no 
harm.  Taking the outer product of both sides with $b$ similarly gets you $\alpha 
= (x \out b)/(a \out b)$, so now we know that
\begin{equation} 
x =  \left( \frac{x \out b}{a \out b} \right) a + \left( \frac{a \out x}{a \out b} \right) b.
\label{xon2dbasis}
\end{equation}
This expression is called Cramer's Rule.  Here I've derived it much more quickly than 
is done in regular vector algebra, it's expressed directly in terms of the vectors instead 
of in components, and the geometric meaning of the coefficients (ratios of areas in the 
plane) is immediately apparent.  Also note that this expression is defined iff $a \out b 
\neq 0$, which is exactly the condition that $a$ and $b$ span the plane.

The generalization from planes to volumes is straightforward; if $a$, $b$, and $c$ span 
the space then
\begin{equation} 
x =  \left( \frac{x \out b \out c}{a \out b \out c} \right) a + 
       \left( \frac{a \out x \out c}{a \out b \out c} \right) b + 
       \left( \frac{a \out b \out x}{a \out b \out c} \right) c
\label{xon3dbasis}
\end{equation}
and so on for higher dimensions.

When you have a linear equation like this, taking the outer product with one of the 
terms, and thus removing that term, is often a handy trick.  Here's another example.
Suppose I have two lines that lie in a plane: The first passes through point $p$ and points 
in direction $a$, while the second passes through point $q$ and points in direction $b$.
Assuming the lines aren't parallel, at what point $x$ do they cross?

If the lines aren't parallel then $a$ and $b$ aren't parallel, so they span the plane.  
Therefore $x$ is a linear combination of $a$ and $b$ as given by Eq.~\eqref{xon2dbasis}.  
That's nice but unhelpful, because this time $x$ is unknown and we're trying to solve for 
it.  But wait; $x$ lies on the line through $p$ pointing along $a$, or
\begin{equation} x = p + \lambda a \end{equation}
for some $\lambda$.  That means that $a \out x = a \out p$.  And the fact that $x$ lies 
on the line through $q$ pointing along $b$ tells me that $x \out b = q \out b$,
so when I put all this in Eq.~\eqref{xon2dbasis} I find that the intersection point $x$ is
\begin{equation}
x =   \left( \frac{q \out b}{a \out b} \right) a +  \left( \frac{a \out p}{a \out b} \right) b,
\end{equation}
expressing the unknown $x$ in terms of the four known vectors defining the two lines.

The solutions to these last two exercises are expressed in an entirely intrinsic, coordinate-free 
way, which means that the results of this calculation can be used as inputs in any further 
calculations.  Once you get to the end, of course, you can certainly use coordinates to
perform the final computations.  To do all this, though, you have to be comfortable with
these new kinds of products and their inverses.  I'm here to help with that.

Now for a little geometry.  I'll start by looking at reflections, like the operation performed 
by a mirror.  How do we perform a mirror reflection on a vector?  Well, we often think 
of a reflection as happening in one of two complementary ways:  either 
\emph{through a plane} (components in the plane are left alone, the 
remaining component gets a minus sign) or \emph{along an axis} (the 
component along the axis gets a minus sign and the other components are 
left alone).  However, these ways of thinking are interchangeable only in 
three dimensions, because only there does any plane have a unique normal.  
I want a picture that works in any number of dimensions, and only the second
one does that, because it works even in one dimension.  So I'll use it from now on.

Let $v$ be the vector we want to reflect and let $n$ be a vector
along the reflection axis.  Then
\begin{align} 
v & = v (nn^{-1}) \nonumber \\
  & = (v n)n^{-1} \nonumber \\
  & = (v \inp n) n^{-1} + (v \out n) n^{-1}. 
\label{vectordecomp}
\end{align} 
The first term looks like the right hand side of Eq.~\eqref{bettervecproj}, so it represents 
the orthogonal projection of $v$ along $n$.  That means the other 
term is the component of $v$ perpendicular to $n$, also called the \emph{orthogonal 
rejection} of $v$ from $n$.  (I'll bet you've never heard that term before.)  
Now let $v'$ be the reflected vector; its component along $n$ has the opposite 
sign, while its perpendicular component is the same, so it is given by
\begin{equation}
v' = -(v \inp n) n^{-1} + (v \out n) n^{-1}.
\end{equation}
Using the symmetry and antisymmetry of the inner and outer products respectively, 
I can recast this as
\begin{align}
v' & = -(n \inp v) n^{-1} - (n \out v) n^{-1} \nonumber \\
    & = -(n \inp v + n \out v) n^{-1} \nonumber \\
    & = -n v n^{-1}.
\label{refdef}
\end{align}
This is a nifty little result; the equation for reflecting a vector along an axis is very tidy.  
Compare that to
\begin{equation} v' = v - 2\frac{n \inp v}{\ |n|^2}n, \end{equation}
which is the simplest one can do with traditional vector algebra.

The appearance of both $n$ and $n^{-1}$ in Eq.~\eqref{refdef} guarantees 
that the result depends neither on the weight (length) nor the orientation of 
$n$, only its attitude (the axis it represents), as it should.

The next operation I'll describe is rotation. First note that the usual 
way one thinks of rotations, as being performed around an axis, works 
only in three dimensions.  In general, it is better to think of a rotation of a
vector as being performed \emph{in a plane}; the component in
the plane is rotated while the components perpendicular to the plane
are left alone.  Again, this picture works perfectly well in any
number of dimensions.

\begin{figure}
\begin{center}
\includegraphics[height=0.4\textheight]{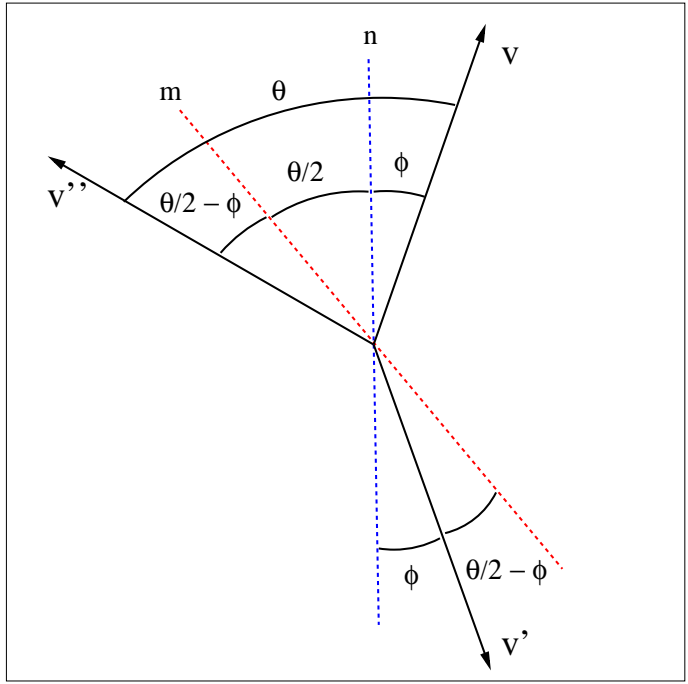}
\end{center}
\caption{The vector $v$ makes an angle $\phi$ with axis $n$.  It is then 
         reflected along $n$, producing vector $v'$, then along axis $m$, 
         producing vector $v''$.  Notice that the angle between vectors 
         $v$ and $v''$ is $\theta$, twice the angle between $n$ and $m$,
         regardless of the value of $\phi$.}
\label{rotfromref}
\end{figure}

Hamilton discovered a great way to perform rotations:  To rotate through 
angle $\theta$ in a plane, perform two reflections in succession 
along any two axes in the plane, as long as (a) the angle between 
the axes is $\theta/2$ and (b) a rotation from the first axis to the second is 
in the same direction as the rotation to be performed.  This is shown in 
Figure \ref{rotfromref}.  
So if I want to rotate vector $v$, then I let $m$ and $n$ be vectors along axes
satisfying the conditions, and the result of the rotation is
\begin{align} 
v' & = -m (-n v n^{-1}) m^{-1} \nonumber \\
   & = (m n) v (m n)^{-1} \nonumber \\
   & = R v R^{-1}
\label{rotdef} 
\end{align}
where $R = mn$.  $R$ is an object called a \emph{rotor}.  
Typically a rotor is the product of unit vectors, in which case $m = m^{-1}$ 
and $n= n^{-1}$ , which means $R^{-1}=nm$.  

These two examples, reflections and rotations, introduce a second theme in 
geometric algebra: elements of the algebra represent both geometric objects 
(vectors, subspaces) and operations on those objects.  There's no need to introduce 
any new elements (e.g.\ matrices) to represent operators.

I'll finish this section by looking at the rotor more closely.
\begin{align}
R & = m n \nonumber \\
  & = m \inp n + m \out n \nonumber \\
  & = m \inp n - n \out m.
\end{align}
I reversed the order of $m$ and $n$ because the sense of rotation of this 
rotor is from $n$ to $m$ (reflection $n$ was applied first).  Since $m$ 
and $n$ are now unit vectors and the angle between the two axes of 
reflection is $\theta/2$, $m \inp n = \cos(\theta/2)$ and 
$(n \out m)^2 = -\sin^2(\theta/2)$.  Therefore the bivector $B = 
(n \out m)/\sin(\theta/2)$ is a unit bivector: $B^2 = -1$.  So now
\begin{align}
R  & = \cos(\theta/2) - \sin(\theta/2) B \nonumber \\
     & = \exp(-B\theta/2) 
\end{align}
where the exponential is defined by its power series; the scalar terms and 
the terms proportional to $B$ can be grouped and summed separately.  
Now I have a rotation operator that explicitly displays the rotation's angle, 
plane (attitude of $B$), and direction (orientation of $B$), and all without 
coordinates.  

Recall how little I started with: a product of vectors with the minimal 
algebraic properties to be useful, plus the extra bit that the symmetric part equals
the inner product.  From only that much, I've gotten a formula for rotating
a vector that looks a lot like the formula for rotating a complex number, 
$z' = e^{i\theta} z$, except that it's double-sided and uses half of the rotation 
angle.  The resemblance to complex numbers is no accident; as I will show later 
on, the complex numbers are contained in the geometric algebra of the real 
Euclidean plane.  Therefore, all of complex algebra and analysis is 
subsumed into and generalized to arbitrary dimensions by geometric algebra.  
As for the half angles, in physics they normally show up in the quantum 
theory of half-integer spin particles, but this suggests that there's nothing 
particularly quantum about them; they arise simply because a rotation 
equals two reflections.

\subsection{Where now?}
\label{wherenow}

When I first read about geometric algebra, examples like these immediately
made me think it might have a lot to offer in terms of conceptual simplicity, 
unifying potential, and computational power.  I was on the lookout for something
like this because I had found standard mathematical physics dissatisfying in
two main ways:
\begin{enumerate}
\item We use a hodgepodge of different techniques, particularly in more advanced
          work, each of which seems to find a home only in one or two specialized 
          branches of theory.  It seems like an unnecessary fragmentation of what should 
          be a more unified subject.
\item As long as you stay in three dimensions and work only with vectors, 
          everything is very concrete, geometrical, and easy to express in intrinsic form without 
          coordinates.  However, none of these desirable features seem to survive in more 
          general situations (say, special relativity).  Either you end up expressing everything 
          in coordinates from the start, as in classical tensor analysis, or you use coordinate-free
          forms like those found in modern differential geometry, which I find hard to calculate
          with and which seem to leave their geometrical roots in favor of some abstract analytic 
          form.  (I put differential forms in this category.)
\end{enumerate}

Despite signs of promise, however, I also have to admit I was taken aback by what 
looked like an enormous proliferation of new objects.  After all, it seems like 
geometric algebra lets you multiply vectors all day long and keep getting
new things, and I had very little sense of how they all related to each other.  (I imagine
this is why geometric algebra lost out to Gibbs' vector algebra in the first place.)  I
was also puzzled about how the rules for using these objects really worked.
For example, if I had just read the previous two sections, I'd have questions like these.
\begin{enumerate}
\item In my first two examples in Section \ref{simpleapps}, I used the inverse 
          of a $2$-blade, $(a \out b)^{-1}$, and I mentioned that in Euclidean space 
          every nonzero $r$-blade has an inverse.  I've shown how to calculate the
          $2$-blade itself: it's the antisymmetrized product.  But how do you calculate 
          the inverse?
\item In Eq.~\eqref{refdef}, I multiplied three vectors, $n v n^{-1}$, and the 
          result was also a vector.  However, you could tell that only by following
          the derivation.  What's the product of three vectors in general?  Is it always 
          a vector?  Is it something else?  How can you tell?
\item Then I multiplied a bivector by a vector, $(v \out n) n^{-1}$.  What's that?  
          In this case the result was another vector, but again you had to follow 
          the derivation to know that.  In addition, I also said it was perpendicular 
          to $n$.  How do I check that?  Presumably I should show that
          \begin{equation} \left[(v \out n) n^{-1}\right] \inp n = 0, \end{equation}
          but that looks scary.
\end{enumerate}
To answer these and other questions for myself, I wrote these notes.  I suspect I'm not
the only one who reacted this way on seeing geometric algebra for the first time, so I 
hope the notes can help others understand geometric algebra and decide for themselves 
whether it's as good as advertised.

The structure of the notes reflects the best way I've found to explain geometric 
algebra to myself.
\begin{itemize}
\item In Section \ref{axioms} I lay out a set of axioms.  I find it helpful to present 
          axioms first, so we can have all of our basic rules in place immediately,
          knowing we won't be surprised later by having to take anything else into account.
\item With the axioms in hand, in Section \ref{whatshere} I answer the first question
          I asked myself when I saw all this: exactly what's in here?  I describe
          both algebraically and geometrically what a generic multivector looks like, 
          and I justify the claims I made at the end of Section \ref{motive} in terms of the 
          axioms.  By this point, a multivector should seem a lot more concrete than just 
          ``lots of vectors multiplied together."
\item Having explained what a general multivector looks like, in Section \ref{productprops}
          I explain what a general product of multivectors looks like.  I also explain how to
          take the inner and outer products of any two multivectors, and I explain what they 
          mean geometrically; this is a natural continuation of the geometrical discussion in 
          Section \ref{whatshere}.  I claimed earlier that geometric algebra lets you take coordinate-free, 
          intrinsic calculations much farther than standard methods; it does this because it
          has a large number of algebraic identities, which I'll start to derive here.  These identities
          make vector algebra start to look a lot more like algebra with real numbers.
\item A handful of additional operations are used all the time in calculations,
          and I collect them in Section \ref{otherops}.  By describing them all together, I
          can show the relationships between them more easily.
\item At this point even I think the reader needs a break, so I pause in Section \ref{eucl} 
          for an ``application" by describing what our favorite vector spaces, two- and three-dimensional 
          real Euclidean space, look like in these terms.  I show how the complex numbers pop up all 
          by themselves in the two-dimensional algebra, and in three dimensions I show how to
          convert back and forth from geometric algebra to the traditional language of cross products,
          triple products, and so on.  
\item With the full algebra at my disposal, in Section \ref{projrefrot} I return with a vengeance 
          to my initial examples: orthogonal projection, reflections, and rotations.  
          Now I really get to show you why this isn't your grandpa's vector algebra.  We can project 
          vectors into subspaces, and even subspaces into other subspaces, far more easily than 
          traditional methods ever made you think was possible.  And wait till you see what rotations 
          look like.  Ever tried to rotate a plane?  In geometric algebra, it's easy.
\item Coordinates do have a role in geometric algebra, although it's vastly reduced, and I
          describe it in Section \ref{framesbases}.
\item Linear algebra looks very different when it's done not just on vector spaces but on 
          geometric algebras; that's the subject of Section \ref{linalg}.  I'll review the basics, but
          even familiar subjects like adjoints and skew symmetric operators take on a new flavor
          and significance in this system.  And eigenvectors will be joined by eigenspaces of any
          dimension.  I'll even show how to act with a linear operator on the whole vector space 
          at once, and the eigenvalue of that operation will be our friend the determinant.
\item Right now, the notes are very light on applications to physics; so far I have included only 
          a brief discussion of classical angular momentum (which is no longer a vector, by the way) 
          and the Kepler problem, which gets a pretty snazzy treatment.  I'll add more 
          applications soon.
\end{itemize}
All the important definitions and relations are listed together in Appendix 
\ref{app:summary}, and the topics I plan to include in future versions are listed in 
Appendix \ref{app:future}.

\subsection{References and comments}
\label{refs}

Although geometric algebra dates from the 19th century, it was recovered in the form
described here only in the 20th century by David Hestenes \cite{STA,CAtoGC,NewFounds}, 
and it is slowly gaining popularity in various math and applied math communities.  
My primary sources are Hestenes' books; Doran and Lasenby \cite{GAforphys}, written 
for physicists; Dorst, Fontijne, and Mann \cite{GAforCS}, written for computer scientists; 
and the introductory linear algebra text by Macdonald \cite{LandGA}, which includes 
geometric algebra alongside traditional linear algebra topics.  You'll see 
their influence everywhere; for example, my axioms were inspired by \cite{CAtoGC}, 
Section \ref{motive} and the second half of Section \ref{simpleapps} come from 
\cite{GAforphys}, and the first half of Section \ref{simpleapps} is lifted from \cite{GAforCS}.
I'll mention other areas where I'm particularly indebted my sources as I come to them.
I follow \cite{GAforCS} in defining two inner products, instead of Hestenes' one, but I 
continue to refer to them as inner products instead of ``contractions" as Dorst \textit{et al.}\ 
do.   Finally, this approach to geometric algebra is far from the only one: Lounesto 
\cite{CAS} describes this one and several others, and he gives a great overview of the 
history that has brought Clifford algebra to this point.

Given all the other introductions to geometric algebra out there, I hope this treatment is
made distinctive by two elements.  First, I have worked everything out in more detail than 
I've seen anywhere else, which I think is very helpful for getting one's initial bearings in 
the subject.  Second, I don't believe this way of organizing the material is found in other 
sources either, and as I said in the previous section, this is the way I've found easiest to 
understand.  I try to convey Hestenes' attitude toward Clifford algebra as not just another 
algebraic system but the natural extension of real numbers to include the geometric idea 
of direction, which I find very attractive.

I also prefer a more general treatment over a more specific one when the
theory seems to be equally easy in either case.  For example, all applications
of geometric algebra I'm familiar with take the scalars to be $\mathbb{R}$, the real
numbers, and an important part of Hestenes' view is that many of the other number 
systems used in mathematics are best understood not separately but as 
subsets of certain real Clifford algebras.  (I dropped a hint about this regarding 
complex numbers in Section \ref{motive}, to which I'll return in Section \ref{eucl}, where 
I'll handle the quaternions too.)  However, I don't force the scalars to be $\mathbb{R}$ 
here, because the majority of results don't actually depend on what the scalars are.  
One thing that does change a bit, however, is the geometrical interpretation.  
For example, suppose the scalars are complex; how does the orientation of a vector
change when you multiply by $i$?  In fact, the two notions of weight and orientation make sense
only for real vector spaces, and as a result they won't have a place in a general
geometric algebra.  They're still important for all those applications, however, so 
I'll make sure to explain them at the right time.  And whenever the scalars have to be real for 
something to be true, I'll say so.

As part of my goal to work everything out in detail but keep the notes easy to follow, 
I've set the theorem proofs off from the rest of the text so they can be easily skipped.  
Nonetheless, I urge you to take a look at the shorter proofs;  I tried to motivate
them well and convey some useful insights.  Even some of the long proofs consist of 
more than just turning the algebra crank.  I like proofs that do more than show that
something is true; they give a sense of why.  I have tried to write those sorts of proofs
here.

Because geometric algebra has found its way into most applied mathematics, albeit 
in a very fragmented way, everything I describe in these notes can be done using some other 
system: matrices, Grassmann algebras, complex numbers, and so on.  The advantage that 
I see here is that one system, a natural extension of elementary vector algebra, can do all 
these things, and so far I've always found I can better understand what's going on when all 
the different results are related through a unified perspective.  

\section{Definitions and axioms}
\label{axioms}

The purpose of this section is to define a geometric algebra completely and 
unambiguously.  This is the rigorous version of the discussion from Section 
\ref{motive}, and you'll see all of the basic ideas from that section reintroduced
more precisely here.

A \emph{geometric algebra} is a set \G\ with two composition laws,
addition and multiplication (also called the \emph{geometric
product}), that obey these axioms.

\begin{axiom}
\G\ is a ring with unit.  The additive identity is called $0$ and 
the multiplicative identity is called $1$.
\label{ring}
\end{axiom}

Axiom \ref{ring} is the short way to say that (a) addition and 
multiplication in \G\ are both associative, (b) both operations 
have identities, (c) every element has an additive inverse, (d) 
addition commutes, and (e) multiplication is left and right 
distributive over addition.  So now I've said it the long way too.

A generic element of \G\ is denoted by a capital Roman letter 
($A$, $B$, etc.) and is called a \emph{multivector}.  Notice that a 
geometric algebra is one big system from the get-go:  all multivectors, 
which will eventually include scalars, vectors, and much more, are 
part of the same set, and addition and multiplication are equally 
available to all.  I'll continue to follow this philosophy as I introduce new 
operations by defining them for all multivectors.  Also, this axiom 
formalizes the first requirement I made of the geometric product in 
Section \ref{motive}; it gives addition and multiplication the minimal properties
needed to be useful.

\begin{axiom}
\G\ contains a field \G[0] of characteristic zero which includes $0$ 
and $1$.
\label{scalars}
\end{axiom}

A member of \G[0] is called a \emph{0-vector}, a \emph{homogeneous} 
multivector of \emph{grade} $0$, or a \emph{scalar}.  Scalars are 
denoted by lower case Greek letters ($\lambda,$ $\mu$, etc.).  Being a 
field means that \G[0] is closed under addition and multiplication, 
it contains all inverses of its elements, and it obeys all the rules that 
\G\ obeys from Axiom \ref{ring} plus the additional rules that (a) 
everything except $0$ has a multiplicative inverse and (b) multiplication 
commutes.  The rational numbers, real numbers, and complex numbers 
are all fields.  The property of having characteristic zero saves me
from getting in trouble in the following way.  Since \G[0] doesn't have to be
$\mathbb{R}$, the integers aren't actually
the usual integers, but sums of terms all equaling $1$, the multiplicative 
identity of \G[0].  (So, for example, by $2$ I literally mean $1+1$.)  If I 
don't specify any further properties of \G[0], then I haven't ruled out  
$1+1=0$, which would be bad when I try to divide by $2$.  (Which I'll
be doing frequently; see Eq.~\eqref{decomp}.)  Having characteristic 
zero means that no finite sum of terms all equaling $1$ will ever add up to $0$, 
so I can divide by integers to my heart's content.

\begin{axiom}
\G\ contains a subset \G[1] closed under addition, and $\lambda \in \G[0], v \in \G[1]$ 
implies $\lambda v = v \lambda \in \G[1]$.
\label{vectors}
\end{axiom}

A member of \G[1] is called a \emph{1-vector}, a \emph{homogeneous} multivector 
of \emph{grade} $1$, or just a \emph{vector}.  Vectors are denoted by lower case 
Roman letters ($a$, $b$, $u$, $v$, etc.).  The axioms imply that \G[1] obeys all the 
rules of a vector space with \G[0] 
as scalars, justifying their names.  However, all is not the same as what
you're used to.  In standard vector algebra, the scalars and vectors are usually
separate sets.  For example, consider the vector space $\mathbb{R}^3$ 
with the real numbers as scalars; the zero scalar is the number $0$, but 
the zero vector is the ordered triple $(0,0,0)$.  In geometric algebra 
this is not the case, and here's why.
\begin{enumerate}
\item $0$ is a scalar by Axiom \ref{scalars}.
\item $0v=0$ for any vector $v$ by Axiom \ref{ring}.
\item A scalar times a vector is a vector by Axiom \ref{vectors}.
\item Therefore, $0$ is also a vector.  
\end{enumerate}
It will turn out that $0$ is a whole lot of other things too.

So far the axioms have told us how to add scalars,
add vectors, multiply scalars, and multiply a scalar and a
vector.  Multiplying vectors is next.

\begin{axiom}
The square of every vector is a scalar.
\label{sqvec}
\end{axiom}

As it was in Section \ref{motive}, this is the most important axiom of the bunch.  
Here's the first consequence: for any vectors $u$ and $v$,
\begin{equation}
\half(uv + vu) = \half\left[(u+v)^2 - u^2 - v^2\right]
\end{equation}
(you can easily prove this by expanding out the right hand side), and the right side 
is a scalar thanks to Axiom \ref{sqvec}, so it follows that the symmetrized product 
of any two vectors is a scalar.  In fact, this is not merely implied by 
Axiom \ref{sqvec}; it's equivalent to it.  (Assume the statement is true. 
Since the square of a vector is its symmetrized product with itself, Axiom \ref{sqvec} 
follows.)  The symmetrized product of two vectors defined above is called their
\emph{inner product} and is denoted either $u \lin v$ or $u \rin v$.  It is 
symmetric and linear in both terms, thus obeying the usual rules for an inner 
product on a real vector space (but not a complex vector space).  Vectors $u$ 
and $v$ are said to be \emph{orthogonal} if $u \lin v = 0$, $u$ is a \emph{unit} 
vector if $u^2=\pm1$, and $u$ is \emph{null} if $u^2=0$.  Notice that vectors 
are orthogonal iff they anticommute.  This turns out to be handy.  Recall my 
earlier comment that if $v$ is non-null, then $v$ is invertible and $v^{-1}=v/v^2$.  

I have two inner products, $\lin$ and $\rin$, instead of just the usual $\inp$, for 
reasons that won't be clear until Section \ref{inneroutergeom}.  However, the two 
products are equal when both factors are vectors, so I can continue
to use the standard terminology of inner products as I please.  The next
axiom is an example.

Axiom \ref{sqvec} by itself is a little too general; for instance, it
would allow the product of any two vectors to be zero.  That seems 
pointless.  To prevent that, I'll add another axiom.

\begin{axiom}
The inner product is nondegenerate.
\label{nondeg}
\end{axiom}

This means that the only vector orthogonal to all vectors, including itself, is $0$.
This axiom is true in every application I can imagine, and I use it to prove some
useful results in Section \ref{dual}.  However, it is possible to replace it with a 
weaker axiom that accomplishes most of the same things; I discuss that in Section 
\ref{dual} too.  So if you ever find yourself reading other treatments of Clifford algebras,
watch out to see whether they use this axiom or put something else in its place.

Now I'll name other elements of \G.  Let $r > 1$; then an \emph{r-blade} or 
\emph{simple r-vector} is a product of  $r$ orthogonal (thus anticommuting) vectors.
A finite sum of $r$-blades is called an \emph{r-vector} or  \emph{homogeneous} 
multivector of \emph{grade} $r$.  (I'll bet you didn't see that coming.)  $2$-vectors are also 
called \emph{bivectors}, $3$-vectors \emph{trivectors}.  The set of $r$-vectors is called \G[r].  
Notice that this 
definition of simple $r$-vectors uses the geometric product of orthogonal vectors, 
not the outer product of arbitrary vectors as I did in Section \ref{motive}.  The definitions are 
equivalent, as I'll show later.

Products of vectors play an important role, so they get their own name.  
An \emph{$r$-versor} is a product of $r$ vectors.  So far we've seen two types
of versor: blades (where the vectors in the product are orthogonal) and rotors,
introduced in Section \ref{simpleapps}.  A rotor was defined to be a product of
two invertible vectors, so a rotor is an invertible biversor.  Later, a rotor will be 
any invertible even versor.

From these definitions and Axiom \ref{vectors} it follows 
that that if $A \in \G[r]$ and $\lambda$ is a scalar,
\begin{equation}
\lambda A = A \lambda \in \G[r].
\end{equation}
So multiplication by a scalar doesn't change the grade of an $r$-vector.  
This in turn implies that (a) each \G[r] is a vector space with \G[0] as 
scalars and (b) $0 \in \G[r]$ for every $r$.  So all the results I gave 
right after Axiom \ref{vectors} generalize fully.

Now we know that \G\ contains all the \G[r], and we know a few things about
how different \G[r] are related.  For example, suppose $u$ and $v$ are orthogonal 
and consider the $2$-blade $uv$.  It anticommutes with both $u$ and $v$, which 
means that it can't have a scalar part, because that part would have commuted with 
all vectors.  In fact, for this same reason no even blade can have a scalar part; and 
no odd blade can either, as long as there's another vector orthogonal to all the factors 
in the blade.  You can continue on this line and deduce a few more results, but it's not 
clear to me that you can use only the axioms so far to show that all the \G[r] are completely 
independent of each other.  So I add one final axiom for cleaning up.


\begin{axiom}
If $\G[0] = \G[1]$, then $\G = \G[0]$.  Otherwise, \G\ is the direct sum of all the \G[r].
\label{dirsum}
\end{axiom}

The first part of the axiom covers a special case: a field by itself, without any vectors, 
can be a geometric algebra.  When there are vectors around, the axiom says that every 
$A \in \G$ may be expressed one and only one way
as $A = \sum_r A_r$ where $A_r \in \G[r]$ and all but finitely many
$A_r$ vanish.  Therefore, every $A \neq 0$ is either an $r$-vector for
only one $r$ or is of mixed grade.

For each $r$, let the grade operator \grade[r]{\,} $: \G \rightarrow
\G[r]$ project each $A \in \G$ onto its unique grade-$r$ component.
Then
\begin{description}
\item[(a)] $A$ is an $r$-vector iff $A = \grade[r]{A}$.
\item[(b)] $\grade[r]{A + B} = \grade[r]{A} + \grade[r]{B}$.
\item[(c)] $\grade[r]{\lambda A} = \grade[r]{A \lambda} = \lambda \grade[r]{A}$.
\item[(d)] $\grade[s]{\grade[r]{A}} = \grade[r]{A}\,\delta_{rs}$.  
     (Thus the \grade[r]{\,} are independent projection operators.)
\item[(e)] $\sum_r \grade[r]{A} = A$ for any $A \in \G$.
     (Thus the \grade[r]{\,} are a complete set of projection operators.)
\end{description}
It will turn out to be convenient to define \grade[r]{\,}
even when $r$ is negative, so let me add one final property:
\begin{description}
\item[(f)] $\grade[r]{A} = 0$ if $r < 0$ for all $A \in \G$.
\end{description}
Because we take the scalar part of multivectors so often, I will let
\grade{\,} mean \grade[0]{\,}.

The notation $A_r$ will usually mean that $A_r$ is an $r$-vector.  The
exception is vectors: $a_1$, $a_2$, etc., are all vectors in a single
enumerated set (not objects of increasing grades).  Sometimes $A_r$ will
represent the grade-$r$ component of multivector $A$, which
is more properly denoted $\grade[r]{A}$, but that notation is cumbersome
so sometimes I drop it.  You can always tell from the context.  A blade is 
indicated by boldface; for example, \bl{A_r} is an $r$-blade.  The exceptions
are scalars ($0$-blades) and vectors ($1$-blades).  I want a special
notation for blades because they have geometric meaning while general 
$r$-vectors don't, as I'll show in Section \ref{whatshere}.

Axiom \ref{dirsum} tells us that the relation 
$\lambda A = A \lambda$, which I proved above for any homogeneous
multivector $A$, is true for any $A \in \G$, homogeneous or not.  
Another consequence of Axiom \ref{dirsum} is that \G\ is the
direct sum of subspaces \G[+] and \G[-] consisting of the even-grade
and odd-grade multivectors respectively.  Since many identities will
contain factors of $(-1)^r$, they will only depend on
whether the multivectors are even or odd.  Also, I'll show in
Section \ref{productprops} that the product of even multivectors is also even; this
means that the even subspace is actually a subalgebra, which will turn 
out to be important.  For these reasons
it's good to extend some of my notation to cover even and odd cases;
the notations $A_+$ and $A_-$ will mean that these objects have only
even-grade or odd-grade terms, respectively, and for any multivector
$A$, \grade[+]{A} (resp.\ \grade[-]{A}) is the 
even-grade (resp.\ odd-grade) part of $A$.

By the way, I haven't actually proved that anything satisfying these axioms exists.
That's done in \cite{elementary}.

\section{The contents of a geometric algebra}
\label{whatshere}

According to Axiom \ref{dirsum}, a geometric algebra consists of $r$-blades and their
sums.  However, the axioms and my comments at the end of Section \ref{motive} give 
us two different pictures of what $r$-blades are.  According to the axioms, an $r$-blade 
is a product of $r$ orthogonal vectors; according to the end of Section \ref{motive}, 
an $r$-blade is an outer product of $r$ arbitrary vectors.  I also said that blades represent 
subspaces, with weights and orientations when the scalars are real.  I'll spend this section 
relating these two pictures; first I'll show the two definitions of $r$-blades are equivalent, 
and then I'll justify the geometric interpretation.  Then we'll have a good intuitive 
feel for what a geometric algebra really is: sums of subspaces, with orientations and 
weights if the algebra is real.  To do this, I'll be using some concepts that I haven't fully 
explained yet.  Everything left hanging here will be fixed up in the next few sections.

First I want to show that outer products of vectors really are $r$-blades in the 
axiomatic sense of Section \ref{axioms}.  To do this, I define the outer product of 
vectors $\{a_i\}_{i=1,\dotsc,r}$ to be their fully antisymmetrized product, or
\begin{equation}
a_1 \out a_2 \out \dotsb \out a_r := \frac{1}{r!} \sum_\sigma 
    (\sgn\sigma)\, a_{\sigma(1)} a_{\sigma(2)} \dotsb a_{\sigma(r)}
\label{manyvwedge}
\end{equation}
where $\sigma$ is a permutation of 1 through $r$, sgn $\sigma$ is
the sign of the permutation (1 for even and $-1$ for odd), and the sum is over
all $r!$ possible permutations.  If $r=2$ this reduces to the outer product of two 
vectors defined previously.  Here's the result I need.
\begin{thm}
The outer product of $r$ vectors is an $r$-blade, and every $r$-blade is the outer 
product of $r$ vectors.
\label{outerisblade}
\end{thm} 
A corollary is that the outer product of two vectors is a $2$-blade, as I said in 
Section \ref{motive}.  This means that I could have used this as
the definition of an $r$-blade in the axioms, but the definition I did
use is more convenient in many situations.  Now, however, I'll use either definition 
as I need to.
\bp
To begin, I'll show that if the $a_i$ all anticommute then
the outer product reduces to the geometric product, so the result is
an $r$-blade.  Let $\{e_i\}_{i=1,\dotsc,r}$ anticommute with one another, and
consider their outer product
\begin{equation}
e_1 \out e_2 \out \dotsb \out e_r = \frac{1}{r!} \sum_\sigma 
    (\sgn\sigma)\, e_{\sigma(1)} e_{\sigma(2)} \dotsb e_{\sigma(r)}.
\end{equation}
In each term, the $e_i$ can be reordered so they're in ascending
numerical order, and each interchange of two $e_i$ introduces a minus
sign.  The end result is a factor of the form $\sgn \sigma$, which
cancels the $\sgn \sigma$ that's already there.  The result is
\begin{align}
e_1 \out e_2 \out \dotsb \out e_r & = \frac{1}{r!} \sum_\sigma 
       e_1 e_2 \dotsb e_r \nonumber \\
 & = e_1 e_2 \dotsb e_r
\end{align}
since there are $r!$ permutations to sum over and all $r!$ terms are
the same.  So when the vectors all anticommute, the wedges can be
retained or dropped as desired.  The $r=2$ version of this result,
\begin{equation} e_1 \out e_2 = e_1 \, e_2, \end{equation}
was already obvious from Eq.~\eqref{def} since $e_1 \inp e_2 = 0$, or 
$e_1 \lin e_2 = 0$ as I would say it now.

Turning to the general case, I can show that this is an $r$-blade
by examining the matrix $M$ with entries $M_{ij} = a_i \lin a_j$.
This is a real symmetric matrix, so it can be diagonalized by an
orthogonal transformation, meaning that there exists an orthogonal
matrix $R$ and a set of vectors $\{e_i\}_{i=1,\dotsc,r}$ such that
\begin{equation}
a_i = \sum_j R_{ij} e_j \quad \text{and} \quad e_i \lin e_j = e_i^2 \, \delta_{ij},
\end{equation}
so the $e_i$ anticommute with each other.  In that case
\begin{align}
a_1 \out a_2 \out \dotsb \out a_r & = \sum_{i,j,\dotsc,m}
       R_{1i}e_i \out R_{2j}e_j \out \dotsb \out R_{rm}e_m \nonumber \\
 & = \det(R) \, e_1 \out e_2 \out \dotsb \out e_r.
\end{align}
Now det($R$)=$\pm 1$, and if it equals $-1$ I interchange $e_1$ and
$e_2$ and relabel them, with the result
\begin{align}
a_1 \out a_2 \out \dotsb \out a_r & = e_1 \out e_2 \out \dotsb 
                                              \out e_r \nonumber \\
 & = e_1 e_2 \dotsb e_r
\end{align}
where the final line relies on the result from the previous paragraph.
So the outer product of $r$ vectors can be re-expressed as the product
of $r$ anticommuting vectors, making it an $r$-blade.  Further, every
$r$-blade is such an outer product (since for anticommuting vectors the
wedges can be added or dropped at will), so an object is an $r$-blade
iff it's the outer product of $r$ vectors.
\ep

Since every multivector is a unique sum of $r$-vectors by Axiom \ref{dirsum}, 
and every $r$-vector is a sum of $r$-blades, I can now say that a multivector 
is a sum of a scalar, a vector, and a bunch of outer products.  Now let's take 
the geometric point of view.  I know what scalars and vectors are geometrically, 
but what are the outer products?  To answer that, I need to look at when they 
vanish.  For example, $a \out a = 0$ for any $a$ by antisymmetry.  The more 
general case is given by this theorem.
\begin{thm}
The simple $r$-vector $a_1 \out a_2 \out \dotsb \out a_r = 0$ iff 
the vectors $\{a_i\}_{i=1,\dotsc,r}$ are linearly dependent.
\label{zeroifffindep}
\end{thm}
\bp
The outer product is clearly antisymmetric under interchange of any pair of 
factors, so it vanishes if any factor repeats.  It is also linear in each of its 
arguments, so if one factor is a linear combination of the others, the outer product 
vanishes.  So if the vectors are dependent, their product vanishes.  The other half 
of the proof, that the product of independent vectors doesn't vanish, is given in
Theorem \ref{indepnonzero}, which we don't have the tools to prove yet, so I'll 
defer it until later.
\ep
So an $r$-blade \bl{A_r} is nonzero exactly when its factors span an 
$r$-dimensional subspace.  Thus I associate \bl{A_r} with that subspace (attitude).  

To solidify the connection between subspaces and $r$-blades, here's a really
cool result.  It uses $a \out \bl{A_r}$, which I  haven't defined yet, but for now let's 
just say that it equals the outer product of $a$ and the factors of \bl{A_r}.
\begin{thm}
If \bl{A_r} is a nonzero $r$-blade with $r \geq 1$, then vector $a$ lies in the 
span of the factors of \bl{A_r} iff $a \out \bl{A_r} = 0$.  
\label{Aissubspace}
\end{thm}
\bp
$a \out \bl{A_r} =  0$ iff $a$ and the factors of \bl{A_r} are linearly dependent.  
Now the factors of \bl{A_r} are themselves independent because \bl{A_r} is nonzero, 
so $a \out \bl{A_r}$ vanishes iff $a$ is a linear combination of the factors of \bl{A_r}.  
\ep
Therefore \bl{A_r} does indeed define a subspace: the set of all vectors $a$ such that
$a \out \bl{A_r} = 0$.  

The proof of this theorem actually shows a bit more.  If $a \out \bl{A_r} \neq 0$, then it's
an $r+1$-blade, and it represents the direct sum of \bl{A_r} and the one-dimensional 
subspace defined by $a$.  I'll use this fact later when I show how to interpret outer products
geometrically. 

Theorem \ref{Aissubspace} implies another useful fact.
\begin{thm}
Two nonzero $r$-blades \bl{A_r} and \bl{B_r} define the same subspace iff each is a 
nonzero multiple of the other.  
\label{AmultofBifsamespace}
\end{thm}
\bp
If $\bl{A_r} = \lambda \bl{B_r}$ for some nonzero $\lambda$, 
then clearly $a \out \bl{A_r} = 0$ iff $a \out \bl{B_r} = 0$, so they represent 
the same subspace.  Conversely, suppose \bl{A_r} and \bl{B_r} represent the same 
subspace; then $\bl{A_r} = a_1 \out a_2 \out \dotsb \out a_r$ and $\bl{B_r} = 
b_1 \out b_2 \out \dotsb \out b_r$ for some linearly independent 
sets of vectors $\{a_i\}_{i=1,\dotsc,r}$ and $\{b_j\}_{j=1,\dotsc,r}$, and each of the 
$b_j$ is a linear combination of the $a_i$.  Substituting those linear
combinations into the expression for \bl{B_r}, removing the terms where any
$a_i$ appears twice, and reordering the factors in each term, I find that
\bl{B_r} equals \bl{A_r} multiplied by some scalar.  This scalar can't be zero
because \bl{B_r} is nonzero, so that completes the proof.
\ep
Frankly, something would be wrong if this weren't true.  In turn, 
Theorem \ref{AmultofBifsamespace} gives me another result I'll use a lot.
\begin{thm}
If \bl{A_r} represents a proper subspace of \bl{A_s}, then \bl{A_r} can be factored out of 
\bl{A_s}; that is, there exists a blade \bl{A_{s-r}} such that $\bl{A_s} = \bl{A_r} \out \bl{A_{s-r}}$.
\label{factorblade}
\end{thm}
\bp
Let $\bl{A_r} = a_1 \out \dotsb \out a_r$; then $\{a_j\}_{j=1,\dotsc,r}$ is a linearly independent
set lying in \bl{A_s}, so it can be extended to a basis $\{a_j\}_{j=1,\dotsc,s}$ of \bl{A_s}.  That means 
$a_1 \out \dotsb \out a_s$ defines the same subspace as \bl{A_s}, so it differs from \bl{A_s}
by a scalar factor; absorb that factor in the newly-added vectors and we have $\bl{A_s} = 
\bl{A_r} \out \bl{A_{s-r}}$, where \bl{A_{s-r}} is the outer product of the newly-added vectors.
\ep
I'll show later that $A_r \out B_s = (-1)^{rs} B_s \out A_r$ for any $r$- and $s$-vectors, so 
\bl{A_r} can be factored out of \bl{A_s} from either side and the other blade doesn't have to 
change by more than a sign.
 
While I'm here, let me also give a necessary and sufficient condition 
for a vector to be orthogonal to a subspace.  This theorem uses $a \lin \bl{A_r}$, the
left inner product of a vector and an $r$-blade, which once again I haven't defined yet.
For now, think of it as taking the inner product of $a$ with each factor of \bl{A_r} 
separately, as in Eq.~\eqref{aArblade} below. (Now that I'm taking the inner product 
of objects of different grades, it matters which of the two products I use; notice that the 
``floor" of the product points toward the vector.)
\begin{thm}
If \bl{A_r} is a nonzero $r$-blade with $r \geq 1$, $a$ is orthogonal to the subspace \bl{A_r} 
iff $a \lin \bl{A_r} = 0$.  
\label{vecorthtoA}
\end{thm}
\bp
To show this, let $\bl{A_r} = a_1 \out a_2 \out \dotsb \out a_r$ 
and let's look at 
\begin{align}
a \lin \bl{A_r}  & = a \lin (a_1 \out a_2 \out \dotsb \out a_r) \nonumber \\
 & = \sum_{j=1}^r (-1)^{j-1} (a \lin a_j)\, a_1 \out \dotsb \out 
              \check{a}_j \out \dotsb \out a_r,
\label{aArblade}
\end{align}
where I used Eq.~\eqref{veclinoutid} in the second line.  (I'll derive it later.  
The check over $a_j$ means it's not included in the outer product.)  
If $a$ is orthogonal to \bl{A_r} then it's orthogonal to all the $a_j$ and $a \lin 
\bl{A_r} = 0$.  If instead $a$ is orthogonal to all the $a_j$ but one, then 
$a \lin \bl{A_r}$ is the product of a nonzero scalar and the outer product 
of the remaining $a_j$, which is nonzero because they're linearly independent.  
So $a \lin \bl{A_r} \neq 0$. The remaining case is $a$ nonorthogonal to multiple 
$a_j$, in which case let $\{a_j\}_{j=1,\dotsc,s}$ where $1 < s \leq r$ be the vectors 
for which $a \lin a_j \neq 0$, and for $j=2,\dotsc,s$ let $b_j$ be defined by
\begin{equation}
b_j = a_j - \left(\frac{a \lin a_j}{a \lin a_1}\right) a_1.
\end{equation}
None of the $b_j$ equal $0$ because the $a_j$ are linearly independent, and 
$a \lin b_j = 0$; better yet, because each of the $b_j$ is just $a_j$ with 
a multiple of $a_1$ added to it, the outer product is unchanged by replacing 
the $a_j$ with the $b_j$:
\begin{align}
\bl{A_r} & = a_1 \out a_2 \out \dotsb \out a_r \nonumber \\
 & = a_1 \out b_2 \out \dotsb \out b_s \out a_{s+1} \out 
     \dotsb \out a_r.
\end{align}
Now I'm back to the previous case where only one vector in \bl{A_r} is nonorthogonal
to $a$, so I get the same result as before.  Therefore if $a$ is not orthogonal 
to \bl{A_r} then $a \lin \bl{A_r} \neq 0$.
\ep

The \emph{orthogonal complement} of a subspace is the set of all vectors orthogonal to
every vector in the subspace.  Theorem \ref{vecorthtoA} says that the orthogonal 
complement of \bl{A_r} is the set of all vectors $a$ satisfying $a \lin \bl{A_r} = 0$. 

Just as with Theorem \ref{Aissubspace}, the proof of Theorem \ref{vecorthtoA} actually 
shows a bit more.  If $a \lin \bl{A_r} \neq 0$, then it's an $r-1$-blade, and it represents 
the subspace of \bl{A_r} that is orthogonal to $a$.  I'll use this fact later when I interpret
inner products geometrically.

So not only do we know that $r$-blade \bl{A_r} represents an $r$-dimensional 
subspace, we have an easy way to tell whether vector $a$ is in that subspace 
($a \out \bl{A_r} = 0$) or orthogonal to it ($a \lin \bl{A_r} = 0$). Theorems 
\ref{Aissubspace} and \ref{vecorthtoA} are also our first examples of a general fact: 
\emph{algebraic relations between multivectors reflect geometric relations between 
subspaces.}  We'll see more advanced examples later.

Now let's suppose the scalars are real, in which case blades are also supposed 
to have orientation and weight.  To give \bl{A_r} an orientation, I note that it's the 
product of $r$ vectors in a given order.  That order defines an orientation: follow 
the vectors in their given order, and then follow their negatives in the same order 
until you're back where you started.  For example, the orientation of the $2$-blade 
$a \out b$ is found by moving along $a$, then $b$, then $-a$, then $-b$ back to 
the beginning.  A little experimentation shows that interchanging any two vectors 
reverses the orientation, and it also changes the sign of the blade.  Therefore there 
are two orientations, and changing orientations is associated with changing the 
sign of the blade or equivalently interchanging vectors.

Now this definition had nothing to do with what the scalars
are; the problem with non-real algebras arises when you try to decide what scalar
multiplication does to the orientation.  In real algebras it's easy: every nonzero scalar
is positive or negative, which either leaves the orientation alone or reverses it.  Other
sets of scalars are not well-ordered like this, so we can't say unambiguously what they
do to the orientation of blades; this is why I define orientation only for real algebras.

All that remains now is to define a blade's weight, which should include the notions of
length, area, and volume, but generalize them to arbitrary dimensions.  That's most 
easily done in Section \ref{scalarprod} on scalar products, so I will defer it until then.  
(I'll also explain why I define it only in the real case.)  Taking that for granted at the 
moment, I can now conclude that a general multivector is a sum of terms that represent 
different subspaces of all dimensions from $0$ on up, with weights and orientations 
if the scalars are real.  This of course is what I was after in the very beginning.

\section{The inner, outer, and geometric products}
\label{productprops}

Before we go on, I'll repeat what we know about the product of two vectors using 
all my new notation from Section \ref{axioms}.
\begin{equation}
uv = u \lin v + u \out v \label{prodsum} \end{equation}
where the two terms are the symmetric and antisymmetric parts of the product.
We also know that the first term is a scalar and the second is a bivector.
That means the inner and outer products can also be written
\begin{align}
u \lin v & = \grade{uv} \nonumber \\
u \out v & = \grade[2]{uv}. \label{vecdecomp}
\end{align}
My next job is to extend this knowledge to the inner, outer, and geometric product 
of any two multivectors at all. Once that's done, I will also 
geometrically interpret the inner and outer products.  Along the way I'll build up a 
set of tools and identities I'll use later to do calculations.

\subsection{The inner, outer, and geometric products of a vector with anything}

As a steppingstone, I first define the inner and outer products of a vector with any
multivector.  Sometimes I will want to explicitly indicate that a particular vector is absent 
from a product; I do this by including the vector anyway with a check $\check{}$ 
over it.  For example, if $\{a_i\}_{i=1,\dotsc,r}$ is a collection of vectors, 
then $a_1 a_2 \dotsb \check{a}_j \dotsb a_r$ is the product of all the 
vectors except for $a_j$.

I define the \emph{left inner product} of vector $a$ and $r$-vector $A_r$ (also 
called the inner product of $a$ \emph{into} $A_r$) to be
\begin{equation} 
a \lin A_r := \half\big[aA_r - (-1)^r A_ra\big],
\label{vecAinner}
\end{equation}
and I also define the \emph{right inner product} of $A_r$ and $a$ (or the 
inner product of $A_r$ \emph{by} $a$) to be
\begin{align}
A_r \rin a & := \half\big[A_ra - (-1)^r aA_r\big] \nonumber \\
           & \ = (-1)^{r-1} a \lin A_r. 
\label{Avecinner}
\end{align}
(Just as in Section \ref{whatshere}, the ``floor" of the inner product always points 
toward the vector.  Later I'll show how to calculate it the other way.)  When $r=1$, 
I recover the inner product of vectors defined earlier.  An equivalent way to write 
the relation between these products under interchange is
\begin{equation}
a \lin A_+ = -A_+ \rin a \ \ \ \text{while} \ \ \ a \lin A_- = A_- \rin a.
\end{equation}
Here's why I define inner products this way.
\begin{thm}
$a \lin A_r$ and $A_r \rin a$ are both $r-1$-vectors, so the left or
right inner product with a vector is a grade lowering operation. 
\label{innerlowersgrade}
\end{thm}
\bp
To show this, I start by proving this relation:  if $a$, $a_1$, $a_2$, 
$\dotsc$, $a_r$ are vectors, then
\begin{equation}
\half\big[a a_1 a_2 \dotsb a_r - (-1)^r a_1 a_2 \dotsb a_r a\big] = 
         \sum_{j=1}^r (-1)^{j-1} (a \lin a_j) a_1 a_2 \dotsb \check{a}_j 
         \dotsb a_r. 
\label{gradelower}
\end{equation}
I proceed by induction.  If $r=1$ the result is true because
it reduces to the definition of the inner product.  Suppose the result
holds for $r-1$, so
\begin{equation}
\half(a a_1 a_2 \dotsb a_{r-1})
 = \half (-1)^{r-1} (a_1 a_2 \dotsb a_{r-1} a) + \sum_{j=1}^{r-1} (-1)^{j-1} 
       (a \lin a_j) a_1 a_2 \dotsb \check{a}_j \dotsb a_{r-1}.
\end{equation}
Then since 
\begin{align}
\half(a a_1 a_2 \dotsb a_r) & = \half(a a_1 a_2 \dotsb a_{r-1}) a_r \nonumber \\
 & = \half(-1)^{r-1}(a_1 a_2 \dotsb a_{r-1} a) a_r + \sum_{j=1}^{r-1} (-1)^{j-1} 
     (a \lin a_j) a_1 a_2 \dotsb \check{a}_j \dotsb a_{r-1} a_r,
\end{align}
we find
\begin{align}
\half\big[a a_1 a_2 & \dotsb a_r - (-1)^r a_1 a_2 \dotsb a_r a\big] \nonumber \\
 & = \half(-1)^{r-1}(a_1 a_2 \dotsb a_{r-1} a a_r + a_1 a_2 \dotsb a_{r-1} 
       a_r a) + \sum_{j=1}^{r-1} (-1)^{j-1} (a \lin a_j) a_1 a_2 \dotsb 
       \check{a}_j \dotsb a_{r-1} a_r \nonumber \\
 & = (-1)^{r-1} (a \lin a_r) a_1 a_2 \dotsb a_{r-1} + \sum_{j=1}^{r-1} 
       (-1)^{j-1} (a \lin a_j) a_1 a_2 \dotsb \check{a}_j \dotsb a_{r-1} a_r 
       \nonumber \\
 & = \sum_{j=1}^r (-1)^{j-1} (a \lin a_j) a_1 a_2 \dotsb \check{a}_j 
       \dotsb a_r,
\end{align}
which is the desired result.  

Now let's look at the special case that the numbered vectors are an anticommuting set:
\begin{equation}
\half\big[a e_1 e_2 \dotsb e_r - (-1)^r e_1 e_2 \dotsb e_r a\big] = 
         \sum_{j=1}^r (-1)^{j-1} (a \lin e_j) e_1 e_2 \dotsb \check{e}_j
         \dotsb e_r. 
\end{equation}
The right hand side is a sum of $r-1$-blades, making it an $r-1$-vector.  Now 
a generic $r$-vector is a sum of $r$-blades, and any $r$-blade can be written 
$e_1 e_2 \dotsb e_r$, so it follows that for a vector $a$ and $r$-vector $A_r$
the quantity $\half\big[aA_r - (-1)^r A_ra\big]$, which is the left inner product, 
is an $r-1$-vector.  Since the right inner product differs only by a sign, it's
an $r-1$-vector too.
\ep
This begins to show why we have two inner products: the vector operates on the 
$r$-vector to lower its grade, not the other way around.  Two products allow the 
vector to do this from either side.  Notice that when $r=0$ (so $A_r=\lambda$) both 
inner products reduce to
\begin{equation} a \lin \lambda = \lambda \rin a = 0. \end{equation}
In retrospect this makes sense: these products lower grade, and the scalars have the 
lowest grade in the algebra, so there's nothing other than zero for them to be.

At this point it can't be much of a surprise that the \emph{outer
product} of vector $a$ and $r$-vector $A_r$ is defined to be
\begin{equation} 
a \out A_r := \half\big[aA_r + (-1)^r A_ra\big],
\label{vecAouter}
\end{equation}
and the outer product with the order reversed is given by
\begin{align}
A_r \out a & := \half\big[A_ra + (-1)^r aA_r\big] \nonumber \\
             & \ = (-1)^r a \out A_r. 
\label{Avecouter}
\end{align}
When $r=1$, of course I recover the outer product of vectors.  
The behavior of the outer product under interchange is the 
opposite of the inner product, and it can also be written
\begin{equation}
a \out A_+ = A_+ \out a \ \ \ \text{while} \ \ \ a \out A_- = -A_- \out a.
\end{equation}
This theorem is probably no surprise either.
\begin{thm}
$a \out A_r$ is an $r+1$-vector, so the outer product with a vector 
is a grade raising operation.
\label{outerraisesgrade}
\end{thm}
\bp
To show this, I again need to prove a preliminary result:
\begin{equation}
\half\big[a (a_1 \out a_2 \out \dotsb \out a_r) + 
      (-1)^r (a_1 \out a_2 \out \dotsb \out a_r) a\big] = 
         a \out a_1 \out a_2 \out \dotsb \out a_r. 
\label{wedgeresult}
\end{equation}

Again I use induction.  If $r=1$ the expression reduces to the
definition of the outer product of two vectors, so suppose it's true
for $r-1$.  Let $a_1 \out a_2 \out \dotsb \out a_r = e_1 e_2
\dotsb e_r$ where any two $e_i$ anticommute; then
\begin{equation}
a \out a_1 \out a_2 \out \dotsb \out a_r = 
a \out e_1 \out e_2 \out \dotsb \out e_r
\end{equation}
because the substitution of the $e_i$ for the $a_i$ yields a factor of
det($R$)=$\pm 1$ which can be eliminated as before, so the preliminary 
result becomes
\begin{equation}
\half\big[a (e_1 e_2 \dotsb e_r) + (-1)^r (e_1 e_2 \dotsb e_r) a\big] = 
         a \out e_1 \out e_2 \out \dotsb \out e_r.
\end{equation}
To prove it, let's begin by looking at the term on the right hand side, which we
know is the sum of $(r+1)!$ permutations.  I want to regroup it into
$r+1$ terms, each of which consists of all permutations that put a
particular one of the $r+1$ vectors in the first position.  Let the
permutations that put $a$ first be called $\pi$, and the ones that put
$e_i$ first be called $\pi_i$; then
\begin{align}
a \out e_1 \out e_2 \out \dotsb \out e_r  & =  
       \frac{1}{r+1} \Bigg[ \frac{1}{r!} \sum_{\pi} (\sgn\pi) a e_{\pi(1)}
       e_{\pi(2)} \dotsb e_{\pi(r)} + \nonumber \\
 & \qquad \qquad \sum_{j=1}^r \frac{1}{r!} \sum_{\pi_j} (\sgn\pi_j) e_j 
       a_{\pi_j(1)} a_{\pi_j(2)} \dotsb a_{\pi_j(r)} \Bigg]
\end{align}
where one of the $a_{\pi_j(i)}$ is $a$ and the others are the $e_i$
other than $e_j$.  Now the $e_{\pi(i)}$ in the first term on the right
hand side can be rearranged, canceling the sgn $\pi$ factors just as
before, so
\begin{equation}
\frac{1}{r+1} \left[ \frac{1}{r!} \sum_{\pi} (\sgn\pi) a e_{\pi(1)}
       e_{\pi(2)} \dotsb e_{\pi(r)} \right] = \frac{1}{r+1} a e_1 e_2 \dotsb e_r.
\end{equation}
As for the other terms, $\pi_j$ as a permutation of $a$ and all the
$e_i$ that puts $e_j$ in the first spot has the same sign as the
corresponding permutation of just $a$ and the other $e_i$ times a
factor $(-1)^j$, because this is the factor gained by moving $e_j$
from its original $j+1$ position to the front.  Therefore with that
factor added each $\pi_j$ may be thought of as a permutation of only
the remaining $r$ vectors, or
\begin{align}
\frac{1}{r!} \sum_{\pi_j} & (\sgn\pi_j) e_j a_{\pi_j(1)} a_{\pi_j(2)} 
             \dotsb a_{\pi_j(r)} \nonumber \\
 & = (-1)^j e_j (a \out e_1 \out e_2 \out \dotsb \out \check{e}_j 
     \out \dotsb \out e_r) \nonumber \\
 & = \half (-1)^j e_j [a e_1 e_2 \dotsb \check{e}_j \dotsb e_r + (-1)^{r-1} 
     e_1 e_2 \dotsb \check{e}_j \dotsb e_r a] \nonumber \\
 & = \half \big[(-1)^j e_j a e_1 e_2 \dotsb \check{e}_j \dotsb e_r + 
              (-1)^r e_1 e_2 \dotsb e_r a\big].
\end{align}
In the second line I used the fact that the relation is assumed true
for $r-1$, and in the second term on the third line I moved $e_j$ past
the first $j-1$ $e_i$.  Now since
\begin{equation} e_j a = 2 a \lin e_j - a e_j,   \end{equation}
the relation above becomes
\begin{align}
\frac{1}{r!} \sum_{\pi_j} & (\sgn\pi_j) e_j a_{\pi_j(1)} 
               a_{\pi_j(2)} \dotsb a_{\pi_j(r)} \\
 & = \half (-1)^{j-1} a e_j e_1 e_2 \dotsb \check{e}_j \dotsb e_r + (-1)^j 
       (a \lin e_j) e_1 e_2 \dotsb \check{e}_j \dotsb e_r + 
       \half (-1)^r e_1 e_2 \dotsb e_r a \nonumber \\
 & = \half a e_1 e_2 \dotsb e_r + \half (-1)^r e_1 e_2 \dotsb e_r a + (-1)^j 
       (a \lin e_j) e_1 e_2 \dotsb \check{e}_j \dotsb e_r.
\end{align}
Putting all this back together, I get
\begin{align}
a \out e_1 & \out e_2 \out \dotsb \out e_r \nonumber \\
 & = \frac{1}{r+1} \Bigg[ a e_1 e_2 \dotsb e_r + \sum_{j=1}^r 
     \bigg(\half a e_1 e_2 \dotsb e_r + \half (-1)^r e_1 e_2 \dotsb e_r a + 
     \nonumber \\
 &   \qquad \qquad (-1)^j (a \lin e_j) e_1 e_2 \dotsb 
     \check{e}_j \dotsb e_r\bigg) \Bigg] \nonumber \\
 & = \left(\frac{1}{r+1} + \frac{r}{2r+2}\right) a e_1 e_2 \dotsb e_r + 
     \frac{r}{2r+2} (-1)^r e_1 e_2 \dotsb e_r a - \nonumber \\
 &   \qquad \frac{1}{2r+2}\big[a e_1 e_2 \dotsb e_r - (-1)^r 
     e_1 e_2 \dotsb e_r a\big] \nonumber \\
 & = \half \big[a e_1 e_2 \dotsb e_r + (-1)^r e_1 e_2 \dotsb e_r a\big],
\end{align}
which proves the preliminary result.  

The right hand side of Eq.~\eqref{wedgeresult} is an $r+1$-blade.  Now a
generic $r$-vector is a sum of $r$-blades, and any $r$-blade is of the form 
$a_1 \out a_2 \out \dotsb \out a_r$, so it follows that for a vector $a$ 
and $r$-vector $A_r$ the quantity $\half\big[aA_r + (-1)^r A_ra\big]$, which is
just the outer product, is an $r+1$-vector.  And that's that.
\ep
You may wonder why I didn't define two outer products.  In this case, it makes equal
sense to think of the vector raising the $r$-vector's grade by $1$ or the $r$-vector 
raising the vector's grade by $r$.  

So now we know that
\begin{equation}
aA_r = a \lin A_r + a \out A_r,
\label{vecrprod}
\end{equation}
and further
\begin{align}
a \lin A_r & = \grade[r-1]{aA_r} \nonumber \\
a \out A_r & = \grade[r+1]{aA_r},
\label{aArgrade}
\end{align}
and similar results hold for $A_ra$:  
\begin{equation}
A_ra = A_r \rin a + A_r \out a,
\label{rvecprod}
\end{equation}
where
\begin{align}
A_r \rin a & = \grade[r-1]{A_ra} \nonumber \\
A_r \out a & = \grade[r+1]{A_ra}.
\label{Aragrade}
\end{align}
These expressions generalize Eqs.~\eqref{prodsum} and \eqref{vecdecomp} and reduce
to them when $r = 1$.  In fact, summing over grades $r$ in Eqs.~\eqref{vecrprod} and 
\eqref{rvecprod} shows that they're true for any multivector $A$.  So I've achieved 
my goal from the beginning of this section, at least  for the special case of multiplying by 
a vector.  The expressions for $\lin$ and $\rin$ in Eqs.~\eqref{aArgrade} and \eqref{Aragrade} 
work even when $r=0$, because back in Section \ref{axioms} I made a point of defining 
all negative-grade components of a multivector to vanish.  This is why.

\subsection{The general inner product, outer product, and geometric product}
\label{geninnerouter}

So far I have shown that the product of a vector and a multivector is the sum 
of two terms; if the multivector is grade $r$, the two terms have grades $r-1$ 
and $r+1$. I've also shown how to calculate each term separately.  
Now I'll use this information to characterize the product of any 
two multivectors, and after that I'll introduce the most general forms for 
the inner and outer products.

Let $A = \sum_r A_r$ and $B = \sum_s B_s$; then $AB = \sum_{r,s} A_r
B_s$, so I'll consider each term separately.
\begin{thm}
$A_r B_s$ consists of $\min\{r,s\}+1$ terms of grades $|r-s|$, $|r-s|+2$,
$|r-s|+4$, $\dotsc$, $r+s$, or
\begin{equation} 
A_r B_s = \sum_{j=0}^{\min\{r,s\}} \grade[|r-s|+2j]{A_r B_s}.
\end{equation}
\label{generalprod}
\end{thm}
\bp
If $r=0$ or $s=0$ this expression is obviously true, so next I'll
consider the case $ 0 < r \leq s$, assume $A_r$ is an $r$-blade \bl{A_r}
(if it's true for a blade it's true for sums of
blades), and proceed by induction on $r$.  If $r=1$ the expression
becomes Eq.~\eqref{vecrprod}, which I've proved already; so assume
it's true for $r-1$.  \bl{A_r} can be written $a \bl{A_{r-1}}$ where $a$ is a
vector and \bl{A_{r-1}} is an $r-1$-blade, so
\begin{align}
\bl{A_r} B_s & = a \bl{A_{r-1}} B_s \nonumber \\
 & = \sum_{j=0}^{\min\{r-1,s\}} a \grade[|r-1-s|+2j]{\bl{A_{r-1}} B_s} \nonumber \\
 & = \sum_{j=0}^{r-1} a \grade[s-r+2j+1]{\bl{A_{r-1}} B_s}
\end{align}
where the second line uses the fact that the relation is assumed true
for $r-1$ and the last line follows from the inequality $r \leq s$.
Now applying Eq.~\eqref{vecrprod},
\begin{align}
\bl{A_r} B_s & = \sum_{j=0}^{r-1} \left[ a \lin \grade[s-r+2j+1]{\bl{A_{r-1}} B_s} + 
     a \out \grade[s-r+2j+1]{\bl{A_{r-1}} B_s} \right] \nonumber \\
 & = a \lin \grade[s-r+1]{\bl{A_{r-1}} B_s} \, + \nonumber \\
 &   \quad \sum_{j=1}^{r-1} \left[a \lin \grade[s-r+2j+1]
     {\bl{A_{r-1}} B_s} + a \out \grade[s-r+2j-1]{\bl{A_{r-1}} B_s} 
     \right] \, + \nonumber \\
 &   \quad a \out \grade[s+r-1]{\bl{A_{r-1}} B_s}.
\label{expandedprod} 
\end{align}
Noting the grades of the various terms in the sum, I identify
\begin{align}
\grade[s-r]{\bl{A_r} B_s} & = a \lin \grade[s-r+1]{\bl{A_{r-1}} B_s} 
    \nonumber \\
\grade[s-r+2j]{\bl{A_r} B_s} & = a \lin \grade[s-r+2j+1]{\bl{A_{r-1}} B_s} 
    + a \out \grade[s-r+2j-1]{\bl{A_{r-1}} B_s} \label{terms} \\
\grade[r+s]{\bl{A_r} B_s} & =  a \out \grade[s+r-1]{\bl{A_{r-1}} B_s}.
     \nonumber
\end{align}
Since $s-r=|r-s|$, $\bl{A_r} B_s$ is now expressed as a sum of terms of
grade $|r-s|$, $|r-s|+2$, $|r-s|+4$, $\dotsc$, $r+s$, which proves the
result for $r$.  The remaining case is $0 < s \leq r$, which is proved
by induction on $s$.
\ep

This proof actually gives somewhat explicit formulas for the terms 
in the product.  To illustrate this I'll consider Eqs.~\eqref{terms} 
for the special case $r=2$, so $\bl{A_2} = e_1 e_2$:
\begin{align}
\bl{A_2} B_s & = e_1 \lin \grade[s-1]{e_2 B_s} + e_1 \lin \grade[s+1]
            {e_2 B_s} + e_1 \out \grade[s-1]{e_2 B_s} 
            + e_1 \out \grade[s+1]{e_2 B_s} \nonumber \\
 & =  e_1 \lin (e_2 \lin B_s) + e_1 \lin (e_2 \out B_s) + 
      e_1 \out (e_2 \lin B_s) + e_1 \out (e_2 \out B_s).
\end{align}
I could have arrived at the same result by writing $\bl{A_2} B_s = e_1 e_2
B_s$, using Eq.~\eqref{vecrprod} to expand $e_2 B_s$, and using 
Eq.~\eqref{vecrprod} again to expand the product of $e_1$ with each term.
The first term is of grade $s-2$, the middle two terms are of grade
$s$ (one grade lowering and one grade raising operation applied to
$B_s$), and the final term is of grade $s+2$.

Theorem \ref{generalprod} tells us something important: while $A_r B_s$ is not an 
$r+s$-vector, every term has grade $r+s-2j$ for some
$j$, so $A_r B_s$ is even if $r+s$ is even and odd if $r+s$ is odd.
That means that the product of two even grade elements is itself an
even grade element, so the even grade subspace of any geometric
algebra (defined at the end of Section \ref{axioms}) is not just a
subspace but a subalgebra.  (This is not true of the odd subspace,
because the product of two odd elements is also even.)  Also, since 
$(-1)^{r+s-2j} = (-1)^{r+s}$, it follows that
\begin{align}
a \lin (A_r B_s) & = \half(a A_r B_s - (-1)^{r+s} A_r B_s a) \nonumber \\
a \out (A_r B_s) & = \half(a A_r B_s + (-1)^{r+s} A_r B_s a)
\label{evenoddinnerouter} 
\end{align}
for any vector $a$, which is kinda nice.  Four identities follow from this.
\begin{thm}
\begin{align}
a \lin (A_r B_s) & = (a \lin A_r) B_s + (-1)^r A_r (a \lin B_s) \nonumber \\
                   & = (a \out A_r) B_s - (-1)^r A_r (a \out B_s) \nonumber \\
a \out (A_r B_s) & = (a \out A_r) B_s - (-1)^r A_r (a \lin B_s) \nonumber \\
                   & = (a \lin A_r) B_s + (-1)^r A_r (a \out B_s).
\label{usefulids}
\end{align}
\end{thm}
\bp
These are all proved the same way, so I'll show only the first one.  Starting with
the first of Eqs.~\eqref{evenoddinnerouter} and then adding and subtracting
$\half (-1)^r A_r a B_s$,
\begin{align}
a \lin (A_r B_s) & = \half(a A_r B_s - (-1)^{r+s}A_r B_s a) \nonumber \\
 & = \half(a A_r B_s - (-1)^r A_r a B_s) + \half(-1)^r (A_r a B_s - 
     (-1)^s A_r B_s a) \nonumber \\
 & = (a \lin A_r) B_s + (-1)^r A_r (a \lin B_s)
\end{align}
where I reassembled the terms into inner products using the first of
Eqs.~\eqref{evenoddinnerouter} again.
\ep

By summing over grades $s$, you can see that these identities are valid even if 
$B$ is a general multivector.  (They're also valid for general $A$ with a little 
tweaking; see Section \ref{gradeinv}.)

An obvious generalization of Eqs.~\eqref{evenoddinnerouter} is
\begin{align}
a \lin (a_1 a_2 \dotsb a_r) & = \half(a a_1 a_2 \dotsb a_r - 
                                 (-1)^r a_1 a_2 \dotsb a_r a) \nonumber \\
a \out (a_1 a_2 \dotsb a_r) & = \half(a a_1 a_2 \dotsb a_r + 
                                 (-1)^r a_1 a_2 \dotsb a_r a)
\label{rvecsdotwedge}
\end{align}
and the first of these equations can be used to prove another handy result.
(I used this result to prove Theorem \ref{vecorthtoA}, you may recall.)
\begin{thm}
\begin{equation}
a \lin (a_1 \out a_2 \out \dotsb \out a_r) = \sum_{j=1}^r (-1)^{j-1} 
    (a \lin a_j) \, a_1 \out a_2 \out \dotsb \out \check{a}_j \out 
    \dotsb \out a_r.
\label{veclinoutid}
\end{equation}
\label{alinoutai}
\end{thm}
\bp
Using the first of Eqs.~\eqref{rvecsdotwedge} and Eq.~\eqref{gradelower}, I can write
\begin{align} 
a \lin (a_1 a_2 \dotsb a_r) & = \half(a a_1 a_2 \dotsb a_r - (-1)^r a_1 a_2 
                                  \dotsb a_r a) \nonumber \\
 & = \sum_{j=1}^r (-1)^{j-1} (a \lin a_j) \, a_1 a_2 \dotsb 
                  \check{a}_j \dotsb a_r.
\label{ainnerprod}
\end{align}
I'll prove just below that the grade-$s$ term in the product of $s$ vectors 
is their outer product (see Eq.~\eqref{wedgehighest}), so by taking the 
$r-1$-grade term of both sides and using that result, the identity follows. 
\ep
Here's a nice mnemonic for remembering the coefficients in this sum.  In the 
$j$th term, $a$ acts on $a_j$, so imagine that $a_j$ first has to be moved 
to the far left side of the outer product, which requires $j-1$ interchanges of adjacent
vectors and introduces a factor of $(-1)^{j-1}$.

I can use the method of proof of this theorem to prove a fact about versors.
\begin{thm}
An $r$-versor $a_1 a_2 \dotsb a_r$ is a linear combination of terms, each of which is
an outer product of some subset of $\{a_j\}_{j=1,\dotsc,r}$.  The number of factors in each
term is even or odd as $r$ is even or odd.  
\label{versorsumofblades}
\end{thm}
\bp
As usual, the proof is by induction.  The result is true if $r=0$, $1$, or $2$, so assume 
it's true for $r-1$; then
\begin{align}
a_1 a_2 \dotsb a_r & = a_1 \lin (a_2 \dotsb a_r) + a_1 \out (a_2 \dotsb a_r) \nonumber \\
                               & = \sum_{j=2}^r (-1)^{j-2} (a_1 \lin a_j) a_2 \dotsb \check{a}_j \dotsb a_r +
                                                              a_1 \out (a_2 \dotsb a_r),
\label{prodofr}
\end{align}
where I used Eq.~\eqref{ainnerprod} to go from the first to the second line.  
The first term is a linear combination of products of $r-2$ of the $a_i$, 
so by the $r-2$ result the first term is a linear combination of outer products.  
The number of factors in each term is even or odd as $r-2$ is even or odd, 
or as $r$ is even or odd.  The second term is the outer product of $a_1$ 
with the product of the remaining $r-1$ vectors.  By the $r-1$ result, 
that product is a linear combination of outer products, and each term is even
or odd as $r-1$ is even or odd.  When you take its outer product with $a_1$, 
it's still a linear combination of outer products, and each term is even or odd as
$r$ is even or odd.
\ep

Now I define for any $A=\sum_r A_r$ and 
$B=\sum_s B_s$
\begin{align}
A \lin B & := \sum_{r,s} \grade[s-r]{A_r B_s} \nonumber \\
A \rin B & := \sum_{r,s} \grade[r-s]{A_r B_s} \nonumber \\
A \out B & := \sum_{r,s} \grade[r+s]{A_r B_s}.
\label{geninoutdefs}
\end{align}
All previous expressions for the inner and outer products of two 
objects are special cases of these definitions.  Further, this defintion for
the outer product of two objects and the definition for the outer product of 
arbitrarily many vectors in Eq.~\eqref{manyvwedge} are consistent with each
other thanks to Eq.~\eqref{wedgeresult}, which shows that
\begin{equation}
a \out (a_1 \out a_2 \out \dotsb \out a_r) = 
a \out a_1 \out a_2 \out \dotsb \out a_r. 
\end{equation}
Some other facts are worth mentioning.
\begin{enumerate}
\item $A_r \lin B_r = A_r \rin B_r = \grade{A_r B_r}$.
\item If $r > s$ then $A_r 
      \lin B_s = B_s \rin A_r = 0$ because all negative-grade multivectors vanish.
\item The lowest grade term in $A_r B_s$ is $A_r \lin B_s$ if $ r \leq s$ and 
      $A_r \rin B_s$ if $r \geq s$.
\item The highest grade term in $A_r B_s$ is $A_r \out B_s$.  
\item For any $\lambda$, $\lambda A = \lambda \lin A = \lambda \out A$ and 
      $A \lambda = A \rin \lambda = A \out \lambda$.  The 
      product of a vector and any multivector is a sum of inner and outer products, as
      shown by Eqs.~\eqref{vecrprod} and \eqref{rvecprod}.  In all other cases there are 
      additional terms of intermediate grades.
\end{enumerate}

These definitions make the inner and outer products much easier to work 
with, because in general the geometric product has nicer algebraic properties.
The main advantage of the inner and outer products over the full product 
is nice behavior under interchange of the factors; as you'll see in Section \ref{reverse},
$A_r B_s$ and $B_s A_r$ are related, but not in a way that's easy to use, while going
from $A_r \lin B_s$ to $B_s \rin A_r$ or from $A_r \out B_s$ to $B_s \out A_r$ is just 
a matter of a sign change (Eqs.~ \eqref{commuteinner} and \eqref{commuteouter}).

The definitions also allow me to deduce three more identities from Eqs.~\eqref{usefulids}; 
I take the $r+s-1$-grade term of the first of Eqs.~\eqref{usefulids}, the 
$r-s+1$-grade term of the third, and the $s-r+1$-grade term of the last, with 
the results
\begin{align}
a \lin (A_r \out B_s) & = (a \lin A_r) \out B_s + (-1)^r A_r \out 
                              (a \lin B_s) \nonumber \\
a \out (A_r \rin B_s) & = (a \out A_r) \rin B_s - (-1)^r A_r \rin 
                              (a \lin B_s) \nonumber \\
a \out (A_r \lin B_s) & = (a \lin A_r) \lin B_s + (-1)^r A_r \lin 
                              (a \out B_s).
\label{contwedgeids}
\end{align}
(Taking an appropriate-grade term of the second identity only yields a special 
case of the third of Eqs.~\eqref{associdents} below.)  Again, these expressions 
are actually valid for general $B$, and also for general $A$ when I use some 
results from Section \ref{gradeinv}.

Like the geometric product, the inner and outer products are
distributive, and they obey these identities.
\begin{thm}
\begin{align}
A \out (B \out C) & = (A \out B) \out C \nonumber \\
A \lin (B \rin C) & = (A \lin B) \rin C \nonumber \\
A \lin (B \lin C) & = (A \out B) \lin C \nonumber \\
A \rin (B \out C) & = (A \rin B) \rin C
\label{associdents}
\end{align}
\label{wedgelinrinidents}
\end{thm}
So the outer product and certain combinations of left and right inner products are
associative, but neither left nor right inner products are associative
by themselves.  In the homogeneous case, the first relation above becomes
\begin{equation}
A_r \out (B_s \out C_t) = \grade[r+s+t]{A_r B_s C_t}.
\label{highgrade}
\end{equation}
An important special case of this is
\begin{equation}
a_1 \out a_2 \out \dotsb \out a_r = \grade[r]{a_1 a_2 \dotsb a_r}.
\label{wedgehighest}
\end{equation}
Associativity of the outer product plus its properties under interchange of 
factors leads to the result
\begin{equation}
a \out A \out b = -b \out A \out a
\label{aoutAoutb}
\end{equation}
where $a$ and $b$ are any vectors and $A$ is any multivector.
\bp
For the first relation, I note that
\begin{align}
A_r \out (B_s \out C_t) & = A_r \out \grade[s+t]{B_s C_t} \nonumber \\
 & = \grade[r+s+t]{A_r \grade[s+t]{B_s C_t}} \nonumber \\
 & = \grade[r+s+t]{A_r B_s C_t} \nonumber \\
 & = \grade[r+s+t]{\grade[r+s]{A_r B_s} C_t} \nonumber \\
 & = (A_r \out B_s) \out C_t.
\end{align}
The crucial step is taken on the third line, where the \grade[s+t]{\,}
is dropped.  This can be done because the only term in $A_r B_s C_t$
that has grade $r+s+t$ is the term that comes from multiplying $A_r$
by the highest grade term in $B_s C_t$.  From this it follows that 
$A \out (B \out C) =  (A \out B) \out C$ for any $A$, $B$, and 
$C$.  The third line also gives me Eq.~\eqref{highgrade}.  For the second
relation, consider
\begin{align}
A_r \lin (B_s \rin C_t) & = A_r \lin \grade[s-t]{B_s C_t} \nonumber \\
 & = \grade[s-(r+t)]{A_r \grade[s-t]{B_s C_t}}.
\end{align}
Now this vanishes automatically unless $r \leq s - t$, in which case 
\begin{equation} 
\grade[s-(r+t)]{A_r \grade[s-t]{B_s C_t}} = \grade[s-(r+t)]{A_r B_s C_t}
\end{equation}
because when the inequality is satisfied, the only term in $A_r B_s C_t$ 
that has grade $s-(r+t)$ is the term that comes from multiplying $A_r$
by the lowest grade term in $B_s C_t$.  Therefore
\begin{align}
A_r \lin (B_s \rin C_t) & = \grade[s-(r+t)]{A_r B_s C_t}
                                \nonumber \\
 & = \grade[(s-r)-t]{A_r B_s C_t} \nonumber \\
 & = \grade[(s-r)-t]{\grade[s-r]{A_r B_s} C_t} \nonumber \\
 & = \grade[s-r]{A_r B_s} \rin C_t \nonumber \\
 & = (A_r \lin B_s) \rin C_t.
\label{linrin} 
\end{align}
This expression also vanishes unless the inequality is satisfied, so $A \lin 
(B \rin C) = (A \lin B) \rin C$ in general.  Finally, 
\begin{align}
A_r \lin (B_s \lin C_t) & = A_r \lin \grade[t-s]{B_s C_t} \nonumber \\
 & = \grade[t-(r+s)]{A_r \grade[t-s]{B_s C_t}}.
\end{align}
Now this vanishes automatically unless $r + s \leq t$, in which case
logic similar to that used above yields
\begin{align}
A_r \lin (B_s \lin C_t) & = \grade[t-(r+s)]{A_r B_s C_t} \nonumber \\
 & = \grade[t-(r+s)]{\grade[r+s]{A_r B_s} C_t} \nonumber \\
 & = \grade[r+s]{A_r B_s} \lin C_t \nonumber \\
 & = (A_r \out B_s) \lin C_t.
\label{linwedge} 
\end{align}
This final expression also vanishes unless the inequality is satisfied; 
therefore $A \lin (B \lin C) = (A \out B) \lin C$ for all $A$, 
$B$, and $C$.  A similar proof shows that $(A \rin B) \rin C = A \rin 
(B \out C)$.  
\ep

It's useful to introduce an order of operations of these
products to cut down on parentheses.  The order is
outer products, followed by inner products, followed by geometric
products.  Thus, for example,
\begin{equation}
A \lin B \out C D = \left\{A \lin (B \out C)\right\} D.
\end{equation}
Despite this convention, I'll occasionally put the parentheses back in for clarity.  
However, I will use it a lot in Section \ref{projrefrot} on projections, rotations, and reflections.  
(This is not the only convention in use; the other one reverses the order 
of inner and outer products. I picked this one.)

Now that I've defined two separate inner products, I should probably say why.  
Generally, we associate inner products with projections, but that turns out not to
be quite right.  Look back at Eq.~\eqref{bettervecproj} for the projection of $v$ along 
$u$; noting that the inner product of a scalar into
a vector is actually their product, you'll see Eq.~\eqref{bettervecproj} can also be written
\begin{equation}  
P_u(v) = (v \lin u) \lin u^{-1}.
\end{equation}
So orthogonal projection is actually a \emph{double} inner product.  The geometric 
meaning of a single inner product, as I'll show in Section \ref{inneroutergeom}, is this:  
for blades \bl{A_r} and \bl{B_s}, $\bl{A_r} \lin \bl{B_s}$ is also 
a blade, and it represents the subspace of vectors \emph{orthogonal to} \bl{A_r} 
and \emph{contained in} \bl{B_s}.  So the roles played by the factors in the inner 
product are not the same; that's why the product is asymmetric and there are 
two of them, so either factor can play either role.

Finally, I can use the new definition of the left inner product to generalize 
Theorem \ref{alinoutai}.
\begin{thm}
If $r \leq s$ then
\begin{equation}
B_r \lin (a_1 \out a_2 \out \dotsb \out a_s) = \sum
    (-1)^{\sum_{j=1}^r (i_j - j)} (B_r \lin a_{i_1} \out a_{i_2} \out 
    \dotsb \out a_{i_r}) \, a_{i_{r+1}} \out \dotsb \out 
    a_{i_s},
\label{mveclinoutid}
\end{equation}
where the sum is performed over all possible choices of $\{a_{i_j}\}_{j=1, \dotsc,r}$ 
out of $\{a_i\}_{i=1,\dotsc,s}$, and in each term $i_1$ through 
$i_r$ and $i_{r+1}$ through $i_s$ separately are in ascending order.  
\label{Brlinoutai}
\end{thm}
The coefficients in this sum can be remembered using the same trick used for
Theorem \ref{alinoutai}.  In that case, you imagine that you need to permute 
each vector to the far left in order to act on it with $a$.  For this theorem, you 
imagine that you need to permute each distinct subset of $r$ vectors to the far left, 
keeping them in their original order, in order to act on them with $B_r$.  In 
both cases, permuting vectors to the far left introduces a power of $-1$ equal to 
the required number of interchanges of adjacent vectors.  

Just as in Theorem \ref{alinoutai}, each inner product in the sum is a scalar.  Since 
each term picks $r$ elements out of a set $s$ elements, the sum has $\binom{s}{r}$ terms.
\bp
If the result is true for an $r$-blade, it is true for any $r$-vector, so 
assume \bl{B_r} is an $r$-blade.  Now I'll proceed by induction on $r$.  
If $r=1$ this becomes Theorem \ref{alinoutai}; so assume it's true for 
$r-1$.  Now $\bl{B_r}=a \out \bl{B_{r-1}}$ where $a$ is a vector and 
\bl{B_{r-1}} is an $r-1$-blade, so using the third of Eqs.~\eqref{associdents} 
I can write
\begin{align}
\bl{B_r} \lin (a_1 \out a_2 \out \dotsb \out a_s) & = (a \out \bl{B_{r-1}}) \lin (a_1 \out a_2 \out \dotsb \out a_s) \nonumber \\
 & = a \lin \left[\bl{B_{r-1}} \lin (a_1 \out a_2 \out \dotsb \out a_s)\right] 
 \nonumber \\
 & = \sum (-1)^{\sum_{j=1}^{r-1} (i_j - j)}
    \bl{B_{r-1}} \lin (a_{i_1} \out a_{i_2} \out \dotsb 
    \out a_{i_{r-1}}) \, \times \nonumber \\
 & \qquad \qquad a \lin \left[ a_{i_r} \out \dotsb \out 
    a_{i_s}\right] 
\end{align}
where $i_1$ through $i_{r-1}$ and $i_r$ through $i_s$ are in
ascending order separately.  Now I use Theorem \ref{alinoutai} to get
\begin{align}
\bl{B_r} \lin (a_1 \out a_2 \out \dotsb \out a_s) & = \sum (-1)^{\sum_{j=1}^{r-1} (i_j - j)} \bl{B_{r-1}} \lin (a_{i_1} 
     \out a_{i_2} \out \dotsb \out a_{i_{r-1}}) \, \times \nonumber \\
 & \qquad \left[\sum_{k=r}^s (-1)^{k-r} a \lin a_{i_k} \, a_{i_r} \out \dotsb \out \check{a}_{i_k} 
     \out \dotsb \out a_{i_s}\right].
\label{fullexpand}
\end{align}
This expression has $(s-r+1)\binom{s}{r-1}=r\binom{s}{r}$ terms, which
is too many by a factor of $r$, so it's time to do some grouping.

First, notice that each term is a scalar calculated using $r$ of the vectors,
multiplied by the outer product of the remaining $s-r$ vectors arranged in 
ascending order, and that every possible choice of $r$ vectors occurs.  That
means that for some choice of scalars $C(a_{i_1}, \dotsc, a_{i_r})$,
\begin{equation}
\bl{B_r} \lin (a_1 \out a_2 \out \dotsb \out a_s) = \sum C(a_{i_1}, \dotsc, a_{i_r}) \, a_{i_{r+1}} \out \dotsc \out a_{i_s} 
\end{equation}
where the sum is over all choices of $r$ vectors out of the set of $s$.
All that remains is to figure out what the coefficients $C(a_{i_1}, \dotsc, a_{i_r})$ 
are.  Well, for a given choice of $a_{i_1}$ through $a_{i_r}$, the coefficient
will include terms in which one of the $a_{i_j}$ is in the inner product with $a$ 
while the others are in the inner product with \bl{B_{r-1}}, or
\begin{equation}
C(a_{i_1}, \dotsc, a_{i_r}) = \sum_{j=1}^r (-1)^{\epsilon_j} \bl{B_{r-1}} \lin 
                                                 (a_{i_1} \out \dotsb \out \check{a}_{i_j} \out \dotsb \out a_{i_r}) \, a \lin a_{i_j}
\label{expandcoeff}
\end{equation}
for some value of $\epsilon_j$ for each $j$.  This sum has $r$ terms, one 
for each of the $a_{i_j}$, which is exactly the number I need, so now I need 
to figure out the exponents $\epsilon_j$.  Remember the mnemonic I've been using:
each vector's contribution to $\epsilon_j$ equals the difference between its position 
in the original outer product and the position to which it is moved to compute the inner 
product.  (You can verify that this is true for every vector in Eq.~\eqref{fullexpand} by 
inspection.)  So all we have to do is figure out those positions.  First consider every 
$a_{i_k}$ in the inner product with \bl{B_{r-1}} where $k < j$: each one is moved from 
position $i_k$ to position $k$, so it contributes $i_k - k$ to $\epsilon_j$.  Now consider 
the $a_{i_k}$ where $k > j$: each one is moved from position $i_k$ to position $k-1$ 
(because position $j$ is empty), so it contributes $i_k - k + 1$.  

Finally, let's take a look at $a_{i_j}$.  If its inner product is taken with $a$, then 
Eq.~\eqref{fullexpand} tells me that it is part of the second group of vectors.  Therefore
vectors $a_{i_{j+1}}$ through $a_{i_r}$ had to be moved to its left, moving it from its
original position $i_j$ ahead to $i_j - j + r$.  It is then moved to position $r$ for the inner 
product with $a$, so its contribution to $\epsilon_j$ is $i_j - j$.  Therefore 
\begin{align}
\epsilon_j & = \sum_{k=1}^{j-1} (i_k - k) + \sum_{k=j+1}^r (i_k - k + 1) + i_j - j \nonumber \\
                   & = \sum_{k=1}^r (i_k - k) + r - j.
\end{align}
Putting this in Eq.~\eqref{expandcoeff} gets me
\begin{align}
C(a_{i_1}, \dotsc, a_{i_r}) & = \sum_{j=1}^r (-1)^{ \sum_{k=1}^r (i_k - k) + r - j} \bl{B_{r-1}} \lin 
                       (a_{i_1} \out \dotsb \out \check{a}_{i_j} \out \dotsb \out a_{i_r}) \, a \lin a_{i_j} \nonumber \\
                                               & = (-1)^{ \sum_{k=1}^r (i_k - k)} (-1)^{r-1} \bl{B_{r-1}} \lin \left[ \sum_{j=1}^r (-1)^{1 - j} 
                       (a \lin a_{i_j}) \, a_{i_1} \out \dotsb \out \check{a}_{i_j} \out \dotsb \out a_{i_r} \right] \nonumber \\
                                               & = (-1)^{ \sum_{k=1}^r (i_k - k)} (-1)^{r-1} \bl{B_{r-1}} \lin \left[a \lin (a_{i_1} \out \dotsb \out a_{i_r})\right]
                       \nonumber \\
                                               & = (-1)^{ \sum_{k=1}^r (i_k - k)} (-1)^{r-1} (\bl{B_{r-1}} \out a) \lin (a_{i_1} \out \dotsb \out a_{i_r})
                       \nonumber \\
                                               & = (-1)^{ \sum_{k=1}^r (i_k - k)} (a \out \bl{B_{r-1}}) \lin (a_{i_1} \out \dotsb \out a_{i_r}) \nonumber \\
                                               & =  (-1)^{ \sum_{k=1}^r (i_k - k)} \bl{B_r} \lin (a_{i_1} \out \dotsb \out a_{i_r}).
\end{align}
Comparing this expression for $C(a_{i_1}, \dotsc, a_{i_r})$ with the statement of the theorem, 
I see that I've proved the result.
\ep
It's easy to verify that
\begin{equation}
\sum_{j=1}^r (i_j - j) = \sum_{j=1}^r (i_j - 1) - \frac{r(r-1)}{2},
\label{expdiff}
\end{equation}
so this is another way to write the exponent of $-1$ in the statement of the theorem.  
I'll use this later.

Now I can answer some questions left hanging in Section \ref{wherenow}.  I asked 
how you could tell what $n v n^{-1}$ and $v \out n n^{-1}$ are without knowing how 
they were derived.  (Notice I'm using order of operations to drop parentheses.)  First 
let's do $n v n^{-1}$; since this is proportional to $n v n$, I'll look at that instead.  We 
know now that the product of three vectors will in general 
by the sum of a vector and a trivector, and the trivector is the outer product of the
factors.  In this case the outer product is $n \out v \out n$, which vanishes 
because $n$ appears twice, so the product must be pure vector.

Next let's look at $v \out n n^{-1}$; again I'll consider $v \out n n$ because 
the answer will be the same.  The easiest thing to do here is expand the product
with the final $n$ into inner and outer products:
\begin{align}
v \out n n & = v \out n \rin n + v \out n \out n \nonumber \\
 & = v \out n \rin n
\end{align}
because the $v \out n \out n$ term vanishes.  Even if you didn't 
know the remaining term had to be a vector, you could figure it out because it starts 
as a vector and has one grade raising and one grade lowering operation applied to it.  

I also asked how to calculate $\left(v \out n n^{-1}\right) \lin n$ to verify that 
$v \out n n^{-1}$ really is perpendicular to $n$.  (Remember that $\inp$ has changed 
to $\lin$ since we got through Section \ref{axioms}.)  Since the inner product of 
vectors is just the scalar part of their geometric product,
\begin{align}
(v \out n n^{-1}) \lin n & = \grade{v \out n n^{-1} n} \nonumber \\
 & = \grade{v \out n} \nonumber \\
 & = 0
\end{align}
since $v \out n$ has no scalar part.  Easy, huh?

\subsection{The geometric meaning of the inner and outer products}
\label{inneroutergeom}

I've now accomplished all I set out to do in this section except for geometric interpretation.
First I'll handle the outer product.
\begin{thm}
Let \bl{A_r} and \bl{B_s} be nonzero blades where $r, s \geq 1$.
\begin{description}
\item[(a)] $\bl{A_r} \out \bl{B_s} = 0$ iff \bl{A_r} and \bl{B_s} share nonzero vectors.
\item[(b)] $\bl{A_r} \out \bl{B_s}$, if nonzero, represents the direct sum of the 
      corresponding subspaces.
\end{description}
\label{outerprodmean}
\end{thm}
\bp
This is true because, by the same reasoning used in Theorem \ref{Aissubspace}, 
$\bl{A_r} \out \bl{B_s} = 0$ iff the factors 
of \bl{A_r} and \bl{B_s} form a linearly dependent set, which is true iff the 
subspaces share a nonzero vector.  The fact that nonzero $\bl{A_r} \out \bl{B_s}$ 
represents the direct sum of \bl{A_r} and \bl{B_s} follows immediately, since
$a \out \bl{A_r} \out \bl{B_s} = 0$ iff $a$ is a linear combination of the 
factors of \bl{A_r} and \bl{B_s}.
\ep
So forming the outer product is equivalent to taking the direct sum, and it's nonzero iff
the direct sum can be taken.

Next let's move on to the inner product.  I already said at the end of Section 
\ref{geninnerouter} that the inner product combines inclusion in one subspace and
orthogonality to the other, and I need some terminology to more conveniently 
describe this.  In Section \ref{whatshere} I defined the orthogonal complement of 
a subspace: it's the set of all vectors orthogonal to every vector in the subspace, 
or equivalently the orthogonal complement of \bl{A_r} is the set of all vectors $a$ 
satisfying $a \lin \bl{A_r} = 0$.  Here's another definition: the \emph{orthogonal 
complement of \bl{A_r} in \bl{B_s}} is the intersection of \bl{B_s} and the orthogonal 
complement of \bl{A_r}.  Algebraically, $a$ is in the orthogonal complement of 
\bl{A_r} in \bl{B_s} iff $a \lin \bl{A_r} = 0$ and $a \out \bl{B_s} = 0$.  Now 
I'm ready to prove the result.
\begin{thm}
Let \bl{A_r} and \bl{B_s} be nonzero blades where $r, s \geq 1$.
\begin{description}
\item[(a)] $\bl{A_r} \lin \bl{B_s} = 0$ iff \bl{A_r} contains a nonzero vector orthogonal to \bl{B_s}.
\item[(b)] If $r < s$ then $\bl{A_r} \lin \bl{B_s}$, if nonzero, is an $s-r$-blade 
      representing the orthogonal complement of \bl{A_r} in \bl{B_s}.
\end{description}
\label{innerprodmean}
\end{thm}
Again, this is why two different inner products are defined; the 
geometric roles played by the two factors in the product aren't the same.  
In contrast, Theorem \ref{outerprodmean} shows that the roles played by the two 
factors in the outer product are the same, which is why there's only one outer 
product.  This also explains geometrically why $\bl{A_r} \lin \bl{B_s} = 0$ when $r > s$; 
as I show in the proof, if one subspace is higher-dimensional than other, the larger 
subspace always contains a nonzero vector that is orthogonal to the smaller one.
\bp
First I'll consider the case $r \leq s$; try to be surprised that the 
proof is by induction on $r$.  The $r=1$ result is taken care of by the proof of 
Theorem \ref{vecorthtoA} (see both the theorem and the discussion right after the proof),
so assume the results have been proved for $r-1$ and consider $\bl{A_r}
\lin \bl{B_s}$.  To prove part (a), let $a_1$ be any vector in \bl{A_r}; then for
some \bl{A_{r-1}} I can write $\bl{A_r} = \bl{A_{r-1}} \out a_1$, which means 
\begin{equation} 
\bl{A_r} \lin \bl{B_s} = \bl{A_{r-1}} \out a_1 \lin \bl{B_s} = \bl{A_{r-1}} \lin (a_1 \lin \bl{B_s}). 
\end{equation}
Suppose \bl{A_r} contains a vector orthogonal to \bl{B_s}; then let that vector 
be $a_1$, so $a_1 \lin \bl{B_s} = 0$, so $\bl{A_r} \lin \bl{B_s} = 0$.
For the converse, assume $\bl{A_r} \lin \bl{B_s} = 0$; then it follows that
$\bl{A_{r-1}} \lin (a_1 \lin \bl{B_s}) = 0$.  There are now three possibilities.  The 
first is $a_1 \lin \bl{B_s}=0$, in which case $a_1$ is orthogonal to \bl{B_s} and 
I'm done.  If not, then by the $r-1$ result \bl{A_{r-1}} contains a vector $a_2$ orthogonal 
to $a_1 \lin \bl{B_s}$; if that vector happens to be orthogonal to all of \bl{B_s} then 
I'm also done.  Now for the third case: $a_2$ is orthogonal to $a_1 \lin \bl{B_s}$ but 
not \bl{B_s}.  The proof of Theorem \ref{vecorthtoA} showed that I can factor \bl{B_s} as 
$b_1 \out \dotsb \out b_s$ where $ a_1 \lin b_1 \neq 0$ while the other $b_j$ are
orthogonal to $a_1$, so
\begin{equation} a_1 \lin \bl{B_s} = (a_1 \lin b_1) b_2 \out \dotsb \out b_s. \end{equation}
The only way $a_2$ can be orthogonal to $a_1 \lin \bl{B_s}$ but not \bl{B_s} is if
$a_2 \lin b_1 \neq 0$ while $a_2$ is orthogonal to all the other $b_j$.  In that case 
consider
\begin{equation} a = a_1 - \left(\frac{a_1 \lin b_1}{a_2 \lin b_1}\right) a_2. \end{equation} 
This vector lies in \bl{A_r}; it's nonzero because $a_1$ and $a_2$ are linearly independent; 
and it's orthogonal to all the $b_j$, including $b_1$, by construction.  Thus it's orthogonal to
\bl{B_s}.  So in all three cases \bl{A_r} contains a vector orthogonal to \bl{B_s}.
To prove part (b), assume $r < s$ and $\bl{A_r} \lin \bl{B_s} = \bl{A_{r-1}} \lin (a_1 \lin \bl{B_s}) 
\neq 0$.  Then by the $r-1$ result $\bl{A_r} \lin \bl{B_s}$ is the space of all vectors in 
$a_1 \lin \bl{B_s}$ that are orthogonal to \bl{A_{r-1}}.  However, by the $r=1$ 
result $a_1 \lin \bl{B_s}$ is the orthogonal complement of $a_1$ in \bl{B_s}, so 
a vector lies in $\bl{A_r} \lin \bl{B_s}$ iff it lies in \bl{B_s} and is orthogonal 
both to $a_1$ and to \bl{A_{r-1}}, and thus to all of \bl{A_r}.  This proves both 
parts for $r \leq s$. 

Now for the case $r > s$; $\bl{A_r} \lin \bl{B_s} = 0$ automatically, so I can forget 
about part (b) and I only need to show that \bl{A_r} always contains a vector 
orthogonal to \bl{B_s}.  Consider $\bl{B_s} \lin \bl{A_r}$; either it vanishes or it 
doesn't.  If it doesn't, then it contains vectors in \bl{A_r} orthogonal to \bl{B_s} and 
I'm done.  If it does vanish, then \bl{B_s} contains vectors orthogonal to \bl{A_r}; let 
\bl{B_p} represent the subspace of all such vectors.  If $p = s$, then \bl{B_s} is 
orthogonal to \bl{A_r} and I'm also done. If $p < s$, then $\bl{B_s} = \bl{B_p} \out 
\bl{B_{s-p}}$ where \bl{B_{s-p}} contains no vectors orthogonal to \bl{A_r}, 
which implies $\bl{B_{s-p}} \lin \bl{A_r} \neq 0$.  Then any vector in 
$\bl{B_{s-p}} \lin \bl{A_r}$ lies in \bl{A_r} and is orthogonal to \bl{B_{s-p}}; but 
by lying in \bl{A_r} the vector is already orthogonal to \bl{B_p}, so it's orthogonal 
to all of \bl{B_s} and the result is proved.
\ep

Incidentally, I'll show in Section \ref{reverse} that $A_r \rin B_s = 
(-1)^{r(s-1)} B_s \lin A_r$ for any $r$- and $s$-vectors, so the geometric 
interpretation of the right inner product is the same as the left product, but 
with the factors reversed, which certainly seems reasonable.

As a fun exercise, at this point you might look back at the identities in Eqs.~
\eqref{associdents} in the special case that $A$, $B$, and $C$ are blades and try 
to figure out the geometrical meaning of each one.

To explain the next few theorems, I need some facts about blade \bl{A} to be 
proved in Section \ref{scalarprod}.
\begin{enumerate}
\item $\bl{A}^2$ is a scalar.
\item \bl{A} is invertible iff $\bl{A}^2 \neq 0$, and $\bl{A}^{-1}=\bl{A}/\bl{A}^2$.  
Therefore \bl{A} and $\bl{A}^{-1}$ represent the same subspace.
\item \bl{A} is invertible iff the inner product is nondegenerate on \bl{A}.
\end{enumerate}

The inner and outer products of blades can shed light on how their subspaces
are related.  For example, if one subspace lies inside another, their blades are 
related as follows.
\begin{thm}
Let \bl{A_r} and \bl{B_s} be nonzero blades where $1 \leq r \leq s$.
\begin{description}
\item[(a)] If \bl{A_r} is a subspace of \bl{B_s}, then $\bl{A_r} \bl{B_s} = \bl{A_r} \lin \bl{B_s}$.
\item[(b)] The converse is true if either (1) $r=1$ or $s=1$ or (2) \bl{A_r} or 
      \bl{B_s} is invertible.
\end{description}
\label{AsubspaceofB}
\end{thm}
\bp
First let $r=1$; then $a \bl{B_s} = a \lin \bl{B_s}$ iff $a \out \bl{B_s} = 0$, which 
is true iff $a$ belongs to \bl{B_s}.  Now assume the result is true for $r-1$ 
and let \bl{A_r} be a subspace of \bl{B_s}; I can write $\bl{A_r} = \bl{A_{r-1}} a$ for some 
vector $a$ such that $a$ and \bl{A_{r-1}} are orthogonal, so the $r=1$ result
lets me write
\begin{equation} 
\bl{A_r} \bl{B_s} = \bl{A_{r-1}} a \bl{B_s} = \bl{A_{r-1}} a \lin \bl{B_s}. 
\end{equation}
Now $a \lin \bl{B_s}$ is the orthogonal complement of $a$ in \bl{B_s}, which means 
$a \lin \bl{B_s}$ contains \bl{A_{r-1}}, so by the $r-1$ result
\begin{align}
\bl{A_r} \bl{B_s} & = \bl{A_{r-1}} a \lin \bl{B_s} = \bl{A_{r-1}} \lin (a \lin \bl{B_s}) \nonumber \\
 & = (\bl{A_{r-1}} \out a) \lin \bl{B_s} \nonumber \\
 & = \bl{A_r} \lin \bl{B_s},
\end{align}
which is the desired result.  (If $a \lin \bl{B_s} = 0$ then $\bl{A_r} \lin \bl{B_s} = 0$ 
also by Theorem \ref{innerprodmean}, 
so if one side vanishes then so does the other.)  I've already shown the 
converse is true when $r=1$, and $s=1$ 
implies $r=1$, so assume \bl{A_r} is invertible; then $\bl{A_r} \bl{B_s} = \bl{A_r} \lin \bl{B_s}$ 
implies $\bl{B_s} = \bl{A_r}^{-1} \bl{A_r} \lin \bl{B_s}$.  By assumption $\bl{B_s} \neq 0$, so 
$\bl{A_r} \lin \bl{B_s} \neq 0$ also.  Therefore if $r=s$, then \bl{B_s} is just a nonzero 
multiple of $\bl{A_r}^{-1}$.  Since \bl{A_r} and $\bl{A_r}^{-1}$ represent the same subspace, 
\bl{A_r} and \bl{B_s} also represent the same subspace.  If $r < s$, then since \bl{B_s} is an
$s$-vector, $\bl{A_r}^{-1} \bl{A_r} \lin \bl{B_s}$ must be too; but only its highest
grade term, its outer product, has grade $s$, so for this relation to
hold the product must equal the outer product, so $\bl{B_s} = \bl{A_r}^{-1}
\out (\bl{A_r} \lin \bl{B_s})$.  Therefore \bl{B_s} is the direct sum of $\bl{A_r}
\lin \bl{B_s}$ and $\bl{A_r}^{-1}$; but $\bl{A_r}^{-1}$ represents the same subspace as
\bl{A_r}, so \bl{A_r} is obviously a subspace of \bl{B_s}.  The proof when \bl{B_s} 
is invertible is similar; $\bl{A_r} \bl{B_s} = \bl{A_r} \lin \bl{B_s}$ implies $\bl{A_r} = 
\bl{A_r} \lin \bl{B_s} \bl{B_s}^{-1}$, so if $r=s$, \bl{A_r} is a nonzero multiple of 
\bl{B_s}.  If $r < s$, then since \bl{A_r} is an $r$-vector, $\bl{A_r} \lin \bl{B_s} 
\bl{B_s}^{-1}$ must be too; but only its lowest grade term, the inner product, 
has grade $r$, so the product must equal the inner product, so $\bl{A_r} = 
(\bl{A_r} \lin \bl{B_s}) \lin \bl{B_s}^{-1}$.  Therefore, \bl{A_r} is a subspace of 
$\bl{B_s}^{-1}$, and thus of \bl{B_s}.
\ep

Another possible relationship is orthogonality: two subspaces are orthogonal if
every vector in one is orthogonal to every vector in the other.  In that case,
their blades are related as follows.
\begin{thm}
Let \bl{A_r} and \bl{B_s} be nonzero blades where $r, s \geq 1$.
\begin{description}
\item[(a)] If \bl{A_r} and \bl{B_s} are orthogonal, then $\bl{A_r} \bl{B_s} = \bl{A_r} \out \bl{B_s}$.
\item[(b)] The converse is true if either (1) $r=1$ or $s=1$ or (2) \bl{A_r} or \bl{B_s} is invertible.
\end{description}
\label{AorthtoB}
\end{thm}
\bp
To begin, I note that \bl{A_r} can be written $a_1 a_2 \dotsb a_r$ 
where the $a_i$ are orthogonal to each other; similarly, \bl{B_s} can be 
expressed $b_1 b_2 \dotsb b_s$ where the $b_j$ are also orthogonal to each 
other.  Now suppose that \bl{A_r} and \bl{B_s} are orthogonal; then all of the 
$a_i$ and $b_j$ are orthogonal to each other as well, so now I can use 
the rule that the product of orthogonal vectors equals their outer product 
to get
\begin{align}
\bl{A_r} \bl{B_s} & = a_1 a_2 \dotsb a_r b_1 b_2 \dotsb b_s \nonumber \\
 & = a_1 \out a_2 \out \dotsb \out a_r \out b_1 \out b_2 \out 
                    \dotsb \out b_s \nonumber \\
 & = \bl{A_r} \out \bl{B_s}.
\end{align}
To prove the converse, first let $r=1$; then $a \bl{B_s} = a \out \bl{B_s}$ iff 
$a \lin \bl{B_s} = 0$, which is true iff $a$ is orthogonal to \bl{B_s}.  Now 
assume \bl{A_r} is invertible and let $\bl{A_r} \bl{B_s} = \bl{A_r} \out \bl{B_s}$; then 
$\bl{B_s} = \bl{A_r}^{-1} \bl{A_r} \out \bl{B_s}$.  Since $\bl{A_r}^{-1}$ represents the same 
subspace as \bl{A_r}, which is a subspace of $\bl{A_r} \out \bl{B_s}$, it follows 
that $\bl{B_s} = \bl{A_r}^{-1} \lin (\bl{A_r} \out \bl{B_s})$, so \bl{B_s} is orthogonal to 
\bl{A_r}.  The proof when $s=1$ or \bl{B_s} is invertible proceeds similarly.
\ep

You may be surprised that the converse parts of these theorems aren't 
generally true; let me give an example to show why not.  Consider an algebra with 
orthogonal vectors $e_1, e_2$, $e_3$, and $e_4$ such that $e_2$ is null 
but the others aren't.  (These four vectors define only a subspace of the 
full space of vectors, or Axiom \ref{nondeg} would be violated.)  Let 
$\bl{A_3} = e_1 e_2 e_3$ and $\bl{B_3} = e_2 e_3 e_4$; then $e_2^2 = 0$ implies 
$\bl{A_3} \bl{B_3} = 0$, so $\bl{A_3} \lin \bl{B_3} = \bl{A_3} \out \bl{B_3} = 0$ also.  
However, neither \bl{A_3} nor \bl{B_3} is a subspace of the other, so the converse 
part of Theorem \ref{AsubspaceofB} doesn't hold, and the subspaces are not orthogonal 
to each other (they both contain non-null $e_3$), so the converse part of Theorem 
\ref{AorthtoB} doesn't hold either.  Null vectors make life hard sometimes.

Incidentally, $A_r B_s = A_r \lin B_s$ implies $B_s A_r = B_s \rin A_r$ for any 
$r$- and $s$-vectors, and the same is true if the inner product is replaced with 
the outer product (this follows from Eq.~\eqref{revex}), so the last two theorems
don't depend on the order of \bl{A_r} and \bl{B_s}.  Really, it would be weird if they did.

Let me end with a result that combines the previous few theorems in an interesting way.
Suppose \bl{A_r} is a subspace of \bl{B_s}; then it seems plausible that \bl{B_s} should be 
the direct sum of \bl{A_r} and its orthogonal complement in \bl{B_s}.  (For example, 
three-dimensional Euclidean space is the direct sum of the $z$ axis and its orthogonal 
complement, the $xy$ plane.)  Using our theorems, that suggests something like $\bl{B_s} = 
\bl{A_r} \out (\bl{A_r} \lin \bl{B_s})$.  Now that can't be right as it stands because the result 
shouldn't depend on the weight of \bl{A_r}, just its attitude.  That's easy to fix, though: 
maybe $\bl{B_s} = \bl{A_r} \out (\bl{A_r}^{-1} \lin \bl{B_s})$ instead.  That turns out to be right,
but with some caveats.  Let me prove a theorem I need first.
\begin{thm}
If $1 \leq r \leq s$ and nonzero blades \bl{A_r} and \bl{B_s} satisfy $\bl{A_r} \bl{B_s} = \bl{A_r} 
\lin \bl{B_s}$, then $\bl{A_r}^2 \bl{B_s} = \bl{A_r} \out (\bl{A_r} \lin \bl{B_s})$.  
\label{miscident}
\end{thm}
\bp
Believe it or not, this result can be proved without induction.  Suppose the 
condition is true; then
\begin{equation}
\bl{A_r}^2 \bl{B_s} = \bl{A_r} (\bl{A_r} \bl{B_s}) = \bl{A_r} (\bl{A_r} \lin \bl{B_s}).
\end{equation}
Since $\bl{A_r}^2$ is a number, the left hand side is an $s$-vector; so the right hand side 
must be also.  Using the same logic as in the proof of Theorem \ref{AsubspaceofB}, the 
product on the right hand side must equal the outer product, and that proves the result.
\ep
To interpret this theorem, suppose \bl{A_r} is a subspace of \bl{B_s}, so $\bl{A_r} \bl{B_s} 
= \bl{A_r} \lin \bl{B_s}$ by Theorem \ref{AsubspaceofB}.  Now either \bl{A_r} is invertible 
or it's not; first suppose it is.  Then $\bl{A_r}^2 \neq 0$, so taking the result of Theorem 
\ref{miscident} and dividing by $\bl{A_r}^2$ yields $\bl{B_s} = \bl{A_r} \out (\bl{A_r}^{-1} 
\lin \bl{B_s})$, which is the result I was after.  Notice I had to assume \bl{A_r} was invertible 
to get this, though.  What if it isn't?  In that case \bl{A_r} contains a nonzero vector 
orthogonal to all of \bl{A_r}, so one of two things can happen: either that vector 
is also orthogonal to all of \bl{B_s}, so $\bl{A_r} \lin \bl{B_s} = 0$, or it isn't, in 
which case that vector also lies in $\bl{A_r} \lin \bl{B_s}$, so $\bl{A_r} \out (\bl{A_r} 
\lin \bl{B_s}) = 0$. Since $\bl{A_r}^2 = 0$, the theorem nicely includes both of 
those cases.

\section{Other operations}
\label{otherops}

Our computational powers have grown by leaps and bounds, but they're not yet complete.
Some more operations will be useful to us later, so I'll describe them all here.  Please be
aware that different authors use different symbols for some of these operations; I've listed 
all my symbol choices in Appendix \ref{app:summary}.

\subsection{Grade involution}
\label{gradeinv}

The first has the formidable name \emph{grade involution}.  
It is represented by an \grinv{} and defined as follows.
\begin{align}
\grinv{\lambda} & := \ \ \lambda \nonumber \\
\grinv{a} & := -a \nonumber \\
\grinv{(AB)} & := \grinv{A} \grinv{B} \nonumber \\
\grinv{(A + B)} & := \grinv{A} + \grinv{B}.
\label{grinvrules} 
\end{align}
This operation takes reflection of vectors through the origin (the second line 
in the definition), sometimes called the parity operation, and extends it to the 
whole algebra.  From these rules it follows that
\begin{equation}
\grinv{(a_1 a_2 \dotsb a_r)} = (-1)^r a_1 a_2 \dotsb a_r,
\end{equation}
which implies 
\begin{equation} \grinv{A_r} = (-1)^r A_r, \label{grinvr} \end{equation}
so grade involution leaves even grades alone while changing the sign 
of odd-grade multivectors, or
\begin{equation} \grinv{A} = \grade[+]{A} - \grade[-]{A}.  \end{equation}
This is equivalent to the occasionally handy result
\begin{equation}
\grade[\pm]{A} = \half (A \pm \grinv{A}).
\label{evenoddfromgrinv}
\end{equation}
Eq.~\eqref{grinvr} tells me that
\begin{equation}
\grinv{\grade[r]{A}} = \grade[r]{\grinv{A}},
\end{equation}
so grade involution commutes with taking the grade-$r$ part, and
\begin{equation} \doublegrinv{A} = A \end{equation} 
for any multivector $A$.  (That's what makes it an involution.)  Suppose $A$ is invertible;
then since $\grinv{(A^{-1})}\grinv{A} = \grinv{(A^{-1} A)} = \grinv{1} = 1$,
\begin{equation} \grinv{(A^{-1})} = (\grinv{A})^{-1}. \end{equation}
By projecting onto terms of appropriate grade, the third rule in the definition becomes
\begin{align}
\grinv{(A \lin B)} & = \grinv{A} \lin \grinv{B} \nonumber \\
\grinv{(A \rin B)} & = \grinv{A} \rin \grinv{B} \nonumber \\ 
\grinv{(A \out B)} & = \grinv{A} \out \grinv{B}.
\label{invprods}
\end{align}
Formulas with factors of $(-1)^r$ are usually simplified by 
grade involution.  For example, Eqs.~\eqref{vecAinner} and \eqref{vecAouter} 
for the inner and outer products of vector $a$ with multivector $A$ become
\begin{align}
a \lin A & = \half(a A - \grinv{A} a) \nonumber \\ 
a \out A & = \half(a A + \grinv{A} a),
\end{align}
the definitions of $A \rin a$ and $A \out a$ from Eqs.~\eqref{Avecinner} and 
\eqref{Avecouter} become
\begin{align}
A \rin  a & = -a \lin \grinv{A} \nonumber \\
A \out a & = \ \ a \out \grinv{A},
\end{align}
and finally the identities from Eqs.~\eqref{usefulids} and \eqref{contwedgeids} can 
now be written
\begin{align}
a \lin (A B) & = (a \lin A) B + \grinv{A} (a \lin B) \nonumber \\
                   & = (a \out A) B - \grinv{A} (a \out B) \nonumber \\
a \out (A B) & = (a \out A) B - \grinv{A} (a \lin B) \nonumber \\
                   & = (a \lin A) B + \grinv{A} (a \out B).
\end{align}
\begin{align}
a \lin (A \out B) & = (a \lin A) \out B + \grinv{A} \out (a \lin B) \nonumber \\
a \out (A \rin B) & = (a \out A) \rin B - \grinv{A} \rin (a \lin B) \nonumber \\
a \out (A \lin B) & = (a \lin A) \lin B + \grinv{A} \lin (a \out B).
\end{align}
Every once in a while I'll use \grinv[r]{} to indicate grade involution taken $r$ 
times.  $\grinv[r]{A}$ equals $A$ if $r$ is even and $\grinv{A}$ if $r$ is odd.

This operation is called ``inversion'' by some authors, but that can be confused 
with the multiplicative inverse, which would be bad because both are important and
are used frequently (sometimes at the same time).

A blade and its grade involution represent the same subspace since 
each is a multiple of the other by Eq.~\eqref{grinvr}.

\subsection{Reversion}
\label{reverse}

The second operation is called \emph{reversion} or \emph{taking the
reverse} and is represented by a \rev{}. It's a little more complicated,
and it's defined as follows.
\begin{align}
\rev{\lambda} & := \lambda \nonumber \\
\rev{a} & := a \nonumber \\
\rev{(AB)} & := \rev{B} \rev{A} \nonumber \\
\rev{(A + B)} & := \rev{A} + \rev{B}.
\label{revrules} 
\end{align}
From this it follows that, for example, 
\begin{equation}
\rev{(a_1 a_2 \dotsb a_r)} = a_r \dotsb a_2 a_1,
\end{equation}
which shows that the reverse of any multivector is found by writing it
as a sum of blades and reversing the order of the vectors in each
blade.  Hence the name.  This also implies
\begin{equation} \doublerev{A} = A \end{equation} 
for any multivector $A$.  So reversion is also an involution.

Let $\{e_i\}_{i=1,\dotsc,r}$ be an anticommuting set; then
\begin{align}
\rev{(e_1 e_2 \dotsb e_r)} & = e_r \dotsb e_2 e_1 \nonumber \\
 & = (-1)^{r(r-1)/2} \, e_1 e_2 \dotsb e_r
\end{align}
because $r(r-1)/2$ interchanges are needed to return the vectors to 
their original order; therefore
\begin{equation} \rev{A_r} = (-1)^{r(r-1)/2} A_r. \label{revA} \end{equation}
If you evaluate this expression for different $r$, you quickly find (a) multivectors 
two grades apart behave oppositely under reversion, and (b) two adjacent 
grades behave the same under reversion iff the lower grade is even (scalars 
and vectors, for example).  Since the effect of either reversion or grade involution 
is to change signs of a multivector grade by grade, these operations commute:  
\begin{equation} \grinvrev{A} = \revgrinv{A}.  \end{equation}
Eq.~\eqref{revA} also shows that
\begin{equation}
\rev{\grade[r]{A}} = \grade[r]{\rev{A}},
\end{equation}
so taking the reverse commutes with taking the grade-$r$ part, just as grade
involution does.  Suppose $A$ is invertible;
then since $\rev{(A^{-1})}\rev{A} = \rev{(A A^{-1})} = \rev{1} = 1$,
\begin{equation} \rev{(A^{-1})} = (\rev{A})^{-1}. \end{equation}

By projecting onto terms of appropriate grade, the third 
rule in the definition becomes
\begin{align}
\rev{(A \lin B)} & = \rev{B} \rin \rev{A} \nonumber \\
\rev{(A \rin B)} & = \rev{B} \lin \rev{A} \nonumber \\
\rev{(A \out B)} & = \rev{B} \out \rev{A}.
\label{revprods}
\end{align}
Among other things, this shows that the right inner product can be defined in terms 
of the left inner product and reversion and is thus technically redundant.  Oh well.

Here's an easy application of Eq.~\eqref{revA}:
\begin{align}
\grade[r]{A B} & = (-1)^{r(r-1)/2} \grade[r]{\rev{(A B)}} \nonumber \\
 & = (-1)^{r(r-1)/2} \grade[r]{\rev{B} \rev{A}}
\end{align}
with the nice special case
\begin{equation} \grade{A B} = \grade{\rev{B} \rev{A}}. \label{invrev} \end{equation}
The even more special case where $A$ and $B$ are homogeneous is also useful.  The
product of two homogeneous multivectors of different grades doesn't have a 
scalar part, so $\grade{A_r B_s} = \grade{B_s A_r} = 0$ if $r \neq s$, and when $r = s$ we get
\begin{align}
\grade{A_r B_r} & = \grade{\rev{B_r} \rev{A_r}} \nonumber \\
 & = (-1)^{r(r-1)/2}\,(-1)^{r(r-1)/2} \grade{B_r A_r} \nonumber \\
 & = \grade{B_r A_r}.
\end{align}
Therefore $\grade{A_r B_s} = \grade{B_s A_r}$ in general.  Since the geometric
product is distributive, I can go all the way to
\begin{equation}
\grade{A B} = \grade{B A}
\label{inv}
\end{equation}
for any $A$ and $B$, or better yet
\begin{equation} 
\grade{A B \dotsb C D} = \grade{D A B \dotsb C}, 
\label{cyclic}
\end{equation}
so the scalar part of a product is cyclic in its factors.  This is
very useful.  Eqs.~\eqref{invrev} and \eqref{inv} together also imply
\begin{equation} \grade{A B} = \grade{\rev{A} \rev{B}}. \end{equation}

Almost every identity involving reverses is proved by successively
applying Eq.~\eqref{revA} and using the rules in Eqs.~\eqref{revrules}.  
For example, let's examine a generic term in $A_r B_s$:
\begin{align}
\grade[r+s-2j]{A_r B_s} & = (-1)^{(r+s-2j)(r+s-2j-1)/2} 
                            \grade[r+s-2j]{\rev{(A_r B_s)}} \nonumber \\
 & = (-1)^{(r+s-2j)(r+s-2j-1)/2} \grade[r+s-2j]{\rev{B_s} \rev{A_r}} 
     \nonumber \\
 & = (-1)^{(r+s-2j)(r+s-2j-1)/2}(-1)^{r(r-1)/2}(-1)^{s(s-1)/2} 
     \grade[r+s-2j]{B_s A_r} \nonumber \\
 & = (-1)^{rs-j} \grade[r+s-2j]{B_s A_r}.
\label{revex}
\end{align}
So multiplication may not commute, but $A_r B_s$ and $B_s A_r$ aren't 
totally unrelated; term by term, they're actually equal up to signs.  (Also, 
successive terms of $A_r B_s$, whose grades differ by $2$, have opposite 
behavior under reversion, which I expected given what I said after Eq.~\eqref{revA}.)
This result also has two important special cases.  First, suppose $r \leq s$; 
then Eq.~\eqref{revex} with $j=r$ refers to the lowest grade term, so
\begin{equation} 
A_r \lin B_s = (-1)^{r(s-1)} B_s \rin A_r.
\label{commuteinner}
\end{equation} 
Notice that this expression also holds when $r > s$ because both sides 
vanish.  So the inner product of an odd-grade multivector into an even-grade
multivector anticommutes (with left changing to right and vice 
versa), as in
\begin{equation} A_- \lin B_+ = -B_+ \rin A_-, \end{equation}
but in all other cases it commutes.  Without any restrictions on $r$ and $s$, 
Eq.~\eqref{revex} when $j=0$ gives for the highest grade term
\begin{equation} 
A_r \out B_s = (-1)^{rs} B_s \out A_r, 
\label{commuteouter}
\end{equation}
so the outer product of two odd-grade multivectors anticommutes like so,
\begin{equation} A_- \out B_- = -B_- \out A_-, \end{equation}
with all other cases commuting.  (These last few results are equivalent to 
Eq.~\eqref{revprods}, by the way.)

The properties of objects under reversion are sometimes helpful in sorting out 
their grades.  As an example, let me reconsider the product $n v n$ of three vectors
from Section \ref{wherenow}.  Notice that $\rev{(n v n)} = n v n$.  Now vectors don't 
change sign under reversion but trivectors do.  Therefore $n v n$ has no trivector
component and is pure vector.   

A blade and its reverse represent the same subspace since each is a multiple of the 
other by Eq.~\eqref{revA}.

\subsection{Clifford conjugation}
\label{cliffordconj}

The third involution in a geometric algebra is called \emph{Clifford conjugation} or 
\emph{taking the Clifford conjugate}.  It's represented by a \clifconj{} and defined as 
follows:
\begin{align}
\clifconj{\lambda} & := \ \ \lambda \nonumber \\
\clifconj{a} & := -a \nonumber \\
\clifconj{(AB)} & := \clifconj{B} \clifconj{A} \nonumber \\
\clifconj{(A + B)} & := \clifconj{A} + \clifconj{B}.
\label{clifconjrules} 
\end{align}
If this looks like a mixture of grade involution and reversion, that's because 
it is; in fact,
\begin{equation} \clifconj{A} = \grinvrev{A}. \end{equation}
This immediately tells us that Clifford conjugation really is an involution,
\begin{equation} \doubleclifconj{A} = A, \end{equation}
that it commutes with taking the grade-$r$ part,
\begin{equation} \clifconj{\grade[r]{A}} = \grade[r]{\clifconj{A}}, \end{equation}
that when $A$ is invertible
\begin{equation} \clifconj{(A^{-1})} = (\clifconj{A})^{-1}, \end{equation}
and finally
\begin{align}
\clifconj{(A \lin B)} & = \clifconj{B} \rin \clifconj{A} \nonumber \\
\clifconj{(A \rin B)} & = \clifconj{B} \lin \clifconj{A} \nonumber \\
\clifconj{(A \out B)} & = \clifconj{B} \out \clifconj{A}.
\end{align}
The Clifford conjugate of an $r$-vector is given by
\begin{align}
\clifconj{A_r} & = \grinvrev{A_r} \nonumber \\
  & = (-1)^r (-1)^{r(r-1)/2} A_r \nonumber \\
  & = (-1)^{r(r+1)/2} A_r.
\label{clifconjA}
\end{align}
This looks a lot like reversion.  If you evaluate this for different $r$ you find that
(a) multivectors two grades apart behave oppositely under Clifford conjugation, just
as with reversion, but (b) two adjacent grades behave the same under Clifford conjugation 
iff the lower grade is \emph{odd}, not even (vectors and bivectors, for example). 
So Clifford conjugation resembles reversion with grades shifted by $1$, so to speak.

A blade and its Clifford conjugate represent the same subspace since each is a multiple of the 
other by Eq.~\eqref{clifconjA}.

\subsection{The scalar product}
\label{scalarprod}

Next is the \emph{scalar product}, defined by
\begin{equation} \scprod{A}{B} := \grade{\rev{A} B}. \end{equation}
(Some authors define $\scprod{A}{B} = \grade{A B}$.  I'll tell you why 
I don't shortly.)  First consider the scalar product of
homogeneous multivectors.  Only the lowest-grade term in the product (the inner 
product) has any chance of being a scalar, so it's certainly true that
\begin{equation}
\scprod{A_r}{B_s} = \grade{\rev{A_r} \lin B_s} = \grade{\rev{A_r} \rin B_s}.
\label{scin}
\end{equation}
Since the scalar product and inner products are distributive by construction, it 
follows that
\begin{equation}
\scprod{A}{B} = \grade{\rev{A} \lin B} = \grade{\rev{A} \rin B}
\end{equation}
for any multivectors.  Now of course we actually know a little more than Eq.~\eqref{scin} 
lets on.  Only the product of two equal-grade homogeneous multivectors has a scalar 
part, so
\begin{equation}
\scprod{A_r}{B_s} = (\rev{A_r} \lin B_s) \, \delta_{rs} = (\rev{A_r} \rin B_s) \, \delta_{rs}.
\end{equation}
Therefore, homogeneous multivectors of different grades are orthogonal under the 
scalar product.  That means the scalar product of two general multivectors may also be
written
\begin{align}
\scprod{A}{B} & = \sum_r \scprod{A_r}{B_r} \nonumber \\
& = \sum_r \rev{A_r} \lin B_r = \sum_r \rev{A_r} \rin B_r.
\label{scalarproddecomp} 
\end{align}  
This also makes it clear that
\begin{equation} \scprod{A}{B} = \scprod{\grinv{A}}{\grinv{B}}. \label{grinvscprod} \end{equation}
My results for reversion and Clifford conjugation also establish some properties of this product; 
for example,
\begin{equation}
\scprod{A}{B} = \scprod{B}{A} = \scprod{\rev{A}}{\rev{B}} = \scprod{\clifconj{A}}{\clifconj{B}}
\label{revscprod}
\end{equation}
The next to last equality shows that an equivalent definition of the scalar product is 
\grade{A\rev{B}}.  The scalar product interacts with the other products we know this way.
\begin{thm}
\begin{align}
\scprod{A}{(BC)} & = \scprod{(\rev{B} A)}{C} \nonumber \\
\scprod{A}{(B \rin C)} & = \scprod{(\rev{B} \rin A)}{C} \nonumber \\
\scprod{A}{(B \lin C)} & = \scprod{(\rev{B} \out A)}{C} \nonumber \\
\scprod{A}{(B \out C)} & = \scprod{(\rev{B} \lin A)}{C}
\label{scprodinnerouter}
\end{align}
\label{scprodinoutidents}
\end{thm}
\bp
The first identity is proved as follows:
\begin{align}
\scprod{A}{(BC)} & = \grade{\rev{A} BC} \nonumber \\
                 & = \scprod{\rev{(\rev{A} B)}}{C} \nonumber \\
                 & = \scprod{(\rev{B} A)}{C}.
\end{align}
The remaining three are proved roughly the same way, so I'll prove 
only the first one.  Using Eqs.~\eqref{scalarproddecomp} and 
\eqref{revprods} and the second of Eqs.~\eqref{associdents},
\begin{align}
\scprod{A}{(B \rin C)} & = \grade{\rev{A} \lin (B \rin C)} \nonumber \\
                       & = \grade{(\rev{A} \lin B) \rin C} \nonumber \\
                       & = \scprod{\rev{(\rev{A} \lin B)}}{C} \nonumber \\
                       & = \scprod{(\rev{B} \rin A)}{C}.
\end{align}
\ep

The last of Eqs.~\eqref{scprodinnerouter} is the basis for a different approach to 
geometric algebra, followed for example in \cite{GAforCS}.  You start by defining the 
outer product and the scalar product; then you decide you'd like to be able to factor the 
term $B$ out of expressions like $\scprod{A}{(B \out C)}$.  You do this by defining an 
inner product that obeys the last of Eqs.~\eqref{scprodinnerouter}.  Then you define 
the geometric product of two vectors to be the sum of their inner and outer products,
and you're off and running.  This has the advantage that it starts with two products that
have clearly separated geometric functions: the outer product builds subspaces out of 
vectors, and the scalar product carries all the metric information.  It's thus more congenial
to the point of view inherent in differential forms, which are built using only an outer 
product and which clearly separate metric and non-metric properties.  Personally, I 
think it's cleaner to start with the fundamental product and define every other product 
directly in terms of it, which is why I follow the approach given here.

I use the scalar product to define the \emph{magnitude} or \emph{norm}
of a multivector by
\begin{equation}
|A|^2 := \scprod{A}{A}.
\end{equation}
$A$ is said to be \emph{null} if $|A|^2=0$ and a \emph{unit} multivector if
$|A|^2=\pm1$.  (Despite the notation, $|A|^2$ can be negative.  In fact, $|A|^2$ can be
all sorts of things, since the scalars aren't necessarily real numbers.)  Eqs.~\eqref{grinvscprod} 
and \eqref{revscprod} imply 
\begin{equation} |A|^2 = |\grinv{A}|^2 = |\rev{A}|^2 = |\clifconj{A}|^2.  \end{equation}
I define $|A|^n$ for other powers $n$ in the obvious way as a power of $|A|^2$, but 
due care should be taken that the power in question is well-defined (for example, 
be careful if $|A|^2$ is negative).  

The squared magnitude of a scalar or vector is just its square, and that result 
can be generalized a bit.  Suppose $A$ is an $r$-versor, so it is a product 
$a_1 a_2 \dotsb a_r$; then
\begin{align}
|A|^2 & = \scprod{A}{A} = \grade{\rev{A} A} \nonumber \\
 & = \grade{\rev{(a_1 a_2 \dotsb a_r)} a_1 a_2 \dotsb a_r} \nonumber \\
 & = \grade{a_r \dotsb a_2 a_1 a_1 a_2 \dotsb a_r} \nonumber \\
 & = a_1^2 a_2^2 \dotsb a_r^2.
\label{prodnorm}
\end{align}
Therefore $|A|^2$ is the product of the squares of its factors.  (This is why I
included the reverse in the definition.)  Notice also that if $A$ is a versor then 
$\rev{A} A$ also equals $|A|^2$.  This gives me a couple of useful results.

First, versors can be factored out of scalar products in an interesting way.
\begin{thm}
Versor $A$ and general multivectors $B$ and $C$ obey
\begin{equation}
\scprod{(AB)}{(AC)} = \scprod{(BA)}{(CA)} = |A|^2 \, \scprod{B}{C}.
\end{equation}
Therefore if either $A$ or $B$ is a versor, 
\begin{equation} |AB|^2 = |A|^2 |B|^2. \end{equation}
\label{facversorfromscprod}
\end{thm}
\bp
\begin{align}
\scprod{(AB)}{(AC)} & = \grade{\rev{(AB)} AC} \nonumber \\
                                    & = \grade{\rev{B} \rev{A} AC} \nonumber \\
                                    & = \grade{|A|^2 \, \rev{B} C} \nonumber \\
                                    & = |A|^2 \, \scprod{B}{C}
\end{align}
and a similar argument using Eq.~\eqref{cyclic} proves the other part of the equation.
The second part follows by setting $B=C$.
\ep

Second, versors are easy to invert.
\begin{thm}
Versor $A$ is invertible iff it's non-null, its inverse is given by
\begin{equation} 
A^{-1} = \frac{\rev{A}}{|A|^2}, 
\label{versorinv}
\end{equation}
and the squared norm of the inverse is given by
\begin{equation} |A^{-1}|^2 = |A|^{-2}. \end{equation}
\label{inviffnonnull}
\end{thm}
\bp
If $|A| \neq 0$, then clearly Eq.~\eqref{versorinv} gives an inverse 
of $A$, so $A$ must be invertible.  Conversely, suppose $A$ is invertible; 
then there exists a $B$ such that $A B = 1$.  Then it follows that 
$\rev{B} \rev{A} = 1$ also, so
\begin{align}
1 & = \rev{B} \rev{A} A B \nonumber \\
  & = |A|^2 \rev{B} B, 
\end{align}
so $|A| \neq 0$; thus a product of vectors is invertible iff it's non-null, 
and its inverse is given by the above expression.  For the squared norm,
just calculate $|A^{-1}|^2$ using Eq.~\eqref{versorinv}.
\ep

An $r$-blade \bl{A_r} is a special type of $r$-versor, so these theorems apply to 
blades too.  But for blades, a few more results are also true.  Since $\rev{\bl{A_r}} = 
(-1)^{r(r-1)/2} \bl{A_r}$, $|A|^2 = \rev{A}A$ becomes $|\bl{A_r}|^2 = (-1)^{r(r-1)/2}\bl{A_r}^2$.  
Therefore the norm of an $r$-blade differs from its square at most by a sign.  
That means unit $r$-blades also satisfy $\bl{A_r}^2 = \pm1$, although that 
$\pm1$ may not be the blade's squared norm.  It also follows that the inverse 
of \bl{A_r} equals the additional expressions
\begin{equation} 
\bl{A_r}^{-1} = (-1)^{r(r-1)/2} \frac{\bl{A_r}}{|\bl{A_r}|^2} = \frac{\bl{A_r}}{\bl{A_r}^2}. 
\label{bladeinv}
\end{equation}
So the inverse of an $r$-blade is a multiple of the original $r$-blade, just as with vectors.  
Therefore they represent the same subspace.

In Section \ref{wherenow}, I asked how you would calculate the inverse
of $2$-blade $a \out b$.  Well, now we know: divide the original blade by 
its square.  I actually calculated $(a \out b)^2$ in Section \ref{motive}, and the 
result was $-a^2b^2\sin^2\theta$.  Therefore
\begin{equation} 
(a \out b)^{-1} = \frac{b \out a}{a^2 b^2 \sin^2\theta}. 
\end{equation}
By the way, this is also the reverse of $a \out b$ divided by its norm squared, 
as it should be.

Next, I give a geometric property of null blades.
\begin{thm}
A nonzero blade is null (and thus noninvertible) iff the inner product is 
degenerate on the subspace it represents.
\label{nulleqdegen}
\end{thm}
\bp
$\bl{A_r} = e_1 e_2 \dotsb e_r$ is null iff $e_i^2=0$ for at least one $i$, in 
which case $e_i$ is orthogonal to every vector in the span of $\{e_j\}_{i=1,\dotsc,r}$, 
which is just \bl{A_r}.  That means that either (a) $e_i=0$ or (b) $e_i \neq 0$ 
but the inner product is degenerate on \bl{A_r}.  Since $\bl{A_r} \neq 0$, none of 
the $e_i$ vanish, so that leaves case (b): the inner product must
be degenerate.  Therefore nonzero \bl{A_r} is null iff the inner product is 
degenerate on \bl{A_r}.
\ep
So every nonzero blade is invertible in a Euclidean space, while in non-Euclidean
spaces things aren't as simple.

And here's an interesting property of products of versors.
\begin{thm}
If $r, s \geq 1$ and nonzero versors $A_r$ and $B_s$ satisfy $A_r\,B_s = 0$, then 
both versors are null.
\label{prodnonzeroiszero}
\end{thm}  
\bp
This one is easy: if $A_r\,B_s = 0$, then $|A_r|^2 B_s = \rev{A_r}A_r\,B_s = 0$ also.  
Now since $B_s$ is assumed nonzero, it follows that $|A_r|^2 = 0$, or $A_r$ is 
null.  Going back to $A_r\,B_s = 0$ and multiplying from the right by \rev{B_s} 
establishes that $B_s$ is also null.
\ep
This also means that the product of a non-null versor and any versor is nonzero.

A special case of this theorem arises if a vector $a$ both lies in blade \bl{A_r}
($a \out \bl{A_r} = 0$) and is orthogonal to it ($a \lin \bl{A_r} = 0$).  In that case 
$a \bl{A_r} = 0$, and we say $a$ \emph{annihilates} \bl{A_r}.  The theorem tells us 
that $a$ must be a null vector and \bl{A_r} must be a null blade, which is clear 
from Theorem \ref{nulleqdegen} since the existence of such a vector makes the
inner product degenerate on \bl{A_r}.

To conclude this section, I assume the scalars are real so I can define the weight 
of an $r$-blade, as I said I would back in Section \ref{whatshere}.  The 
weight is supposed to be a higher-dimensional generalization of volume, and one 
way to get that is the following: express \bl{A_r} as a product of $r$ orthogonal 
vectors, and define the weight to be the product of the lengths of those vectors.  
Then the weight is the volume of an $r$-dimensional parallelepiped that spans the 
correct subspace.  That's what the norm gives us, as Eq.~\eqref{prodnorm} shows, 
so I define
\begin{equation} \weight(\bl{A_r}) := \sqrt{||\bl{A_r}|^2|}. \label{weightdef} \end{equation}
The extra $|\,|$ is under the square root because, as I've repeatedly mentioned,
the squared norm can be negative.  By this definition when $r=0$, the weight of a 
scalar is its absolute value.  This definition only works on scalars for which an 
absolute value and square root are defined, which is why I'm defining it only for 
real algebras.

When I get to integral calculus on geometric algebras, I'll be using the weights of 
blades not to define the theory but to interpret parts of it.  Thus integration will be
defined on any geometric algebra, but some of its meaning will apply only to real 
algebras.  Since all of our applications will be on real algebras, I think we'll be fine.

\subsection{The dual}
\label{dual}

The next operation is called a \emph{duality transformation} or \emph{taking the dual}.  
Let \bl{A_r} be an invertible $r$-blade; then the dual of any multivector $B$ by
\bl{A_r} is $B \lin \bl{A_r}^{-1}$.  (Duality gets its own symbol only in a special case, 
which I'll describe below.)  To understand what taking the dual does, let $B$ be a 
$s$-blade \bl{B_s}.
\begin{enumerate}
\item If $s > r$, the dual of \bl{B_s} vanishes.
\item If $s=r$, the dual of \bl{B_s} is a scalar which is zero iff \bl{B_s} contains 
          a vector orthogonal to \bl{A_r} (Theorem \ref{innerprodmean}).
\item If $s < r$, the dual of \bl{B_s} is either zero or an $r-s$-blade representing 
          the orthogonal complement of \bl{B_s} in \bl{A_r} (Theorem \ref{innerprodmean} 
          again).  If \bl{B_s} was inside \bl{A_r} to begin with, the dual of \bl{B_s} is just 
          $\bl{B_s}\bl{A_r}^{-1}$ (Theorem \ref{AsubspaceofB}).  
\end{enumerate}
The dual of an arbitrary $B$ is a sum of these results.  Duality 
transformations are useful both for taking orthogonal complements of 
blades (based on the observations above) and for performing orthogonal 
projections into subspaces, as I'll show in Section \ref{proj}.

Although one can take the dual by any invertible blade, one class of blades is by far
the most important: those that represent the entire vector space.  The dual by 
these blades is very useful and also has simpler properties than the dual in general.

Let the dimension of the vector space be $n$; then all $n$-blades either vanish identically
(if the factors are dependent) or represent the same subspace (namely the whole space);
therefore Theorem \ref{AmultofBifsamespace} says that all $n$-blades are multiples of one 
another.  Since the inner product on the whole space is nondegenerate by Axiom \ref{nondeg}, 
Theorem \ref{nulleqdegen} says that all nonzero $n$-blades are also invertible and thus 
non-null, so I define a \emph{volume element} \I\ to be a unit $n$-blade.  This determines \I\ 
to within a sign.  (Some people call a volume element a \emph{pseudoscalar}, but I won't.)  
In fact, I can calculate $|\I|^2$ explicitly.  Let $\{e_i\}_{i=1,\dotsc,r}$ be an orthonormal basis,
and suppose $p$ of the $e_i$ square to $-1$ while the rest square to $1$.  Let $\I = 
e_1 e_2 \dotsb e_n$; then using Eq.~\eqref{prodnorm}, 
\begin{equation} |\I|^2 = e_1^2 e_2^2 \dotsb e_n^2 = (-1)^p. \label{Inormsq} \end{equation}
Therefore $|\I|^2 = 1$ in any Euclidean space, while $|\I|^2 = -1$ in Minkowski space.
Since $|\I|^2 = \rev{\I}\I$, this implies
\begin{equation} \I^2 = (-1)^{n(n-1)/2 + p}. \label{Isq} \end{equation}

Given Theorem \ref{nulleqdegen}, we can now see that Axiom \ref{nondeg} is just 
another way to say ``volume elements are invertible."  I could have used instead a weaker 
axiom that implies only ``volume elements are nonzero," and that would have
been enough to prove some foundational results, like this one that I've been promising for 
some time.
\begin{thm}
The outer product of linearly independent vectors is nonzero.  
\label{indepnonzero}
\end{thm}
\bp
Let \bl{A_r} be the outer product of linearly independent vectors.  Since volume elements are 
nonzero, \bl{A_r} must lie in a subspace represented by a nonzero blade \bl{A_s}; then by 
Theorem \ref{factorblade} there exists a blade \bl{A_{s-r}} such that $\bl{A_r} \out \bl{A_{s-r}} = 
\bl{A_s}$.  Thus \bl{A_r} is a factor of a nonzero blade, so \bl{A_r} is nonzero too.
\ep
This theorem is actually equivalent to ``volume elements are nonzero'' because each implies
the other.  Because of this, some authors take this weaker statement as an axiom instead of 
my Axiom \ref{nondeg}.  I still like my axiom, though, because if \I\ is invertible, taking the dual 
by \I\ is also invertible.  This makes the dual much more useful, as you'll see below.

Unless otherwise specified, ``the dual of $A$" means ``the dual of $A$ by \I" and is
denoted \dual{A}.  Let's reconsider the three ways the dual of a blade can turn out when we're 
taking the dual by \I.
\begin{enumerate}
\item There are no $s$-blades for $s > n$, so the first option can't happen.
\item Any $n$-blade $\bl{B_n} = \lambda \I$ for some $\lambda$, in which case 
          the dual of \bl{B_n} is just $\lambda$.
\item If $s < n$, then \bl{B_s} represents a subspace of the full space, so the dual of \bl{B_s} 
          is just $\bl{B_s}\I^{-1}$.   It cannot be zero; if it were, then Theorem \ref{innerprodmean}
          would say that \bl{B_s} contains a nonzero vector orthogonal to the whole space, which 
          Axiom \ref{nondeg} doesn't allow.  The theorem also tells me that the dual of \bl{B_s} 
          represents the orthogonal complement of \bl{B_s}.
\end{enumerate}
So the general formula for the dual of multivector $A$ is
\begin{equation} \dual{A} := A \lin \I^{-1} = A \I^{-1} \label{dualdef} \end{equation}
and the dual of a blade represents its orthogonal complement.  (Hence the choice of symbol.)
Taking the dual by $\I^{-1}$ instead of \I\ is the inverse operation; it's denoted by \invdual{A}.  
Since $\I^{-1} = \I / \I^2$, \dual{A} and \invdual{A} differ only by a factor of $\I^2$, so the duality 
transformation is its own inverse up to at most a sign.

Since the product of any multivector with \I\ is an inner product, it's true for any $A_r$ that
\begin{align} 
A_r \I & = (-1)^{r(n-1)} \I A_r \nonumber \\
          & = \I \grinv[(n-1)]{A_r},
\label{commuteIAr}
\end{align} 
so for any multivector $A$,
\begin{equation} A \I = \I \grinv[(n-1)]{A}. \label{commuteI} \end{equation}
This has several consequences.
\begin{enumerate}
\item Even multivectors commute with \I\ regardless of the value of $n$, so their duals can 
          be taken be taken from either side with no difference.  
\item In odd-dimensional spaces, the dual of any multivector can be taken from either side.
\item In even-dimensional spaces, the dual of an odd multivector can still be taken from 
          either side, and the results differ only by a sign.
\item The first of Eqs.~\eqref{commuteIAr} is true even if $A_r$ is an $r$-versor because all 
          terms in $A_r$ are even or odd as $r$ is even or odd.
\item In even-dimensional spaces Eqs.~\eqref{evenoddfromgrinv} and \eqref{commuteI} can be 
          used to separate the pure even and pure odd parts of a multivector:
          \begin{equation} \grade[\pm]{A} = \half(A \pm \I A \I^{-1})\ \ \ \text{if $n$ is even.} \end{equation}
\end{enumerate}

Duality lets me prove a surprising result.
\begin{thm}
If the vector space is $n$-dimensional, every $n-1$-vector is an $n-1$-blade.  
\label{vectorisblade}
\end{thm}
\bp
Let $A_{n-1}$ by an $n-1$-vector.  The dual of $A_{n-1}$ is a vector, so $A_{n-1}$ 
is the dual of a vector.  But vectors are $1$-blades, and the dual of a blade is also 
a blade, so $A_{n-1}$ is an $n-1$-blade.  
\ep
One corollary of this is that in dimensions below four, all $r$-vectors are actually
$r$-blades.  $0$-vectors and $1$-vectors are always blades, $n$-vectors
are always blades (which takes care of bivectors in two dimensions and trivectors in
three), and bivectors in three dimensions are $n-1$-vectors and thus blades.

The dual is distributive over addition, and it's easy to show that
\begin{equation} \dual{(AB)} = A\dual{B}. \end{equation}
Taking appropriate-grade terms also shows that
\begin{align}
\dual{(A \out B)} & = A \lin \dual{B} \nonumber \\
\dual{(A \lin B)} & = A \out \dual{B}.
\label{dualprods}
\end{align} 
Thus the dual relates the inner and outer products.  (Here's another way to prove these results:  
start with the third of Eqs.~\eqref{associdents} and set $C=\I^{-1}$.  That gets you the first equation. 
Then replace $B$ with $\dual{B}$ and take the inverse dual of both sides; that gets you the other 
equation.)  A special case of this is $\dual{(a \lin \bl{A_r})} = a \out \dual{\bl{A_r}}$, which means 
vector $a$ is orthogonal to subspace \bl{A_r} iff $a$ lies in \dual{\bl{A_r}}.  That's further confirmation 
that duals represent orthogonal complements.  It also shows that any subspace has a \emph{direct}
representation (all $a$ such that $a \out \bl{A_r} = 0$) and a \emph{dual} representation (all $a$
such that $a \lin \dual{\bl{A_r}} = 0$).  These two representations are both useful in different situations.

In the discussion around Theorem \ref{miscident}, I said that if \bl{A_r} is invertible, 
then any space that contains \bl{A_r} is the direct sum of \bl{A_r} and its orthogonal 
complement.  This is certainly true for the whole space, and it's nicely expressed in 
terms of duals.
\begin{thm}
The whole space is the direct sum of \bl{A_r} and \dual{\bl{A_r}} iff \bl{A_r} 
is invertible.
\label{Aandorthspan}
\end{thm}
\bp
\begin{equation}
\bl{A_r} \out \dual{\bl{A_r}} = \dual{(\bl{A_r} \lin \bl{A_r})} = \bl{A_r}^2 \I^{-1} = \frac{\bl{A_r}^2}{\I^2} \I.
\label{AoutdualA}
\end{equation}
Now Theorem \ref{inviffnonnull} and Eq.~\eqref{bladeinv} tell me \bl{A_r} is invertible iff 
$\bl{A_r}^2 \neq 0$.  So if \bl{A_r} is invertible, Eq.~\eqref{AoutdualA} shows that \I\ is the 
direct sum of \bl{A_r} and its orthogonal complement; and if \bl{A_r} is not invertible, the 
equation shows that \bl{A_r} and its dual have vectors in common, so they don't even 
have a direct sum.
\ep

If $A$ is invertible, so is $\dual{A}$:
\begin{align}
(\dual{A})^{-1} & = (A\I^{-1})^{-1} \nonumber \\
                   & = \I A^{-1}.
\label{dualinv}
\end{align}
The dual of a grade involution is given by
\begin{align}
\dual{(\grinv{A})} & = \grinv{A} \I^{-1} \nonumber \\
                    & = (-1)^{n} \grinv{A} \grinv{(\I^{-1})} \nonumber \\
                    & = (-1)^n \grinv{(A \I^{-1})} \nonumber \\
                    & = (-1)^n \grinv{(\dual{A})}. 
\end{align}
The dual of a reverse is 
\begin{align}
\dual{(\rev{A})} & = \dual{\left[\rev{(\dual{A}\I)}\right]} \nonumber \\
                   & = \dual{\left[\rev{\I} \rev{(\dual{A})}\right]}.
\end{align}
Combining these results,  the dual of a Clifford conjugate is
\begin{equation}
\dual{(\clifconj{A})} =  \dual{\left[\clifconj{\I} \clifconj{(\dual{A})}\right]}.
\end{equation}

Finally, Theorem \ref{facversorfromscprod} and the second part of Theorem \ref{inviffnonnull}
show that the dual almost preserves scalar products:
\begin{align}
\scprod{\dual{A}}{\dual{B}} & = \scprod{(A \I^{-1})}{(B \I^{-1})} \nonumber \\
                                                & = |\I^{-1}|^2 \, \scprod{A}{B} \nonumber \\
                                                & = |\I|^{-2} \, \scprod{A}{B}.
\end{align}
So taking the dual preserves scalar products up to a scale factor.

Occasionally it's convenient to take the dual by volume elements that aren't normalized.  
In that case, the dual and its inverse may differ by more than a sign, but the difference is 
still only a scalar multiple.  All the results in this section are valid regardless of the 
normalization of \I.

\subsection{The commutator}
\label{commute}

The final operation is called the \emph{commutator},
defined as follows.
\begin{equation} \commute{A}{B} := \half(AB - BA). \end{equation}
Notice the factor of $\half$, which is not present in the usual
definition of the commutator, used for example in quantum mechanics.
The commutator obeys the identity
\begin{equation}
\commute{A}{(BC)} = (\commute{A}{B})C + B(\commute{A}{C}),
\label{commident}
\end{equation}
which is easily verified by expanding out the commutators.  This shows that 
the commutator is a derivation on the algebra (it obeys the Leibnitz rule).  
Use this identity to expand $\commute{A}{(BC)}$ and $\commute{A}{(CB)}$ 
and take half the difference; the result is the \emph{Jacobi identity}
\begin{equation}
\commute{A}{(\commute{B}{C})} = \commute{(\commute{A}{B})}{C} + \commute{B}{(\commute{A}{C})}.
\end{equation}
The presence of the second term on the right hand side shows that the 
commutator is not associative.  This identity is often given in the cyclic form
\begin{equation}
\commute{A}{(\commute{B}{C})} + \commute{B}{(\commute{C}{A})} + \commute{C}{(\commute{A}{B})} = 0.
\end{equation}
From the defining properties of the three involutions it's easy to see that
\begin{align}
\grinv{(\commute{A}{B})} & = \commute{\grinv{A}}{\grinv{B}} \nonumber \\
\rev{(\commute{A}{B})} & = \commute{\rev{B}}{\rev{A}} \nonumber \\
\clifconj{(\commute{A}{B})} & = \commute{\clifconj{B}}{\clifconj{A}}.
\end{align} 
The commutator of any multivector with a scalar clearly vanishes, and the 
commutator with a vector can be expressed nicely by decomposing a general
multivector as $A = \grade[+]{A} + \grade[-]{A}$ and recalling Eqs.~
\eqref{vecAinner} and \eqref{vecAouter} for the inner and outer products.  
The result is
\begin{align}
\commute{a}{A} & = a \lin \grade[+]{A} + a \out \grade[-]{A} \nonumber \\
\commute{A}{a} & = \grade[+]{A} \rin a + \grade[-]{A} \out a.
\label{commutewithvec}
\end{align}
This lets me prove an important result about commuting multivectors.
\begin{thm}
The following statements are equivalent.
\begin{enumerate}
\item $A$ commutes with all multivectors.
\item $A$ commutes with all vectors.
\item $A = \lambda + \mu \grade[-]{\I}$.
\end{enumerate}
\label{commutewithvecs}
\end{thm}
Item $3$ is my sneaky way of saying $A$ equals $\lambda$ in even-dimensional vector 
spaces and $\lambda + \mu\I$ in odd-dimensional spaces. 
\bp
Since scalars commute with everything, I won't mention them again.  If $A$ commutes with all
multivectors then it obviously commutes with all vectors.  On the other hand, if $A$ commutes
with all vectors then it commutes with all blades, since these are products of vectors.  Therefore
$A$ commutes with all sums of blades, and thus all multivectors.

Now for item $3$.  The first of Eqs.~\eqref{commutewithvec} tells me that $\commute{a}{\I} = 
a \out \I$ if the vector space is odd-dimensional and $a \lin \I$ if the space is even-dimensional.  
Now $a \out \I = 0$ and $a \lin \I \neq 0$ for all $a$, because every vector lies in \I\ and 
no vector is orthogonal to it; therefore all vectors commute with \I\ in odd-dimensional spaces and 
no vectors commute with \I\ in even-dimensional spaces.  To finish off, let \bl{A_r} be an $r$-blade 
where $0 < r < n$.  If $r$ is even, then $\commute{a}{\bl{A_r}} = a \lin \bl{A_r}$, and if this 
vanished for all $a$ then \bl{A_r} would be orthogonal to the whole space, in violation of 
Axiom \ref{nondeg}.  If $r$ is odd, then $\commute{a}{\bl{A_r}} = a \out \bl{A_r}$.  Since 
$r < n$ there certainly exists a vector $a$ outside \bl{A_r}, in which case 
$a \out \bl{A_r} \neq 0$.
\ep

The most interesting of all is the commutator with a bivector.
\begin{thm}
\begin{equation}
\commute{A_2}{A_r} = \grade[r]{A_2 A_r},
\end{equation}
so the commutator of a bivector and an $r$-vector is an $r$-vector; commutation 
with a bivector is a grade preserving operation.  
\end{thm}
\bp
To show this, I note that
\begin{align}
A_2 A_r & = A_2 \lin A_r + \grade[r]{A_2 A_r} + A_2 \out A_r \nonumber \\
A_r A_2 & = A_2 \lin A_r - \grade[r]{A_2 A_r} + A_2 \out A_r.
\end{align}
The first equation is obvious when $r \geq 2$; for $r<2$, recall that in such
cases $A_2 \lin A_r = 0$.  The second equation follows from the first because 
of the properties of the inner and outer products under interchange and 
Eq.~\eqref{revex} when $j=1$.  Subtracting these equations yields
\begin{equation}
\commute{A_2}{A_r} = \half(A_2 A_r - A_r A_2) = \grade[r]{A_2 A_r}.
\end{equation}
\ep

In particular, the set of bivectors is closed under commutation.  That means 
the bivectors form a Lie algebra with the commutator serving as the Lie product.  
That will be important later when I show how to use geometric algebra to describe
Lie groups and Lie algebras.

Since commutation with a bivector is grade preserving, the identity in
Eq.~\eqref{commident} still holds if $A=A_2$ and I replace all
geometric products with either inner or outer products:
\begin{align}
\commute{A_2}{(B \lin C)} & = (\commute{A_2}{B}) \lin C + B \lin (\commute{A_2}{C}) \nonumber \\
\commute{A_2}{(B \rin C)} & = (\commute{A_2}{B}) \rin C + B \rin (\commute{A_2}{C}) \nonumber \\ 
\commute{A_2}{(B \out C)} & = (\commute{A_2}{B}) \out C + B \out (\commute{A_2}{C}).
\label{wedgeident}
\end{align}
The last of these relations can be generalized in this way.
\begin{thm}
\begin{equation}
\commute{A_2}{(a_1 \out a_2 \out \dotsb \out a_r)} = \sum_{j=1}^r a_1 
    \out a_2 \out \dotsb \out (A_2 \rin a_j) \out \dotsb \out a_r.
\label{biveccommblade}
\end{equation}
\end{thm}
\bp
As usual, I use induction.  The result is true when $r=1$
because the commutator with a vector is the same as the right inner product,
and the $r=2$ result follows from the last of Eqs.~\eqref{wedgeident}, so let's 
assume the result is true for $r-1$.  Then by associativity of the outer product
\begin{equation}
\commute{A_2}{(a_1 \out a_2 \out \dotsb \out a_r)} = 
\commute{A_2}{(\bl{B_{r-1}} \out a_r)}
\end{equation}
where $\bl{B_{r-1}} = a_1 \out a_2 \out \dotsb \out a_{r-1}$.  Applying the
last of Eqs.~\eqref{wedgeident} and the $r-1$ result yields
\begin{align} 
\commute{A_2}{(a_1 \out a_2 \out \dotsb \out a_r)}
 & = (\commute{A_2}{\bl{B_{r-1}}}) \out a_r +  \bl{B_{r-1}} \out (\commute{A_2}{a_r}) \nonumber \\
 & = \sum_{j=1}^{r-1} \big[a_1 \out a_2 \out \dotsb \out (A_2 \rin a_j) 
     \out \dotsb \out a_{r-1} \out a_r\big] \, + \nonumber \\
 & \qquad \qquad a_1 \out a_2 \out \dotsb \out a_{r-1} \out 
     (A_2 \rin a_r) \nonumber \\
 & = \sum_{j=1}^r a_1 \out a_2 \out \dotsb \out (A_2 \rin a_j) \out \dotsb 
     \out a_r,
\end{align}
which completes the proof.
\ep

I expand the order of operations to include all of these new operations as follows: 
perform the involutions, then duals, then outer, then inner, then geometric products, 
then scalar products, and finally commutators.  Following this convention, the parentheses 
in Eqs.~\eqref{scprodinnerouter} and the left hand sides of Eqs.~\eqref{commident}, 
\eqref{wedgeident}, and \eqref{biveccommblade} (but not the right hand sides) may be omitted.

\section{Geometric algebra in Euclidean space}
\label{eucl}

Now let's apply everything I've done so far to some familiar cases.  I'll work through 
the algebras of two- and three-dimensional real Euclidean space explicitly, revealing 
some neat surprises along the way.

\subsection{Two dimensions and complex numbers}
\label{2d}

First I'll consider the real plane $\mathbb{R}^2$ with the
Euclidean scalar product; this is often denoted $\mathbb{E}^2$.  It
has an orthonormal basis $\{e_1, e_2\}$, which produces a
geometric algebra spanned by the elements $1$, $e_1$, $e_2$, and $e_1
e_2$.  That last element satisfies
\begin{equation}
|e_1 e_2|^2 = \rev{(e_1 e_2)} e_1 e_2 = e_2 e_1 e_1 e_2 = 1.
\end{equation}
Therefore it qualifies as a volume element \I.  Since bivectors change sign under 
reversion, it also satisfies $\I^2=-1$.  It defines a right-handed orientation, and 
a few examples show that all vectors anticommute with \I.  This is consistent with 
Eq.~\eqref{commuteI}.

Now for the geometric products.  We know what a scalar times anything and a 
vector times a vector look like; all that remains is the product of a vector and a 
bivector, or equivalently the product of a vector and \I.  To see what that does, 
notice that
\begin{align} 
\I e_1 & = -e_2 \nonumber \\
\I e_2 & = \ \ e_1, 
\label{dual2d} 
\end{align}
so multiplication of a unit vector by \I\ results in an orthogonal
unit vector.  (Which it should, since multiplying by \I\ takes the dual to 
within a sign.)  Eq.~\eqref{dual2d} actually tells us a bit more: left
multiplication by \I\ rotates a vector clockwise through $\pi/2$.
Similarly, right multiplication rotates a vector counterclockwise
through the same angle.  So $\I^2=-1$ means that two rotations in the
same sense through $\pi/2$ have the same effect as multiplying by
$-1$.  Of course, this is true only in two dimensions.

The even subalgebra of any geometric algebra is always of interest, so
let's take a moment to look at it.  A generic even multivector can be written 
$Z = x + \I y$ where $x$ and $y$ are real numbers and $\I^2=-1$; in other 
words, the even subalgebra of $\mathbb{E}^2$ is isomorphic to the algebra of
complex numbers.  Now this may be a little bit of a surprise, because the even 
subalgebra represents scalars and areas, while we normally think of  
complex numbers as vectors in the Argand plane.  But there's another
way to think of complex numbers: the polar form $z=re^{i\theta}$ reminds us
that $z$ also represents a rotation through angle $\theta$ followed by a
dilatation by $r$.  How do these two interpretations of complex numbers
relate?

It works out because we're in two dimensions.  Then and only then, the even 
subalgebra is isomorphic to the space of vectors; a generic vector in 
$\mathbb{E}^2$ takes the form $z = x e_1 + y e_2$ where $x$ and $y$ are 
real numbers, and there's a natural
isomorphism between the vectors and the even subalgebra of the form
\begin{align} Z & = e_1 z \nonumber \\  z & = e_1 Z. \end{align} 
This isomorphism maps a vector in the $e_1$ direction onto a pure
``real'' number, so $e_1$ plays the role of the real axis.  It also maps a 
vector in the $e_2$ direction onto a pure ``imaginary" number, so $e_2$ is
the imaginary axis.  Now think about complex conjugation: it leaves the real
part alone while changing the sign of the imaginary part.  Therefore
complex conjugation is a reflection along the $e_2$
axis, which takes $z$ to $z' = -e_2 z e_2$.  What happens to the
corresponding even element?  It gets mapped to
\begin{align}
Z' & = e_1 z' \nonumber \\
    & = -e_1 (e_2 z e_2) \nonumber \\
    & = -e_1 e_2 (e_1 Z) e_2 \nonumber \\
    & = -\I e_1 (x + \I y) e_2 \nonumber \\
    & = x - \I y \nonumber \\
    & = \rev{Z},
\end{align}
where I used the fact that \I\ anticommutes with all vectors.  Therefore 
complex conjugation corresponds to taking the reverse.  Now let $w$ 
and $z$ be vectors with corresponding even elements $W$ and $Z$; it follows that
\begin{align}
wz & = e_1 W e_1 Z \nonumber \\
      & = e_1 (x + \I y) e_1 Z \nonumber \\
      & = (x - \I y) Z \nonumber \\
      & = \rev{W} Z.
\end{align}
Now let's look at this.  The right hand side is the product of one complex 
number with the conjugate of another.  That has two terms: the real part 
equals the dot product of the corresponding vectors, while the magnitude of the
imaginary part equals the magnitude of the cross product of the vectors.  
The left hand side is the geometric product of the vectors, which is exactly the same thing.
One of the goals of geometric algebra was to take the complex product,
which combines the two-dimensional dot and cross products naturally, and
generalize it to any number of dimensions.  (That was a goal of quaternions too.
I'll show how that worked out in the next section.)

Now for rotations.  An element $W$ of the even subalgebra has a polar 
form $r\exp(-\I\theta)$ for some $r$ and $\theta$.  Letting $r=1$, multiplication 
by a vector $z$ produces the vector
\begin{align}
z' & = W z \nonumber \\
    & = \exp(-\I\theta) z \nonumber \\
    & = \exp(-\I\theta/2) \exp(-\I\theta/2) z \nonumber \\
    & = \exp(-\I\theta/2) z \exp(\I\theta/2) \nonumber \\
    & = R z R^{-1}
\end{align}
where I defined $R=\exp(-\I\theta/2)$.  Clearly $R$ is a rotor, so
multiplication by $W$ performs a counterclockwise rotation through
$\theta$.  What is the corresponding transformation of $Z$?
\begin{align}
Z' & = e_1 z' \nonumber \\
    & = e_1 W z \nonumber \\
    & = wz \nonumber \\
    & = \rev{W} Z \nonumber \\
    & = \exp(\I\theta) Z.
\end{align}
Thus vector $z$ is rotated counterclockwise through $\theta$ when the
corresponding even element $Z$ is multiplied by $\exp(\I\theta)$,
exactly as you'd expect.

In conclusion, the complex numbers are the even subalgebra 
of the geometric algebra of the Euclidean plane; the identification
with vectors is just an accident in two dimensions, just as
identifying planes with normal vectors works only in three
dimensions.  Now while complex algebra is useful, so is 
complex analysis; we use its techniques to perform many ostensibly
real integrals, for example.  If geometric algebra generalizes 
complex algebra to any dimension, then perhaps calculus of
geometric algebras could generalize complex analysis too.
I'll describe geometric calculus later, and I'll show how it generalizes 
the Cauchy integral theorem and other useful results.

\subsection{Three dimensions, Pauli matrices, and quaternions}
\label{3d}

Much that is true in two dimensions carries over to three:
$\mathbb{E}^3$ has an orthonormal basis $\{e_1, e_2, e_3\}$, so its
geometric algebra is spanned by 1, $e_1$, $e_2$, $e_3$, $e_1 e_2$,
$e_1 e_3$, $e_2 e_3$, and the volume element $e_1 e_2 e_3$.  This
volume element is also denoted \I, defines a right-handed
orientation, satisfies $|\I|^2 = \rev{\I} \I = 1$, and squares to $-1$.  
Unlike the two-dimensional case, a few examples show that all vectors 
and bivectors commute with \I, as required by Theorem \ref{commutewithvecs}.  
Now I've repeatedly
mentioned that only in three dimensions can you identify planes with normal
vectors, which is why the cross product works there.  The map between
planes and normal vectors should be a duality transformation, so the
cross product should be the dual of something.  Well, it is.  If $a$ and $b$
are vectors, then
\begin{equation} 
a \times b = \dual{(a \out b)}.
\label{crossprod} 
\end{equation}
So cross products are easily converted into outer products and vice versa.  
Yay.  This shows why the cross product is not associative even though the outer 
product is; the dual gets in the way.    Since duality is just multiplication by $-\I$ 
and vectors and bivectors commute with \I, I can use Eq.~\eqref{crossprod} to write
\begin{equation} a b = a \lin b + \I a \times b, \end{equation}
which is the three-dimensional analog of the product $\rev{W}Z$ of complex
numbers $W$ and $Z$.  Another popular product in traditional
vector algebra is the triple product $a \inp b \times c$, which relates to 
geometric algebra by
\begin{equation} a \inp b \times c = \dual{(a \out b \out c)}. \end{equation}
This form makes the cyclic property of the triple product obvious.  The triple cross
product $a \times (b \times c)$ is also pretty common, and it can be expressed in
geometric algebra as
\begin{equation} a \times (b \times c) = -a \lin (b \out c). \end{equation}
From here, it's easy to see that the BAC-CAB rule for expanding this product is 
really just a special case of Theorem \ref{alinoutai}.

I'd like to say a little more about cross products.  Since $\dual{(a \out b)} = a \lin \dual{b}$
and in three dimensions \dual{b} is a bivector, it follows that $a \times b$ is also
the inner product of $a$ and a bivector.  You may recall that in classical mechanics, 
linear and angular velocity are related by a cross product: $\bm{v} = \bm{\omega} 
\times \bm{r}$.  The geometric algebra equivalent is $v = \Omega \rin r$, where $\Omega = 
\dual{\bm{\omega}}$ is an angular velocity bivector in the instantaneous plane 
of rotation.  (Bivectors figure prominently in rotational dynamics, as I'll show in Section 
\ref{classicalmech}.)  You may also recall that the magnetic part of the Lorentz force 
on a point charge is $\bm{F} = q \bm{v} \times \bm{B}$, where $\bm{B}$ is the magnetic 
field vector.  In geometric algebra this becomes $F = q v \lin B$, where $B = \dual{\bm{B}}$ 
is the magnetic field bivector.  I'll show later on that the bivector representation of $B$ 
is more physically motivated than the vector version.

A consequence of Theorem \ref{vectorisblade}, 
which I mentioned at the time, is that in dimensions under four, every $r$-vector 
is actually an $r$-blade.  In two dimensions that was obviously true; we had 
only scalars, vectors, and multiples of \I.  In three dimensions the scalars, 
vectors, and trivectors are obviously blades (the trivectors are multiples of \I), and 
I can show using geometry that all bivectors are $2$-blades.  Consider two 
$2$-blades \bl{A_2} and \bl{B_2}; each represents a plane passing through the
origin, and any two such planes in three dimensions share a common line.  
Therefore $\bl{A_2} = a \out b$ and $\bl{B_2} = a \out c$ where $a$ is a vector 
along the line shared by the planes.  This means that
\begin{equation} 
\bl{A_2} + \bl{B_2} = a \out b + a \out c = a \out (b+c) 
\end{equation}
is also a $2$-blade.  Thus any bivector in three dimensions is a
$2$-blade, as Theorem \ref{vectorisblade} demands.

Now for the products.  As before, we know what a scalar times anything or 
a vector times a vector looks like; next I'll do a vector times a bivector.  Let 
$a$ be a vector and \bl{B} be a
bivector; then $a = a_\parallel + a_\perp$ where $a_\parallel$ lies in
the plane determined by \bl{B} and $a_\perp$ is perpendicular to it.  In
that case there exists a vector $b$ perpendicular to $a_\parallel$
such that $\bl{B} = a_\parallel b$, so
\begin{align}
a \bl{B} & = (a_\parallel + a_\perp) a_\parallel b \nonumber \\
 & = a_\parallel^2 b + a_\perp a_\parallel b \nonumber \\
 & = a_\parallel^2 b + a_\perp \out a_\parallel \out b.
\end{align}
So the product of $a$ and \bl{B} is the sum of two terms: a
vector in the plane of \bl{B} perpendicular to $a$, and the 
trivector defined by \bl{B} and the component of
$a$ perpendicular to it.  Clearly the vector is $a \lin \bl{B}$ and 
the trivector is $a \out \bl{B}$.  The trivector can also be written
\begin{equation} 
\pm |a_\perp| |a_\parallel| |b| \I = \pm |a_\parallel| |\bl{B}| \I 
\end{equation} 
where $|\,|$ is the magnitude of a multivector defined in Section 
\ref{scalarprod}.  The $\pm$ is there because we don't know the 
orientation of the system defined by the three vectors.  

Vector times trivector is even easier.  If $a$ is a vector and \bl{T} 
is a trivector, then $\bl{T} = a b c$ where $b$
and $c$ are perpendicular to each other and to $a$, so
\begin{equation} a\bl{T} = a^2 b c = a^2 b \out c. \end{equation}
So $a \bl{T}$ is a bivector representing the plane to which $a$ is
perpendicular.  This is clearly $a \lin \bl{T}$.

The product of two bivectors looks like this:  since all bivectors are 
$2$-blades representing planes, let vector $a$ lie along the direction 
shared by bivectors \bl{A_2} and \bl{B_2}, so $\bl{A_2}=ba$ and 
$\bl{B_2}=ac$ where $b$ and $c$ are perpendicular to $a$ but not
necessarily to each other; then
\begin{align}
\bl{A_2} \bl{B_2} & = b a a c \nonumber \\
 & = a^2 b c \nonumber \\
 & = a^2 b \lin c + a^2 b \out c.
\end{align}
So the product of two bivectors is a scalar plus a bivector
representing the plane normal to their intersection line.  The first term
is $\bl{A_2} \lin \bl{B_2} = \bl{A_2} \rin \bl{B_2}$ and the second term is 
$\commute{\bl{A_2}}{\bl{B_2}}$.

Next, a bivector times a trivector: if bivector $\bl{B}=ab$ where $a$ and $b$ are
perpendicular, then there exists vector $c$ perpendicular to $a$ and
$b$ such that trivector $\bl{T}=bac$, in which case
\begin{equation} \bl{B}\bl{T} = b a a b c = a^2 b^2 c, \end{equation}
so the product of a bivector and a trivector is a vector perpendicular
to the plane of the bivector.  This is also $\bl{B} \lin \bl{T}$.

The product of two trivectors is just a number. In fact, it's the product of 
the volumes defined by the two trivectors, with the sign determined 
by their relative orientations.

The general multiplication rule for the basis vectors can be written
as
\begin{equation} e_i e_j = \delta_{ij} + \sum_k \I \epsilon_{ijk} e_k, \end{equation} 
which is exactly the multiplication rule for the Pauli matrices.
Therefore the Pauli matrices are just a matrix representation of the
basis vectors of three dimensional space.  It is well known that the
Pauli matrices form a Euclidean Clifford algebra, but the idea that
they are literally matrix representations of $\bm{\hat{x}}$, $\bm{\hat{y}}$, 
and $\bm{\hat{z}}$ is not so familiar.

Finally, the even subalgebra of the geometric algebra on
$\mathbb{E}^3$ has some surprises for us too.  Let the unit bivectors
be labeled
\begin{equation} 
\bl{B_1} = e_2 e_3, \text{ } \bl{B_2} = e_1 e_3, \text{ and } \bl{B_3} = e_1 e_2.
\end{equation}
Notice that this definition is not consistently right-handed because
of \bl{B_2}.  These objects satisfy the relations
\begin{equation} \bl{B_1}^2 = \bl{B_2}^2 = \bl{B_3}^2 = -1 \end{equation}
and
\begin{equation} \bl{B_1} \bl{B_2} \bl{B_3} = -1, \end{equation}
so the even subalgebra of the algebra on $\mathbb{E}^3$, which is
spanned by 1 and the \bl{B_i}, is isomorphic to the quaternions.  The 
quaternions were created to generalize the complex numbers to three
dimensions, of course, so something like this was expected; but the 
quaternions as Hamilton conceived them were intended to correspond 
to the three unit directions, not three planes.  The map between them 
works differently in two dimensions and three, so while complex numbers
can be thought of consistently as either vectors or bivectors, quaternions
can be mapped from one to the other only by introducing an inconsistency
in the handedness, as I've done here.

\section{More on projections, reflections, and rotations}
\label{projrefrot}

In Sections \ref{motive} and \ref{simpleapps} I introduced 
projections along vectors, reflections along vectors, and rotations in planes.  My purpose was
to get you interested in geometric algebra by showing how well it handled all three
operations compared to traditional vector algebra.  Well, there's more.  It turns out that these 
operations can be defined on subspaces just as well as vectors; for example, rotating a
subspace means rotating all the vectors in it.  As I'll show, in geometric algebra this is 
very easy, and the resulting formulas are almost the same as the formulas for vectors.

\subsection{Orthogonal projections and rejections}
\label{proj}

Let's restate what I did in Sections \ref{motive} and  \ref{simpleapps} a little differently.  
Let $u$ and $v$ be vectors; then the orthogonal projection of $v$ along $u$ is given by
\begin{equation} 
P_u(v) = v \lin u u^{-1} = (v \lin u) \lin u^{-1} 
\label{projvec}
\end{equation}
and the orthogonal rejection of $v$ from $u$ is given by
\begin{equation} 
R_u(v) = v \out u u^{-1} = v \out u \rin u^{-1}. 
\label{rejvec}
\end{equation}
(The second parts of each equation are easy to verify.)  $P_u(v)$ is parallel to $u$,
$R_u(v)$ is orthogonal to $u$, and $P_u(v) + R_u(v) = v$.  These operations require
$u$ to be invertible, so it can't be a null vector.  I promised in Section \ref{simpleapps} 
that this would have geometrical meaning, and now we're about to see what it is.

\subsubsection{Projecting a vector into a subspace}

Let's take a moment to consider the general notion of projection into a subspace.  
Let $S$ be a subspace ($S$ is not a blade this time; it really is the subspace itself) 
and let $a$ be a vector not in $S$.  Then for any $v \in S$ I can write $a = v + (a - v)$, 
which is the sum of a vector in $S$ and a vector not in $S$.  So which $v$ is the
``projection'' of $a$ into $S$?  We can't say without further information.  
For example, consider two subspaces $S_1$ and $S_2$ that share only the zero 
vector; then if a vector lies in their direct sum, it can be expressed only one way as a 
vector from $S_1$ plus a vector from $S_2$, and thus has a unique projection into 
either subspace.  Projection into a subspace is specified not only by the subspace 
itself but also by the subspace the rest of the vector will belong to, 
and the operation is well-defined only if the two subspaces share only the zero vector.

Now consider orthogonal projection as an example of this.  The idea is
to express a vector as a sum of two terms, one in subspace $S$ and one
in $S^\perp$, the orthogonal complement of $S$.  This works only if
$S$ and $S^\perp$ have no nonzero vectors in common, which is true iff
the inner product is nondegenerate on $S$.  Thus orthogonal projection
is well-defined only for a subspace with an invertible
blade.  In that case, I get this result.
\begin{thm}
If $a$ is a vector and \bl{A_r} is an invertible blade, then the orthogonal projection 
of $a$ into and the orthogonal rejection of $a$ from subspace \bl{A_r} are given
by
\begin{align} 
P_{\bl{A_r}}(a) & = a \lin \bl{A_r} \bl{A_r}^{-1} = (a \lin \bl{A_r}) \lin \bl{A_r}^{-1} \nonumber \\
R_{\bl{A_r}}(a) & = a \out \bl{A_r} \bl{A_r}^{-1} = a \out \bl{A_r} \rin \bl{A_r}^{-1}.
\end{align}
\label{vectorprojrej}
\end{thm}
\bp
First, it's clear that $P_{\bl{A_r}}(a) + R_{\bl{A_r}}(a) = a$.  Now, 
$a \lin \bl{A_r}$ is the dual of $a$ by $\bl{A_r}^{-1}$; it is zero 
if $a$ is orthogonal to \bl{A_r}, and otherwise it is an $r-1$-blade 
representing the subspace of \bl{A_r} orthogonal to $a$.  In that case 
its product with $\bl{A_r}^{-1}$ equals its inner product with $\bl{A_r}^{-1}$, 
which is just the dual by \bl{A_r}; the result is a vector that lies 
in \bl{A_r}.   On the other hand, $a \out \bl{A_r}$ is 
zero if $a$ lies in \bl{A_r}, and otherwise it is an $r+1$-blade 
that contains \bl{A_r}.  In that case the product with $\bl{A_r}^{-1}$ 
equals the right inner product, and is the dual of $\bl{A_r}^{-1}$ by $(a \out \bl{A_r})^{-1}$, 
so the result is a vector orthogonal to \bl{A_r}.  Both formulas give vectors, 
they sum to $a$, the first lies in \bl{A_r} and vanishes iff $a$ is orthogonal to \bl{A_r}, 
and the second is orthogonal to \bl{A_r} and vanishes iff $a$ lies in \bl{A_r}.
Therefore the two expressions are obviously the 
orthogonal projection of $a$ into and the orthogonal rejection of $a$ 
from \bl{A_r}.
\ep

So projecting into a subspace is the same as projecting onto a vector; you just replace 
the vector with the blade representing the subspace.  We'll see several more examples 
of this idea below.

I can demonstrate directly that $P_{\bl{A_r}}(a)$ lies in \bl{A_r}:
\begin{equation}
P_{\bl{A_r}}(a) \out \bl{A_r} = (a \lin \bl{A_r} \bl{A_r}^{-1}) \out \bl{A_r} = 
\grade[r+1]{a \lin \bl{A_r} \bl{A_r}^{-1} \bl{A_r}} =  \grade[r+1]{a \lin \bl{A_r}} = 0.
\end{equation}
Similarly, I can show that $R_{\bl{A_r}}(a)$ is orthogonal to \bl{A_r} as follows:
\begin{equation}
R_{\bl{A_r}}(a) \lin \bl{A_r} = (a \out \bl{A_r} \bl{A_r}^{-1}) \lin \bl{A_r}  =
\grade[r-1]{a \out \bl{A_r} \bl{A_r}^{-1} \bl{A_r}} = \grade[r-1]{a \out \bl{A_r}} = 0.
\end{equation}

This result applies to the Gram-Schmidt process for producing an orthogonal 
set of vectors from a linearly independent set with the same span.  Let $\{a_j\}_{j=1,\dotsc,r}$
be linearly independent; then we build the orthogonal set $\{b_j\}_{j=1,\dotsc,r}$
as follows.  Let $b_1 = a_1$ to start with. Then $b_2$ equals $a_2$ minus its 
projection onto $b_1$, or equivalently the orthogonal rejection of $a_2$ from $b_1$.
Next, $b_3$ equals the orthogonal rejection of $a_3$ from the span of $b_1$ and 
$b_2$, and so on through all of the $a_j$.  Therefore we proceed as follows.
\begin{enumerate}
\item Let $b_1 = a_1$.
\item For each $j$ starting with $1$, let $\bl{B_j} = b_1 \out \dotsb \out b_j$.
\item Then let $b_{j+1} = a_{j+1} \out \bl{B_j} \bl{B_j}^{-1}$.
\end{enumerate}
This procedure will work only if each \bl{B_j} is invertible, which is why it is normally used
only in Euclidean spaces.

If \bl{A_r} is a blade, then \dual{\bl{A_r}} represents the orthogonal complement of \bl{A_r}.  
That means that orthogonal projection into \dual{\bl{A_r}} should equal orthogonal 
rejection from \bl{A_r}.  Using Eqs.~\eqref{dualprods} and \eqref{dualinv}, this is easy to 
show directly.
\begin{align}
a \lin \dual{\bl{A_r}} (\dual{\bl{A_r}})^{-1} & = \dual{(a \out \bl{A_r})} (\dual{\bl{A_r}})^{-1} 
        \nonumber \\
 & = a \out \bl{A_r} \I^{-1} \I \bl{A_r}^{-1} \nonumber \\
 & = a \out \bl{A_r} \bl{A_r}^{-1}.
\end{align}

If \bl{A_r} and \bl{B_s} are orthogonal, then the projection of a vector into
their direct sum should be the sum of the projections into the subspaces individually.  
(For example, the projection of a vector into the Euclidean $xy$ plane should be the 
sum of the projections onto the $x$ and $y$ axes separately.)  This can also be shown
directly.  By Theorem \ref{AorthtoB}, $\bl{A_r} \out \bl{B_s} = \bl{A_r}\bl{B_s}$, so using 
the first of Eqs.~\eqref{usefulids} I find
\begin{align}
P_{\bl{A_r} \out \bl{B_s}}(a) & = P_{\bl{A_r}\bl{B_s}}(a) \nonumber \\
                                                  & = a \lin (\bl{A_r}\bl{B_s}) (\bl{A_r}\bl{B_s})^{-1} \nonumber \\ 
                                                  & = [(a \lin \bl{A_r}) \bl{B_s} + (-1)^r \bl{A_r} (a \lin \bl{B_s})] \bl{B_s}^{-1} \bl{A_r}^{-1} \nonumber \\
                                                  & =  a \lin \bl{A_r} \bl{A_r}^{-1} + (-1)^r \bl{A_r} (a \lin \bl{B_s}) \bl{B_s}^{-1} \bl{A_r}^{-1}.
\label{projAoutBstep}
\end{align}
Now let's work on that last term.  If \bl{A_r} and \bl{B_s} are orthogonal, \bl{A_r} and 
$a \lin \bl{B_s}$ are too, so their product is an outer product, so I can interchange them 
and pick up a factor of $(-1)^{r(s-1)}$.  And since $\bl{A_r}^{-1}$ and $\bl{B_s}^{-1}$ 
are multiples of \bl{A_r} and \bl{B_s}, their product is also an outer product, so 
I can interchange them and pick up a factor of $(-1)^{rs}$.  Putting all this in 
Eq.~\eqref{projAoutBstep},
\begin{align}
P_{\bl{A_r} \out \bl{B_s}}(a) & = a \lin \bl{A_r} \bl{A_r}^{-1} + (-1)^{r + r(s-1) + rs} (a \lin \bl{B_s}) \bl{A_r} \bl{A_r}^{-1} \bl{B_s}^{-1} \nonumber \\
                                                  & = a \lin \bl{A_r} \bl{A_r}^{-1} + a \lin \bl{B_s} \bl{B_s}^{-1} \nonumber \\
                                                  & = P_{\bl{A_r}}(a) + P_{\bl{B_s}}(a).
\end{align}

\subsubsection{Projecting a multivector into a subspace}

Now that I can project a vector into a subspace, how about projecting one subspace 
into another?  As I suggested above, this seems simple enough: project subspace 
\bl{B_s} into subspace \bl{A_r} by taking every vector in \bl{B_s}, projecting it into 
\bl{A_r}, and seeing what subspace you get.  The rejection should be similar: just 
reject all the vectors individually.  However, if I am a bit more precise, I discover a 
wrinkle.  I define
\begin{align} 
P_{\bl{A_r}}(b_1 \out \dotsb \out b_s) & := P_{\bl{A_r}}(b_1) \out \dotsb \out  P_{\bl{A_r}}(b_s) 
\nonumber \\
R_{\bl{A_r}}(b_1 \out \dotsb \out b_s) & := R_{\bl{A_r}}(b_1) \out \dotsb \out  R_{\bl{A_r}}(b_s).
\label{projrejwedgeprod}
\end{align}
Now suppose the set $\{b_j\}_{j=1,\dotsb,s}$ is linearly 
independent but their projections are not.  That would happen necessarily if, for example,
I projected a plane into a line.  In that case, the projection defined this way vanishes.  Instead 
of objecting to this wrinkle, I decide that it provides useful extra information.  If by chance 
the projections of the members of \bl{B_s} do not form an $s$-dimensional space, so be it; 
I accept that the projection is zero. 

These formulas make geometric sense, but they aren't very easy to use.  However, they
can be made simpler, and extended to all multivectors to boot.  Here's how.
\begin{thm}
For any invertible blade \bl{A_r} and vectors $\{b_j\}_{j=1,\dotsc,s}$,
\begin{align}
P_{\bl{A_r}}(b_1) \out \dotsb \out  P_{\bl{A_r}}(b_s) & = (b_1 \out \dotsb \out b_s) 
\lin \bl{A_r}  \bl{A_r}^{-1} \nonumber \\
R_{\bl{A_r}}(b_1) \out \dotsb \out  R_{\bl{A_r}}(b_s) & = (b_1 \out \dotsb \out b_s) 
\out \bl{A_r} \bl{A_r}^{-1}.
\end{align}
\label{projblade}
\end{thm}
\bp
I start with the first equation.  Since each $b_j = P_{\bl{A_r}}(b_j) + R_{\bl{A_r}}(b_j)$, the 
outer product $b_1 \out \dotsb \out b_s$ can be written as a sum of terms, one of which 
equals $P_{\bl{A_r}}(b_1) \out \dotsb \out  P_{\bl{A_r}}(b_s)$ while each of the others 
contains at least one $R_{\bl{A_r}}(b_j)$.  Consider what happens to each term when 
you take the inner product with \bl{A_r} and multiply by $\bl{A_r}^{-1}$.  The term 
$P_{\bl{A_r}}(b_1) \out \dotsb \out  P_{\bl{A_r}}(b_s)$ lies inside \bl{A_r}, so by Theorem 
\ref{AsubspaceofB} the inner product becomes a product, so the \bl{A_r} and $\bl{A_r}^{-1}$ 
cancel out and you're left with $P_{\bl{A_r}}(b_1) \out \dotsb \out  P_{\bl{A_r}}(b_s)$.  On the 
other hand, each of the other terms contains a factor orthogonal to \bl{A_r}, so the inner 
product with \bl{A_r} vanishes.  Thus the first equation is valid.

For the second equation, I again write $b_1 \out \dotsb \out b_s$ as a sum of terms,
but this time I note that one of them equals $R_{\bl{A_r}}(b_1) \out \dotsb \out  R_{\bl{A_r}}(b_s)$ 
while each of the others contains at least one $P_{\bl{A_r}}(b_j)$.  Consider what happens to
each term when you take the outer product with \bl{A_r} and multiply by $\bl{A_r}^{-1}$.  The term
$R_{\bl{A_r}}(b_1) \out \dotsb \out  R_{\bl{A_r}}(b_s)$ is orthogonal to \bl{A_r}, so by 
Theorem \ref{AorthtoB} the outer product becomes a product, so the \bl{A_r} and $\bl{A_r}^{-1}$ 
cancel out and you're left with $R_{\bl{A_r}}(b_1) \out \dotsb \out  R_{\bl{A_r}}(b_s)$.
On the other hand, each of the other terms contains a factor that lies in \bl{A_r}, so
the outer product with \bl{A_r} vanishes.  Thus the second equation is valid too.
\ep

Therefore $P_{\bl{A_r}}(\bl{B_s}) = \bl{B_s} \lin \bl{A_r} \bl{A_r}^{-1}$ and 
$R_{\bl{A_r}}(\bl{B_s}) = \bl{B_s} \out \bl{A_r} \bl{A_r}^{-1}$ for any blade \bl{B_s}.  
Taking the obvious step, I define the orthogonal projection and rejection of any 
multivector to be
\begin{align}
P_{\bl{A_r}}(B) & := B \lin \bl{A_r} \bl{A_r}^{-1} = (B \lin \bl{A_r}) \lin \bl{A_r}^{-1} \nonumber \\
R_{\bl{A_r}}(B) & := B \out \bl{A_r} \bl{A_r}^{-1} = B \out \bl{A_r} \rin \bl{A_r}^{-1}.
\label{anyprojrej}
\end{align}
You might be surprised that both projection and rejection leave scalars untouched:
\begin{equation}
P_{\bl{A_r}}(\lambda) = R_{\bl{A_r}}(\lambda) = \lambda.
\end{equation}
This had to happen for reasons I'll explain in Section \ref{linearextensions}.  Projecting into 
and rejecting from \I\ do what you think they should (except for that odd bit with scalars):
\begin{align}
P_{\I}(B) & = B \nonumber \\  R_{\I}(B) & = \grade{B}.
\end{align}
Running it the other way around, here's what happens when you project and reject \I:
\begin{align}
P_{\bl{A_r}}(\I) & = \I \, \delta_{rn} \nonumber \\ R_{\bl{A_r}}(\I) & = \I \, \delta_{r0}.
\label{projrejI}
\end{align}
Again, this makes sense; only the whole space is big enough to project \I\ into, and only
zero-dimensional spaces are small enough to reject \I\ from.

With a little relabeling and rearranging, the first parts of Eqs.~\eqref{anyprojrej} become
\begin{align} 
A \lin \bl{B_s} & = P_{\bl{B_s}}(A) \bl{B_s}  \nonumber \\
A \out \bl{B_s} & = R_{\bl{B_s}}(A) \bl{B_s}.
\label{inofbladeisprodofproj}
\end{align}
This shows that the inner or outer product of a multivector and a blade can also be
expressed as a geometric product, as long as the blade is invertible so projection is defined. 
Using Theorem \ref{facversorfromscprod}, this also shows that the norm squared of 
$A \lin \bl{B_s}$ equals the norm squared of $P_{\bl{B_s}}(A)$ times the 
norm squared of \bl{B_s}, and a similar result holds for the outer product.

Comparing Eqs.~\eqref{projvec} and \eqref{rejvec} with Eqs.~\eqref{anyprojrej}, the 
level of generality achieved is astounding.  Starting with the projection 
of one vector along another, I've shown that the projection of any multivector into 
a subspace is meaningful and is given by the \emph{same expression}, with 
the multivector and blade put in place of the two vectors.  The rejection of one 
vector from another follows the same pattern.  It is true that we've lost one property: we no
longer have $P_{\bl{A_r}}(B) + R_{\bl{A_r}}(B) = B$ in general.  This makes geometric
sense, however, if you look at the proof of Theorem \ref{projblade}: neither a projected
blade nor a rejected blade includes all the terms that are partly projected and 
partly rejected, so to speak.

\subsection{Reflections}
\label{refl}

To start, I'll review reflections from Section \ref{simpleapps}.  I defined the reflection of 
vector $v$ along axis $n$ as follows: the projection of $v$ along $n$ gets a minus
sign, while the rejection of $v$ from $n$ is unchanged.  If the reflection is denoted $v'$,
then
\begin{equation} v' = - nvn^{-1}. \label{refdefagain} \end{equation}
What I didn't show in Section \ref{simpleapps} is that reflections preserve inner products,
which I'll show now.  Using Eq.~\eqref{refdefagain}, the definition of the inner product, and the 
cyclic property of the scalar part of a product,
\begin{align}
a' \lin b' & = \grade{a' b'} \nonumber \\
 & = \grade{n a n^{-1} n b n^{-1}} \nonumber \\
 & = \grade{abn^{-1}n} \nonumber \\ 
 & = \grade{ab} \nonumber \\
 & = a \lin b.
\label{refinnerprod}
\end{align}

\subsubsection{Reflecting a vector in a subspace}
\label{refvecinspace}

Just as I used projection and rejection along an axis to define reflection
along an axis, I can use projection and rejection in a subspace to define reflection in a 
subspace.  The reflection of $a$ in \bl{A_r} is constructed by giving the projection of $a$
into \bl{A_r} a minus sign and leaving the rejection of $a$ from \bl{A_r} alone.  Using
Theorem \ref{vectorprojrej}, Eq.~\eqref{commuteinner}, and Eq.~\eqref{commuteouter}, I find
\begin{align}
a' & := -P_{\bl{A_r}}(a) + R_{\bl{A_r}}(a) \nonumber \\
    & \ = -a \lin \bl{A_r} \bl{A_r}^{-1} + a \out \bl{A_r} \bl{A_r}^{-1} \nonumber \\
    & \ = -(-1)^{r-1}  \bl{A_r} \rin a \bl{A_r}^{-1} + (-1)^r  \bl{A_r} \out a \bl{A_r}^{-1} \nonumber \\
    & \ = (-1)^r \bl{A_r} a \bl{A_r}^{-1} \nonumber \\
    & \ = \bl{A_r} \grinv[r]{a} \bl{A_r}^{-1}.
\label{refainAr}
\end{align}

In the last line, $\grinv[r]{a}$ means $a$ is grade involuted $r$ times; I introduced the notation back 
in Section \ref{gradeinv}.  (You may wonder why I did this instead of just leaving in the $(-1)^r$.  
It will make sense in the next section.)  Another way to arrive at this formula is to write 
$\bl{A_r} = a_1 a_2 \dotsb a_r$ and reflect $a$ along each of the $a_j$ in succession.  Once 
again, an expression in terms of vectors generalizes to subspaces with only minimal change.  
Reflections in subspaces also preserve inner products; the proof is very similar to 
Eqs.~\eqref{refinnerprod}.

\subsubsection{Reflecting a multivector in a subspace}
\label{refmultivector}

Now that I can reflect vectors, I can reflect subspaces too: the reflection of subspace \bl{B_s}
in subspace \bl{A_r} is found by taking every vector from \bl{B_s}, reflecting it in \bl{A_r}, and 
seeing what subspace you get.  That would mean something like
\begin{align} 
(b_1 \out \dotsb \out b_s)' & := b_1' \out \dotsb \out b_s' \nonumber \\
                                              & \ = \left(\bl{A_r} \grinv[r]{b_1} \bl{A_r}^{-1}\right) \out 
                                    \dotsb \out \left(\bl{A_r}^{-1} \grinv[r]{b_s} \bl{A_r}^{-1}\right).
\end{align}
Again, this is geometrically sensible but not easy to use.  Fear not; I can fix that.  To start with,
notice what Eq.~\eqref{refainAr} shows: conjugating a vector by an invertible $r$-blade gives 
you back a vector.  A more general version of that is also true.
\begin{thm}
If $A$ is a versor and $B_s$ is an $s$-vector, then
\begin{equation} A B_s \rev{A} = \grade[s]{A B_s \rev{A}}, \end{equation}
so conjugation by an invertible versor is grade preserving.
\end{thm}
\bp
The theorem is true for versors if it's true for vectors, so I'll look at $a B_s a$.  Using 
Eqs.~\eqref{vecrprod} and \eqref{rvecprod}, I can write
\begin{equation} 
a B_s a = (a \lin B_s) \rin a + (a \lin B_s) \out a + (a \out B_s) \rin a + a \out B_s \out a.
\end{equation}
The first term is grade $s-2$, the middle two terms are grade $s$, and the last term is grade $s+2$,
so I'm done if I can ditch the first and last terms.  The last term vanishes because $a$ appears
twice in the outer product (compare Eq.~\eqref{aoutAoutb}), and the first term vanishes because 
it can be rewritten as $(-1)^{s-1} (B_s \rin a) \rin a = (-1)^{s-1} B_s \rin (a \out a) = 0$.  Since 
Theorem \ref{inviffnonnull} tells me that the inverse of a versor, if it has one, is its reverse divided by 
its norm squared, conjugation by an invertible versor preserves grade too.
\ep
I'll use this to get the result I really want.
\begin{thm}
If $A$ is a versor, then 
\begin{equation} 
(A B \rev{A}) \out (A C \rev{A}) = |A|^2 \, A (B \out C) \rev{A}.  
\end{equation}
Therefore if $A$ is invertible, $(A B A^{-1}) \out (A C A^{-1}) = A (B \out C) A^{-1}$.
\end{thm}
\bp
The result is true for general $B$ and $C$ if it's true for $B_s$ and $C_t$, and I've already shown that
versor conjugation preserves grades, so
\begin{align}
(A B_s \rev{A}) \out (A C_t \rev{A}) & = \grade[s+t]{A B_s \rev{A} A C_t \rev{A}} \nonumber \\
                                                            & = |A|^2 \grade[s+t]{A B_s C_t \rev{A}} \nonumber \\
                                                            & = |A|^2 \, A \grade[s+t]{B_s C_t} \rev{A} \nonumber \\
                                                            & = |A|^2 \, A (B_s \out C_t) \rev{A}.
\end{align}
If $A$ is invertible, then $|A|^2 \neq 0$, so dividing both sides by $|A|^4$ yields the desired result. 
\ep
Now for reflections.  If $\bl{B_s} =  b_1 \out  \dotsb \out b_s$, then 
\begin{align}
\bl{B_s}' & := b_1' \out \dotsb \out b_s' \nonumber \\
 & \ = \left(\bl{A_r} \grinv[r]{b_1} \bl{A_r}^{-1}\right) \out \dotsb \out \left(\bl{A_r} 
        \grinv[r]{b_s} \bl{A_r}^{-1}\right) \nonumber \\
  & \ = \bl{A_r}  (\grinv[r]{b_1} \out \dotsb \out \grinv[r]{b_s}) \bl{A_r}^{-1} \nonumber \\
  & \ = \bl{A_r}  \grinv[r]{(b_1 \out \dotsb \out b_s)} \bl{A_r}^{-1} \nonumber \\
  & \ = \bl{A_r} \grinv[r]{\bl{B_s}} \bl{A_r}^{-1}.
\end{align}
Taking the obvious next step, I define the reflection of multivector $B$ in subspace \bl{A_r} to be
\begin{equation} B' := \bl{A_r} \grinv[r]{B} \bl{A_r}^{-1}. \end{equation}
So reflection in \bl{A_r} is done by grade involuting $r$ times and then conjugating by \bl{A_r}.  
This is a little more complicated than the reflection of a vector 
along an axis that we started with, Eq.~\eqref{refdefagain}, but not much.  And of course it reduces
to Eq.~\eqref{refdefagain} when \bl{A_r} and $B$ are vectors.

The reflection of \I\ in a subspace is
\begin{align} 
\I' & = \bl{A_r} \grinv[r]{\I} \bl{A_r}^{-1} \nonumber \\
    & = (-1)^{nr} \bl{A_r} \I \bl{A_r}^{-1} \nonumber \\
    & =  (-1)^{nr} (-1)^{r(n-1)} \I \bl{A_r} \bl{A_r}^{-1} \nonumber \\
    & = (-1)^r \I .
\label{reflectI}
\end{align}
This makes sense because $r$ directions in the space were reflected.  So
the orientation changes iff $r$ is odd.

You may have noticed that I now have two ways to reflect a vector around the origin.  The first
is grade involution, and the second is to reflect the vector in a volume element.  Since
both operations have been extended to the whole algebra in a way that respects 
products, they ought to be equal not just for vectors but for any multivector, or 
\begin{equation} 
\grinv{A} = \I \grinv[n]{A} \I^{-1}.
\label{explicitspaceref} 
\end{equation}
To show that this really is true, start with Eq.~\eqref{commuteI}, grade involute both sides,
and use $\grinv{\I}=(-1)^n \I$.  Then multiply both sides by $\I^{-1}$ on the right and voil\`{a}.

Finally, I can relate reflection in \bl{A_r} and reflection in \dual{\bl{A_r}}.  Reflection of vector 
$a$ in \bl{A_r} gives the component in \bl{A_r} a minus sign and leaves the component in 
\dual{\bl{A_r}} alone, while reflection in \dual{\bl{A_r}} does the opposite.  Therefore one reflection
should be the negative of the other, or
\begin{equation} 
\dual{\bl{A_r}} \grinv[(n-r)]{a} (\dual{\bl{A_r}})^{-1} = -\bl{A_r} \grinv[r]{a} \bl{A_r}^{-1}.
\end{equation}
Extending this to general multivectors, I expect
\begin{equation} 
\dual{\bl{A_r}} \grinv[(n-r)]{B} (\dual{\bl{A_r}})^{-1} = \grinv{(\bl{A_r} \grinv[r]{B} \bl{A_r}^{-1})}.
\end{equation}
And indeed that's what I find:
\begin{align}
\dual{\bl{A_r}} \grinv[(n-r)]{B_s} (\dual{\bl{A_r}})^{-1} 
     & = (-1)^{s(n-r)} \bl{A_r} \I^{-1} B_s (\bl{A_r} \I^{-1})^{-1} \nonumber \\
     & = (-1)^{s(n-r)} (-1)^{s(n-1)} \bl{A_r} B_s \I^{-1} \I \bl{A_r}^{-1} \nonumber \\
     & = (-1)^s (-1)^{rs} \bl{A_r} B_s \bl{A_r}^{-1} \nonumber \\
     & = \grinv{(\bl{A_r} \grinv[r]{B_s} \bl{A_r}^{-1})}.
\end{align}

\subsection{Rotations}
\label{rot}

After all this work, rotations are fairly anticlimactic.  Once again, I start with a review 
of Section \ref{simpleapps}.  I showed there that a rotation in a plane is the product of two 
reflections along vectors in that plane, so
\begin{equation} v' = R v R^{-1} \end{equation}
where $R$ is the product of two invertible vectors, also called a biversor or a rotor.  A rotation
clearly preserves inner products since it's just two reflections in succession, but you
can show it directly by an argument very much like Eqs.~\eqref{refinnerprod}.

\subsubsection{Rotating a multivector in a plane}
 
The rotation of a subspace is as simple to understand as the reflection.  In fact, it's the
example I started this whole section with: you rotate a subspace by rotating all the 
vectors in it.  The argument is identical to the argument for reflections: if $\bl{A_r} =  
a_1 \out \dotsb \out a_r$, then $\bl{A_r}' = a_1' \out \dotsb \out a_r'$.  Therefore the 
rotation by $R$ is
\begin{align}
\bl{A_r}' & := a_1' \out \dotsb \out a_r' \nonumber \\
 & \ = \left(R a_1 R^{-1}\right) \out \dotsb \out \left(R a_r R^{-1}\right) \nonumber \\
 & \ = R (a_1 \out \dotsb \out a_r) R^{-1} \nonumber \\
 & \ = R \bl{A_r} R^{-1}.
\end{align}
The grade inversion of \bl{A_r} is absent because it is performed twice, once for each 
factor in the rotor.  Therefore the rule for rotating any multivector is
\begin{equation} A' := R A R^{-1}, \label{generalrot} \end{equation} 
which is exactly the same as the formula for vectors.

Since \I\ commutes with even multivectors (Eq.~\eqref{commuteI}), rotations leave \I\ alone,
\begin{equation} R \I R^{-1} = \I, \end{equation}
as expected.

When I first discussed rotations in Section \ref{simpleapps}, I said that any two axes in the
same plane separated by the same angle would generate the same rotation.  That 
means that if I take the two vectors in $R$ and rotate them the same amount in the plane 
of $R$, the resulting rotor should perform the same rotation.  Therefore, if $R$ and $S$ are 
rotors in the same plane, $S R S^{-1}$ should represent the same rotation as $R$.  You can
show this directly:  $R$ and $S$ are both scalars plus multiples of the same area element, so 
they commute.  Therefore $S R S^{-1} = R S S^{-1} = R$.  

Every linear transformation of vectors can be extended to the entire geometric
algebra; I'll describe that process later.  These three transformations extend 
in a particularly compact way, but not all transformations do.  Rotations and reflections
behave as well as they do because they are orthogonal transformations, and 
geometric algebra is particularly well-suited to represent them.  In fact, it's a good 
idea to pause and notice just how good a job it does; compare Eq.~\eqref{generalrot} 
to the increasingly complicated expressions you get when
you rotate tensors of ever-increasing rank.  One of the great strengths
of geometric algebra is its ability to extend orthogonal transformations to the whole
algebra in such a simple fashion.

\subsubsection{Rotations in three dimensions}

In a real three dimensional space, rotations have an interesting property that is easy to 
understand using geometric algebra: the product of two rotations is another
rotation.  If $R_1$ represents the first rotation and $R_2$ the second, then their
product is $R = R_2 R_1$.  We lose no generality by demanding that both $R_1$ 
and $R_2$ are unit rotors; and that means $R$ is a unit even versor.  
In three dimensions the only even grades are zero and two, so $R$ is actually a scalar 
plus a bivector: $R = \grade{R} + \grade[2]{R}$.  Therefore $|R|^2 = 1$ becomes 
$\grade{R}^2 + |\grade[2]{R}|^2 = 1$.  That tells me that $R = \cos (\theta/2) - B \sin (\theta/2)$
for some $\theta$ and unit bivector $B$.  And in three dimensions every bivector is 
a $2$-blade, so \bl{B} represents some plane, and thus $R = \exp(-\bl{B} \theta/2)$, 
which is a rotation through $\theta$ in plane \bl{B}.  As soon as I climb the ladder to four 
dimensions, though, I lose this result, because $R_2 R_1$ could have a $4$-vector part.

\section{Frames and bases}
\label{framesbases}

Now I'll consider a geometric algebra $\G^n$ in which the space of
vectors has finite dimension $n$.  Let $\{a_i\}_{i=1,\dotsc,n}$ be a
basis for the vector space, which I will also call a \emph{frame}.  (The $a_i$ are not
assumed orthogonal.)  Then a generic element of the algebra will be the
sum of a scalar and terms of the form $a_{i_1} a_{i_2} \dotsb a_{i_r}$
for $r \le n$.  Theorem \ref{versorsumofblades} tells me that any such element is a linear
combination of blades made up of the $\{a_{i_j}\}$; therefore the scalar $1$ and the 
blades $a_{i_1} \out a_{i_2} \out \dotsb \out a_{i_r}$ generate the 
whole geometric algebra.  I'll now show that they actually form a basis, 
and I'll also show how to calculate the components of an arbitrary multivector 
in this basis.

\subsection{Reciprocal frames}
\label{recipframe}

Given a frame $\{a_i\}_{i=1,\dotsc,n}$, another
frame $\{a^i\}_{i=1,\dotsc,n}$ is called a \emph{reciprocal frame} to
the first if it satisfies
\begin{equation} a^i \lin a_j = \delta^i_j. \label{recip} \end{equation}
If such a set of vectors exists, it is a frame because Eq.~
\eqref{recip} guarantees that the $a^i$ are linearly independent, so
they form a basis.  To construct such vectors, consider their definition: 
$a^j$ should be orthogonal to all of the $a_i$ except for $a_j$, so an 
obvious way to make it is to take the outer product of all of the $a_i$ 
except for $a_j$ and then take its dual, which is what I'll do.

Let $a_N = a_1 \out a_2 \out \dotsb \out a_n$; then $a_N$ is a 
(possibly unnormalized) volume element.  (Even though $a_N$ is a blade,
I'm not denoting it with capital letters or boldface; you'll see why in the 
next section.)  Then I define
\begin{equation}
a^i := (-1)^{i-1} (a_1 \out a_2 \out \dotsb \out \check{a}_i \out 
           \dotsb \out a_n) a_N^{-1}.
\label{definerecipvec}
\end{equation}
$\{a^i\}$ is a reciprocal frame because, using the first of Eqs.~\eqref{dualprods},
\begin{align}
a_i \lin a^j & = (-1)^{j-1} a_i \lin (a_1 \out a_2 \out \dotsb \out 
    \check{a}_j \out \dotsb \out a_n a_N^{-1}) \nonumber \\
 & = (-1)^{j-1} (a_i \out a_1 \out a_2 \out \dotsb \out 
    \check{a}_j \out \dotsb \out a_n) a_N^{-1}
\end{align}
Now if $i \ne j$ then $a_i$ equals one of the other vectors in the
outer product, so the whole thing vanishes.  If $i = j$, I move $a_i$ past
the first $i-1$ vectors to its original spot, which cancels out the $(-1)^{j-1}$ 
prefactor.  Therefore
\begin{align}
a_i \lin a^j & = (a_1 \out \dotsb \out a_n) a_N^{-1} \delta_i^j \nonumber \\
 & = \delta_i^j.
\end{align}
This definition exactly expresses the geometrical idea I started with;
$a_N$ was chosen to perform the duality transform because it gets
the normalization right.

Since both $\{a_i\}$ and $\{a^j\}$ are bases for the vectors, any
vector $v$ can be written $v=\sum v^i a_i$ or $v=\sum v_j a^j$.  In
fact, it's obvious that $v^i=v \lin a^i$ and $v_j = v \lin a_j $, so
the components of $v$ on either basis are easily calculated using the
other basis.  Using the definition of $a^i$ and the first of 
Eqs.~\eqref{dualprods} again, I find that
\begin{align}
v^i & = v \lin a^i \nonumber \\
 & = (-1)^{i-1} v \lin (a_1 \out a_2 \out \dotsb \out 
    \check{a}_i \out \dotsb \out a_n a_N^{-1}) \nonumber \\
 & = (-1)^{i-1} (v \out a_1 \out a_2 \out \dotsb \out 
    \check{a}_i \out \dotsb \out a_n) a_N^{-1} \nonumber \\
 & = (a_1 \out \dotsb \out a_{i-1} \out v \out a_{i+1}
    \out \dotsb \out a_n) a_N^{-1}.
\end{align}
Compare this with Eqs.~\eqref{xon2dbasis} and \eqref{xon3dbasis} back in 
Section \ref{simpleapps}.

Since $v$ is a vector, the expressions for its components can be
written $v^i=\scprod{v}{a^i}$ and $v_j=\scprod{v}{a_j}$, where $\scprod{}{}$ 
is the scalar product.  These forms for the components can be generalized 
a long way, as I'll show in the next section.

The inner product of any two vectors follows easily from their
components:
\begin{equation} 
b \lin c = \sum_{i,j} b_i c^j (a^i \lin a_j) = \sum_i b_i  c^i, 
\label{inpuvcomps}
\end{equation}
and switching the frames on which I expand $b$ and $c$ gives me an 
equally valid result in terms of the components $b^i$ and $c_i$.

A frame and its reciprocal satisfy a useful identity.
\begin{thm}
\begin{equation} \sum_i a_i\,a^i = \sum_i a^i\,a_i = n. \end{equation}
\label{aiaieqn}
\end{thm}
\bp
\begin{align} 
\sum_i a_i \, a^i & = \sum_i a_i \lin a^i + \sum_i a_i \out a^i \nonumber \\
                             & = n + \sum_i a_i \out a^i. 
\end{align}
To evaluate the second term, expand $a^i$ on the original frame to get 
$a^i = \sum_j (a^i \lin a^j) a_j$, so
\begin{align}
\sum_i a_i \out a^i & = \sum_i a_i \out \left( \sum_j a^i \lin a^j a_j \right) \nonumber \\
                                  & = \sum_{i, j} (a_i \out a_j) (a^i \lin a^j) \nonumber \\
                                  & = 0
\end{align}
because $a^i \lin a^j$ is symmetric in $i$ and $j$ while $a_i \out a_j$ is antisymmetric.
The proof that $\sum_i a^i \, a_i = n$ is the same except for exchanging superscripts 
and subscripts.
\ep

\subsection{Multivector bases}
\label{bases}

Before I continue, I need some fancy new notation.  Let $I$ be a 
string of indices $i_1, i_2, \dotsc, i_r$, and given a string $I$ let 
$a_I$ be defined by
\begin{equation} 
a_I := a_{i_1} \out a_{i_2} \out \dotsb \out a_{i_r}, 
\end{equation}
and similarly for $a^I$.  I will use the symbol $N$ only to refer to the string
$1, 2, \dotsc, n$, to be consistent with $a_N$ in the previous section.  I also allow $I$ 
to be the ``empty'' sequence, in which case I define $a_I = a^I = 1$.  Then I 
immediately know several things:
\begin{enumerate}
\item $a_I = a^I = 0$ iff the string $I$ contains at least one index twice.
\item If $I$ and $J$ contain the same elements but in a different order, 
          then $a_I = (\sgn\sigma) a_J$ and $a^I = (\sgn\sigma) a^J$, where 
          $\sigma$ is the permutation that changes $I$ to $J$.
\item Theorem \ref{versorsumofblades} tells me that given a frame $\{a_i\}$ 
          for the vectors, the set $\{a_I\}$ (or $\{a^I\}$) where $I$ ranges over all 
          \emph{increasing} sequences generates $\G^n$.  ($I$ is an increasing 
          sequence if $i_1 < i_2 < \dotsb < i_r$.)
\end{enumerate}
To show either set forms a basis, I'll use this result.
\begin{thm}
\begin{equation} \scprod{a^I}{a_J} = \delta^I_J \end{equation}
where $\delta^I_J$ vanishes if either $I$ or $J$ repeats indices or if $I$ is not a 
permutation of $J$ (including having a different length), and otherwise it equals
the sign of the permutation that takes $I$ to $J$.
\label{recipscprod}
\end{thm}
\bp
If either string repeats indices then both sides vanish, and both sides also vanish 
when the lengths of $I$ and $J$ are different (the right side by definition, the left side
because $a^I$ and $a_J$ have different grades); to get the other results, I let 
$I = i_1 < i_2 < \dotsb < i_r$ and $J = j_1 < j_2 < \dotsb < j_r$ and use 
Eq.~\eqref{scalarproddecomp} to find
\begin{align}
\scprod{a^I}{a_J} & = \grade{\rev{(a^{i_1} \out a^{i_2} \out \dotsb \out a^{i_r})}
                                       (a_{j_1} \out a_{j_2} \out \dotsb \out a_{j_r})} \nonumber \\
                                & = \grade{(a^{i_r} \out \dotsb \out a^{i_2} \out a^{i_1})(a_{j_1} \out 
                                       a_{j_2} \out \dotsb \out a_{j_r})} \nonumber \\
                                & = (a^{i_r} \out \dotsb \out a^{i_2} \out a^{i_1}) \lin 
                                       (a_{j_1} \out  a_{j_2} \out \dotsb \out a_{j_r}).
\end{align}
Now consider the case where the $i_k$ equal the $j_k$; using the third of
Eqs.~\eqref{associdents} and Eq.~\eqref{veclinoutid},
\begin{align}
\scprod{(a^{i_r} & \out \dotsb \out a^{i_2} \out a^{i_1})}{(a_{i_1} \out 
         a_{i_2} \out \dotsb \out a_{i_r})} \nonumber \\
 & = (a^{i_r} \out \dotsb \out a^{i_2}) \lin \big[a^{i_1} 
     \lin (a_{i_1} \out a_{i_2} \out \dotsb \out a_{i_r})\big] \nonumber \\
 & = (a^{i_r} \out \dotsb \out a^{i_2}) \lin \left[\sum_{j=1}^r 
     (-1)^{j-1} a^{i_1} \lin a_{i_j} \, a_{i_1} \out a_{i_2} \out \dotsb 
     \out \check{a}_{i_j} \out \dotsb \out a_{i_r}\right] \nonumber \\
 & = (a^{i_r} \out \dotsb \out a^{i_2}) \lin (a_{i_2} \out 
     \dotsb \out a_{i_r}),
\end{align}
which can be repeated for the $i_2$ term and for each successive term
until the final result
\begin{equation}
\scprod{(a^{i_r} \out \dotsb \out a^{i_2} \out a^{i_1})}{(a_{i_1} \out a_{i_2} 
\out \dotsb \out a_{i_r})} = 1
\end{equation}
is reached.  Now suppose that $i_k$ equals none of the $j_l$; then
when the evaluation of the scalar product as shown above reaches the
$k$th iteration, all of the $a^{i_k} \lin a_{j_l}$ terms will vanish,
and so will the scalar product.  This establishes the result for $I$
and $J$ increasing; the general result follows from the properties of
$a^I$ and $a_J$ under rearrangement of elements.
\ep

From this result it's pretty obvious that for any multivector $A$,
\begin{equation}
A = \sum_I A^I a_I  \quad \text{where} \quad A^I = \scprod{A}{a^I}
\end{equation}
and the sum extends over all increasing sequences $I$ (including the
null sequence to pick up the scalar part).  Therefore, given a frame
$\{a_i\}$, the elements $a_I$ form a true basis for the geometric
algebra, and the equation above shows how to expand any multivector on
this basis.  (Incidentally, the roles of the frame $\{a_i\}$ and the reciprocal 
frame $\{a^j\}$ can be exchanged in this expansion, just as vectors can be 
expanded on either set with the other used to compute the coefficients.)  
Since the number of distinct $r$-blades in a basis for each $r$ is 
$\binom{n}{r}$, it follows that
\begin{equation}
\dim \G^n = \sum_{r=0}^n \dim \G[r]^n = \sum_{r=0}^n \binom{n}{r} = 2^n.
\end{equation}
I can also express the scalar product of any two multivectors $B$ and $C$ in 
terms of their components:
\begin{equation} 
\scprod{B}{C} = \sum_{I,J} B_I C^J (\scprod{a^I}{a_J}) = \sum_I B_I C^I, 
\label{scprodBCcomps}
\end{equation}
and switching the bases on which I expand $B$ and $C$ gives me an 
equally valid result in terms of the components $B^I$ and $C_I$.

A consequence of all this is the following theorem.
\begin{thm}
Multivector $A$ is uniquely determined by either of the following.
\begin{enumerate}
\item \scprod{A}{B} for every multivector $B$.
\item \grade{A} and $a \lin A$ for every vector $a$.
\end{enumerate}
\label{Afromscprods}
\end{thm}
\bp
Part $1$ is obvious.  In fact, it's overkill; \scprod{A}{B} for all $B$ in a basis for the 
algebra will do.  By the distributive property, part $2$ is equivalent to this statement: 
if $\grade{A}=0$ and all $a \lin A = 0$, then $A=0$.  So that's what I'll prove.  To do 
this, I assume $A$ is an $r$-vector $A_r$; if the result is true for $r$-vectors then it's 
true for general multivectors too.  If $r=0$ or $1$ then I'm done, so let $r > 1$.  Let 
$\{a_i\}$ be a frame and $\{a^i\}$ its reciprocal frame; then a component of $A_r$ 
on the basis defined by $\{a_i\}$ is $(a^{i_r} \out \dotsb \out a^{i_2} \out a^{i_1}) 
\lin A_r$ for some strictly ascending choice of $i_1$ through $i_r$.  However, 
\begin{align}
(a^{i_r} \out \dotsb \out a^{i_2} \out a^{i_1}) \lin A_r 
                   & = (a^{i_r} \out \dotsb \out a^{i_2}) \lin (a^{i_1} \lin A_r) \nonumber \\
                                                                       & = 0
\end{align}
since $a^{i_1} \lin A_r=0$.  So all the components of $A_r$ vanish, so $A_r=0$.

This proves the theorem in every finite-dimensional algebra, but it's usually 
true in infinite-dimensional spaces too.  In fact, extra structures are usually 
imposed on infinite-dimensional spaces for exactly this purpose, and I will 
happily assume henceforth that this has always been done.
\ep
This result can be extended \textit{ad nauseum}: $A$ is uniquely determined by 
\grade{A}, \grade[1]{A}, and $A_2 \lin A$ for every bivector $A_2$, and so on.

While any $r$-vector can be expanded using the frame $r$-vectors, it can also
be expanded using only the frame vectors; but now the coefficients
aren't necessarily scalars.
\begin{thm}
\begin{equation} \sum_i a^i \, a_i \lin A_r = \sum_i a^i \out (a_i \lin A_r) = r A_r, \end{equation}
and the same is true if the frame and its reciprocal are interchanged.
\label{expandArinvecs}
\end{thm}
\bp
For the first equality, note that
\begin{align}
\sum_i a^i \, a_i \lin A_r  & = \sum_i a^i \lin (a_i \lin A_r) + \sum_i a^i \out (a_i \lin A_r) \nonumber \\
                                            & = \sum_i (a^i \out a_i) \lin A_r + \sum_i a^i \out (a_i \lin A_r) \nonumber \\
                                            & = \left( \sum_i a^i \out a_i \right) \lin A_r + \sum_i a^i \out (a_i \lin A_r) \nonumber \\
                                            & = \sum_i a^i \out (a_i \lin A_r)
\end{align}
because I showed in the proof of Theorem \ref{aiaieqn} that $\sum_i a^i \out a_i =0$.

The second equality is true for all $r$-vectors if it's true for all members of a basis for $r$-vectors, 
so I have to prove it only on a basis; and I know just the basis to use.  Let $A_r = a^{j_1} 
\out \dotsb \out a^{j_r}$ for some increasing sequence of indices; then 
\begin{align}
\sum_i a^i \out (a_i \lin A_r) & = \sum_i a^i \out \left[a_i \lin (a^{j_1} \out \dotsb \out a^{j_r})\right] \nonumber \\
                                                  & = \sum_{i, k} (-1)^{k-1} (a_i \lin a^{j_k}) \, a^i \out a^{j_1} \out \dotsb \out \check{a}^{j_k} \out \dotsb \out a^{j_r} \nonumber \\
                                                  & = \sum_k (-1)^{k-1} a^{j_k} \out a^{j_1} \out \dotsb \out \check{a}^{j_k} \out \dotsb \out a^{j_r}.
\end{align}
In each term of the sum, I move $a^{j_k}$ past $k-1$ other vectors to return it to its original spot, which 
cancels the $(-1)^{k-1}$ factor, so
\begin{align}
\sum_i a^i \out (a_i \lin A_r) & = \sum_k a^{j_1} \out \dotsb \out a^{j_r} \nonumber \\
                                                  & = r A_r.
\end{align}
That completes the first half of the proof, and exchanging superscripts and 
subscripts provides the other half.
\ep

Just as the original frame has a volume element $a_N$, the reciprocal
frame has a volume element $a^N$ defined in an analogous way: 
$a^N = a^1 \out \dotsb \out a^n.$  Now $a_N$ and $a^N$ have to be scalar 
multiples of each other, and since Theorem \ref{recipscprod} shows 
that $\scprod{a_N}{a^N}=1$, I conclude that
\begin{equation} a^N = \frac{a_N}{|a_N|^2}. \label{recipvol} \end{equation}
A quick calculation shows
\begin{equation} |a^N|^2 = |a_N|^{-2}, \end{equation}
so for real algebras, the weights of the volume elements of a frame and 
its reciprocal are themselves reciprocals.

Given a frame $\{a_i\}$, I have Eq.~\eqref{definerecipvec} for the members 
$a^i$ of the reciprocal frame, but I don't have an equally direct formula for the 
reciprocal multivectors $a^I$; all I can do right now is take the outer product of 
Eq.~\eqref{definerecipvec} several times.  However, I can get a nicer formula 
for $a^I$ using the same logic that got me the reciprocal vectors in the first place.  By 
construction, $a^i$ is orthogonal to the outer product of all the frame vectors 
except $a_i$, and similarly for $a^j$; therefore $a^i \out a^j$ is orthogonal to the 
outer product of all the frame vectors except $a_i$ and $a_j$.  Therefore $a^i \out a^j$
is dual to $a_1 \out \dotsb \out \check{a}_i \out \dotsb \out \check{a}_j \out \dotsb \out a_n$.  
To make this easier to write out, for any string of indices $I$ let me define $I^c$ 
to be its ascending complement, so $I^c$ includes exactly the indices not in $I$ 
in ascending order.  In these terms, $a^I$ is dual to $a_{I^c}$.  To be more precise,
I have
\begin{thm}
If $I$ represents ascending $i_1$ through $i_r$,
\begin{equation} 
a^I = (-1)^{\sum_{j=1}^r (i_j - 1)} a_{I^c} \, a_N^{-1}. 
\label{definerecipmultivec}
\end{equation}
\end{thm}
Notice that this includes Eq.~\eqref{definerecipvec} as a special case when $I$ has 
only one index $i$.
\bp
To prove this, I'll calculate $a^I a_N$.
\begin{align}
a^I a_N & = a^I \lin a_N \nonumber \\
               & = (a^{i_1} \out \dotsb \out a^{i_r}) \lin a_N \nonumber \\
               & = (a^{i_1} \out \dotsb \out a^{i_{r-1}}) \lin (a^{i_r} \lin a_N).
\label{aiaN}
\end{align}
To calculate $a^{i_r} \lin a_N$ I use Eq.~\eqref{veclinoutid}:
\begin{align}
a^{i_r} \lin a_N & = a^{i_r} \lin (a_1 \out \dotsb \out a_n) \nonumber \\
                            & = \sum_j (-1)^{j-1} (a^{i_r} \lin a_j) a_1 \out \dotsb \out \check{a}_j \out \dotsb \out a_n \nonumber \\
                            & = (-1)^{i_r-1} a_1 \out \dotsb \out \check{a}_{i_r} \out \dotsb \out a_n \nonumber \\
                            & = (-1)^{i_r-1} a_{i_r^c}.
\end{align}
Now Eq.~\eqref{aiaN} becomes
\begin{align}
a^I a_N & = (-1)^{i_r-1} (a^{i_1} \out \dotsb \out a^{i_{r-1}}) \lin a_{i_r^c} \nonumber \\
               & =  (-1)^{i_r-1} (a^{i_1} \out \dotsb \out a^{i_{r-2}}) \lin (a^{i_{r-1}} \lin a_{i_r^c}).
\end{align}
When I evaluate $a^{i_{r-1}} \lin a_{i_r^c}$ using Eq.~\eqref{veclinoutid} again, I remove 
$a_{i_{r-1}}$ and multiply by $(-1)^{i_{r-1}-1}$.  (This is why I put the indices of $I$ in 
ascending order; $i_r$ is later than $i_{r-1}$, so $a_{i_{r-1}}$ is still in position $i_{r-1}$ in 
$a_{i_r^c}$.)  Thus each step removes a factor $a_{i_j}$
from $a_N$ and multiplies by $(-1)^{i_j-1}$, with the final result
\begin{equation} a^I a_N = (-1)^{\sum_{j=1}^r (i_j-1)} a_{I^c}. \end{equation}
Now I'll take care of the special cases on each extreme: $I$ is empty and $I = N$.  When 
$I$ is empty, Eq.~\eqref{definerecipmultivec} reduces to $a^I = a_N a_N^{-1} = 1$, which 
is correct, and when $I = N$, $I^c$ is empty so $a_{I^c} = 1$, so Eq.~\eqref{definerecipmultivec}
becomes
\begin{align} 
a^N & = (-1)^{\sum_{j=1}^n (j-1)} a_N^{-1} \nonumber \\
         & = (-1)^{n(n-1)/2} \, \frac{a_N}{{a_N}^2} \nonumber \\
         & = \frac{a_N}{|a_N|^2},
\label{framevols}
\end{align}
where I used Eq.~\eqref{bladeinv} in the last step.  Since this matches Eq.~\eqref{recipvol}, it too
is correct and the theorem is proved.
\ep

To wrap up this part, let me consider the special case where $\{a_i\}$ is an orthonormal 
frame, which I'll denote $\{e_i\}$.  Then the reciprocal frame is a lot easier to find: it's clear 
on inspection that $e^i = e_i^{-1} = e_i^{-2} e_i$ fits the bill.  Since the frame vectors
are normalized, $e_i^2 = \pm1$, so $e^i = \pm e_i$.  For any $I$, let $I_m$ be the 
number of elements in the product for $e_I$ that have negative square; then $e^I = 
(-1)^{I_m} e_I$, so the multivector basis and its reciprocal differ at most by signs.  
On such a basis, Eq.~\eqref{scprodBCcomps} for the scalar product becomes 
\begin{equation} 
\scprod{B}{C} = \sum_I (-1)^{I_m} B_I \, C_I = \sum_I (-1)^{I_m} B^I C^I.
\end{equation}
If the space of vectors is Euclidean, then $I_m = 0$ for any $I$, so
the magnitude is positive definite, so the entire geometric algebra is 
also a Euclidean space under the scalar product.  If the space of vectors
is non-Euclidean, then the algebra has (very) mixed signature under
the scalar product.

\subsection{Orthogonal projections using frames}
\label{projframes}

In traditional vector algebra, the orthogonal projection of a vector into a subspace is
given as a sum of projections onto a basis for the subspace.  Although 
we don't need to do that in geometric algebra, we still can.  Let $\bl{A_r} = a_1 \out 
\dotsb \out a_r$; then
\begin{align}
P_{\bl{A_r}}(a) & = a \lin \bl{A_r} \bl{A_r}^{-1} \nonumber \\
                           & = a \lin (a_1 \out \dotsb \out a_r) \bl{A_r}^{-1} \nonumber \\
                           & = \sum_{j=1}^r (-1)^{j-1} a \lin a_j (a_i \out \dotsb \out \check{a}_j \out \dotsb \out a_r)  \bl{A_r}^{-1}.
\end{align}
Now \bl{A_r} is a volume element for its subspace, so comparing with 
Eq.~\eqref{definerecipvec} shows me that the vectors in the sum above are the 
reciprocal frame to $\{a_j\}$, or
\begin{equation} 
P_{a_1 \out \dotsb \out a_r}(a) = \sum_{j=1}^r (a \lin a_j) a^j. 
\label{projvecsubspace}
\end{equation}
Since the reciprocal frame volume element $\bl{A^r} = a^1 \out \dotsb \out a^r$ 
equals $\bl{A_r}/|\bl{A_r}|^2$ (cf.\ Eq.~\eqref{framevols}), projection using either 
volume element gives the same result; had I used \bl{A^r}, I'd have ended up with 
Eq.~\eqref{projvecsubspace} with $a_j$ and $a^j$ interchanged.

I can do the same thing with any $s$-vector $B_s$.  Let \bl{A_r} be defined as before;
then using Eq.~\eqref{mveclinoutid} from Theorem \ref{Brlinoutai} I find that
\begin{align}
P_{\bl{A_r}}(B_s) & = B_s \lin \bl{A_r} \bl{A_r}^{-1} \nonumber \\
                                & = B_s \lin (a_1 \out \dotsb \out a_r) \bl{A_r}^{-1} \nonumber \\
                                & = \sum (-1)^{\sum_{j=1}^s (i_j-j)} (B_s \lin a_{i_1} \out \dotsb \out a_{i_s}) (a_{i_{s+1}} \out \dotsb \out a_{i_r}) \bl{A_r}^{-1} \nonumber \\
                                & = (-1)^{s(s-1)/2} \sum (-1)^{\sum_{j=1}^s (i_j-1)} (B_s \lin a_{i_1} \out \dotsb \out a_{i_s}) (a_{i_{s+1}} \out \dotsb \out a_{i_r}) \bl{A_r}^{-1} \nonumber \\
                                & = \sum (-1)^{\sum_{j=1}^s (i_j-1)} (\rev{B_s} \lin a_{i_1} \out \dotsb \out a_{i_s}) (a_{i_{s+1}} \out \dotsb \out a_{i_r}) \bl{A_r}^{-1},
\end{align}
where in the next to last line I used Eq.~\eqref{expdiff}.  If I now let $I$ be the sequence $i_1$ through
$i_s$ and use Eq.~\eqref{definerecipmultivec} for the reciprocal multivector basis, I find
\begin{align}
P_{\bl{A_r}}(B_s) & = \sum_I  (\rev{B_s} \lin a_I) (-1)^{\sum_{j=1}^s (i_j-1)}\, a_{I^c} \, \bl{A_r}^{-1} \nonumber \\
                                & = \sum_I (\scprod{B_s}{a_I}) a^I.
\end{align}
This expression is still true if I let the sum run over increasing sequences of any length, since all
the additional terms vanish.  Therefore for any multivector $B$
\begin{equation}
P_{a_1 \out \dotsb \out a_r}(B) = \sum_I (\scprod{B}{a_I}) a^I,
\label{projmultivecsubspace}
\end{equation}
where the sum runs over all increasing sequences, and the expression is still true 
if the bases $\{a_I\}$ and $\{a^I\}$ are interchanged.

\section{Linear algebra}
\label{linalg}

Now that I've said so much about linear spaces, let's take the next step and 
put some linear functions on them.  If $U$ and $V$ are vector spaces with 
the same set of scalars, a function $F: U \rightarrow V$ is said to be \emph{linear} if
$F(\alpha u + \beta v) = \alpha F(u) + \beta F(v)$, so linear functions respect linear combinations.
Linear functions have a very well-developed theory, and they're important all over 
applied mathematics; in fact, when a function isn't linear, one of the first things 
we do is consider its local linear approximation, the derivative.

In this section I'll hit the highlights of linear algebra using the tools and perspective 
of geometric algebra.  I'll start by reviewing some basic properties of linear functions, 
and then I'll introduce the adjoint.  I'll use it to describe three special types of functions: 
symmetric, skew symmetric, and orthogonal, each of which relates to its adjoint in 
a certain way.  All three have special forms in geometric algebra, which I'll consider 
in detail.  After that, I'll take a giant step into geometric algebra proper by showing 
how to take a linear function on vectors and extend it in a very natural way to every 
multivector in the whole algebra.  This is where geometric algebra really starts to shine, 
because it lets me see old topics in new and useful ways.  For example, our old friend the
eigenvector will be joined by eigenplanes, eigenvolumes, and more, and I'll show how
to use them to describe linear functions.  (It's immediately clear that a rotation has an 
eigenplane with eigenvalue $1$, for example.)  I'll also give a very easy and intuitive 
definition of the determinant, and I'll show how easy determinants are to calculate in 
geometric algebra.

I'm going to focus on functions that take vectors to vectors, and their
extensions to the whole algebra will be grade-preserving.  To some of you that might 
seem rather timid; since a geometric algebra is itself a (big) vector space, why not just jump in
with both feet and go right for linear functions from multivectors to multivectors, whether 
they preserve grade or not?  Well, of course you can, and we already have; duality does that, for 
example.  General tensors will also do that, and I'll consider them in due course.


\subsection{Preliminaries}
\label{prelims}

If $F$ and $G$ are linear, then so are their linear combinations, and so are their inverses if
they exist.  If $F: U \rightarrow V$, then $U$ and $V$ are called the \emph{domain} and 
\emph{codomain} of $F$ respectively.  (Some authors call the codomain the \emph{range}.)  
$F$ singles out two special subspaces: the \emph{kernel} of $F$, or $\Ker(F)$, is a subspace 
of the domain consisting of all the vectors that $F$ maps to $0$, and the \emph{range} of 
$F$, or $\Range(F)$, is a subspace of the codomain containing all the vectors that $F$ maps 
something to.  (The range is sometimes called the \emph{image}, presumably by the same 
folks who've already used the word range to mean the codomain.)  It's suggestive to think of 
$\Ker(F)$ as $F^{-1}(0)$ and $\Range(F)$ as $F(U)$.  $F$ is one-to-one iff $\Ker(F)=\{0\}$, 
and $F$ is onto iff $\Range(F)=V$.  The dimension of $\Ker(F)$ is called the \emph{nullity}
of $F$, or $\nullity(F)$, and the dimension of $\Range(F)$ is called the \emph{rank} of $F$, or
$\rank(F)$.  If the dimension of the domain $U$ is finite, the \emph{rank-nullity theorem} says 
\begin{equation} \rank(F) + \nullity(F) = \dim U. \end{equation}
So if both domain and codomain have the same finite dimension, $F$ is one-to-one iff it's also 
onto.  Therefore to show invertibility, you only have to show either one-to-one or onto, and the 
other part follows automatically.  If $W$ is a subspace of the domain of $F$, then the restriction 
of $F$ to $W$ is well-defined and also linear; it's denoted $F_W$.  Since blades represent 
subspaces, I'll sometimes write $F_{\bl{A}}$ for the restriction of $F$ to the subspace 
\bl{A} represents.

Often we care specifically about linear functions from $U$ to itself, which I'll call \emph{linear 
transformations} or \emph{linear operators}.  A pretty popular linear operator on any space is the 
identity; I denote the identity on $U$ by $\Id_U$.

Since our subject is geometric algebra, I will assume that all vector spaces have
inner products and belong to geometric algebras.  With that, recall that Theorem \ref{Afromscprods} in 
Section \ref{framesbases} shows that any multivector is determined uniquely by its scalar products 
with all multivectors.  Looking only at vectors, that means $u$ is uniquely fixed if one knows 
\scprod{u}{v} (or equivalently $u \lin v$) for all $v$.  This has two useful consequences.  The 
first lets me reconstruct linear functions.
\begin{thm}
A linear function $F: U \rightarrow V$ is completely determined by knowledge of \scprod{F(u)}{v} for all 
$u \in U$ and $v \in V$.
\label{Ffrominp}
\end{thm}
\bp 
In the finite-dimensional case, $F$ can be constructed explicitly: $F(u) = \sum_i \left[\scprod{a^i}{F(u)}\right] 
a_i$, where $\{a_i\}_{i=1, \dotsc, n}$ is any frame in $V$ and $\{a^j\}_{j=1, \dotsc, n}$ is its reciprocal 
frame.  Since we know all the \scprod{a^i}{F(u)}, we know $F(u)$.  We can use any frame for this 
construction since the set of all inner products determines $F(u)$ uniquely for each $u$ by Theorem
\ref{Afromscprods}. 

In the infinite-dimensional case, it's not obvious we can perform this construction, but for all applications
I know of, the space is rigged in some way to allow something like this to be done.  So I'll assume I 
can do it.
\ep  
Any linear transformation $F$ defines a new bilinear product on vectors by $a \lin F(b)$.  The second 
consequence of Theorem \ref{Afromscprods} lets me go the other way: I start with the product and define $F$.
\begin{thm}
If $\circ$ is a bilinear function from vectors to scalars, there's unique a linear transformation $F$ such that
$u \circ v = u \lin F(v)$ for all $u$ and $v$.
\label{prodhasF}
\end{thm}
\bp
Again, in finite dimensions the proof is by construction: Let $F(u) = \sum_i (a^i \circ u)a_i$.
This defines a linear function because the product $\circ$ is bilinear, and it satisfies
\begin{align}
u \lin F(v) & = u \lin \left( \sum_i (a^i \circ v) \, a_i\right) \nonumber \\ 
                  & = \sum_i (a^i \circ v) \, (u \lin a_i) \nonumber \\
                  & = \left( \sum_i (u \lin a_i) \, a^i \right) \circ v \nonumber \\
                  & = u \circ v.
\end{align}
Since $F$ is determined by $u \lin F(v) = u \circ v$ for all $u$ and $v$, this function is unique.

In the infinite-dimensional case, I will assume that whatever structure is needed to make
this result true has been added.
\ep
This is useful, because any given vector space can support many different inner products, but only
one inner product at a time can be encoded into a geometric algebra. This theorem tells me I have a way
to use the other products if I decide I need to.  Also, there are bilinear products that I want to use that
can't be inner products because they aren't symmetric.  This theorem lets me include those 
products too, although of course the corresponding function $F$ will have different properties.  
What those properties are will be the subject of future sections.

\subsection{The adjoint}
\label{adj}

If $F: U \rightarrow V$ is linear, then its \emph{adjoint} is the unique linear function $\adj{F}: V 
\rightarrow U$ defined by
\begin{equation} 
\scprod{\adj{F}(v)}{u} := \scprod{v}{F(u)} \quad \text{for all $u \in U$ and $v \in V$.} 
\label{definevecadj} 
\end{equation} 
Notice that \adj{F} switches domain and codomain compared to $F$ and that \scprod{}{} is 
interchangeable with either $\lin$ or $\rin$ in this definition.  The adjoint of the identity is pretty 
easy: $\adj{\Id_U} = \Id_U$.  Theorem \ref{Ffrominp} tells me how to construct \adj{F} explicitly 
in the finite-dimensional case:
\begin{align}
\adj{F}(v) & = \sum_i \left[\scprod{a^i}{\adj{F}(v})\right] a_i \nonumber \\
                  & = \sum_i \left[\scprod{F(a^i)}{v}\right] a_i.
\end{align}
(Notice that for this to make sense, $v \in V$ while the frame $\{a_i\} \subset U$.)  The bilinearity
of the inner product shows that taking the adjoint is itself a linear operation:
\begin{equation} \adj{\alpha F + \beta G} = \alpha \adj{F} + \beta \adj{G}.  \end{equation}
The relationship between $F$ and \adj{F} is symmetric, so each is the adjoint of the other, or 
equivalently
\begin{equation} \doubleadj{F} = F. \end{equation}
Suppose $F: U \rightarrow V$ and $G: V \rightarrow W$, and let $GF: U \rightarrow W$ denote the 
composition of $F$ and $G$.  Then
\begin{align}
\scprod{w}{GF(u)} & = \scprod{\adj{G}(w)}{F(u)} \nonumber \\
                     & = \scprod{\adj{F}\,\adj{G}(w)}{u},
\end{align}
which tells me that
\begin{equation} \adj{GF} = \adj{F}\,\adj{G}. \end{equation}  
A special case of this arises if $F$ is a operator on $U$, in which case $F^n$ is defined for any 
$n$ and 
\begin{equation} \adj{F^n} = (\adj{F})^n. \end{equation}
Now suppose $F: U \rightarrow V$ is invertible, so there's an $F^{-1}: V \rightarrow U$ such that 
$F^{-1}\,F=\Id_U$ and $F\,F^{-1} = \Id_V$.  In that case, since $\adj{F}\,\adj{F^{-1}} = \adj{F^{-1}\,F} = 
\adj{\Id_U} = \Id_U$, and similarly $\adj{F^{-1}}\,\adj{F}=\Id_V$,
\begin{equation} (\adj{F})^{-1} = \adj{F^{-1}}. \end{equation}

The special subspaces defined by a linear function and its adjoint are related in interesting ways.
\begin{thm}
If $F: U \rightarrow V$ is linear,
\begin{equation} \Ker(F) = \Range(\adj{F})^\perp. \end{equation}
If in addition $U$ is finite-dimensional,
\begin{equation} \rank(F) = \rank(\adj{F}). \end{equation}
\label{adjkerran}
\end{thm}
\bp
For the first part,
\begin{align}
u \in \Ker(F) & \quad \text{iff} \quad F(u) = 0 \nonumber \\
                      & \quad \text{iff} \quad F(u) \lin v = 0 \quad \text{for all $v \in V$} \nonumber \\
                      & \quad \text{iff} \quad u \lin \adj{F}(v) = 0 \quad \text{for all $v \in V$} \nonumber \\
                      & \quad \text{iff} \quad u \in \Range(\adj{F})^\perp. \nonumber
\end{align}
For the second part, we start with the rank-nullity theorem and the result of the first part:
\begin{align}
\rank(F) & = \dim U - \nullity(F) \nonumber \\
               & = \dim U - \dim \Ker(F) \nonumber \\
               & = \dim U - \dim \Range(\adj{F})^\perp.  \nonumber
\end{align}
Now $\Range(\adj{F})$ and $\Range(\adj{F})^\perp$ are duals, so their
dimensions add up to the dimension of $U$; so picking up where I left off,
\begin{align}
\rank(F) & = \dim U - \dim \Range(\adj{F})^\perp \nonumber \\
               & = \dim \Range(\adj{F}) \nonumber \\
               & = \rank(\adj{F}).
\end{align}
\ep
 
\subsection{Normal operators}
\label{normal}

In the next few sections I'll be considering operators that 
commute with their adjoints: $F \adj{F} = \adj{F} F$.  These are called \emph{normal 
operators}, and they have properties that I'll describe here so I can use them later.

\begin{thm} Any power of a normal operator is also normal. \end{thm}
\bp
If $F$ is normal, then $F^n (\adj{F})^n$ is easily transformed to $(\adj{F})^n F^n$ by moving 
all the \adj{F} factors past all the factors of $F$.
\ep

\begin{thm}
$F$ is normal iff $F(u) \lin F(v) = \adj{F}(u) \lin \adj{F}(v)$ for any $u$ and $v$.
\end{thm}
\bp
First assume $F$ is normal.  Then
\begin{align}
F(u) \lin F(v) & = \adj{F}F(u) \lin v \nonumber \\
                       & = F\adj{F}(u) \lin v \nonumber \\
                       & = \adj{F}(u) \lin \adj{F}(v).
\end{align}
Now assume the relation holds.  Then
\begin{align}
F \adj{F}(u) \lin v & = \adj{F}(u) \lin \adj{F}(v) \nonumber \\
                              & = F(u) \lin F(v) \nonumber \\
                              & = \adj{F}F(u) \lin v,
\end{align}
so by Theorem \ref{Ffrominp} $F\adj{F} = \adj{F}F$ and $F$ is normal.
\ep

\begin{thm}
$F$ is normal iff $F(u)^2 = \adj{F}(u)^2$ for all $u$.
\end{thm}
\bp
Since
\begin{equation}
F(u) \lin F(v) = \half\left[F(u+v)^2 - F(u)^2 - F(v)^2\right],
\end{equation}
$F(u)^2 = \adj{F}(u)^2$ for all $u$ implies $F(u) \lin F(v) = \adj{F}(u) \lin \adj{F}(v)$ for all $u$ and $v$.
On the other hand,   $F(u) \lin F(v) = \adj{F}(u) \lin \adj{F}(v)$ for all $u$ and $v$ implies $F(u)^2 = 
\adj{F}(u)^2$ for all $u$ just by considering the case $u=v$.  So squares are equal iff inner products are
equal, which takes us back to the previous theorem. 
\ep
\begin{thm}
If $F$ is normal and the inner product is nondegenerate on both $\Range(F)$ and $\Range(\adj{F})$, 
then $\Ker(F) = \Ker(\adj{F})$.  If in addition the domain of $F$ is finite-dimensional, 
$\Range(F) = \Range(\adj{F})$.
\end{thm}
\bp
For the first part,
\begin{align}
u \in \Ker(F) & \Longrightarrow F(u) = 0 \nonumber \\
                      & \Longrightarrow \adj{F}F(u) = 0 \nonumber \\
                      & \Longrightarrow F\adj{F}(u) = 0 \nonumber \\
                      & \Longrightarrow \adj{F}(u) \in \Ker(F).
\end{align}
By Theorem \ref{adjkerran}, $\Ker(F) = \Range(\adj{F})^\perp$, so $\adj{F}(u) \in \Range(\adj{F})^\perp$.  
But wait a second: $\adj{F}(u) \in \Range(\adj{F})$ by definition, and $\Range(\adj{F})$ is nondegenerate, 
so it must be that $\adj{F}(u) = 0$, so $u \in \Ker(\adj{F})$.  Therefore $\Ker(F) \subset \Ker(\adj{F})$.  
The same argument with $F$ and \adj{F} interchanged shows $\Ker(\adj{F}) \subset \Ker(F)$, so 
$\Ker(F) = \Ker(\adj{F})$.

For the second part, since the domain of $F$ is finite-dimensional  any subspace and its orthogonal
complement are duals, so each is the orthogonal complement of the other.  That and Theorem 
\ref{adjkerran} tell me that  $\Range(F) = \Ker(\adj{F})^\perp$ and $\Range(\adj{F}) = \Ker(F)^\perp$.  But 
I just showed that $\Ker(F) = \Ker(\adj{F})$, so $\Range(F) = \Range(\adj{F})$ too.
\ep
So if the conditions of this theorem are satisfied, normal $F$ and \adj{F} are both one-to-one (or onto) 
or neither one is.

\subsection{Symmetric and skew symmetric operators}
\label{symmskew}

A linear operator is \emph{symmetric} if it equals its adjoint, $\adj{F} = F$, and \emph{skew symmetric} 
or \emph{skew} if it is the negative of its adjoint, $\adj{F} = -F$.  The names come from this theorem.
\begin{thm}
The bilinear product $a \circ b := a \lin F(b)$ is (anti)symmetric iff $F$ is (skew) symmetric.
\label{meaningsymmskew}
\end{thm}
\bp
Since $a \lin F(b) = b \lin \adj{F}(a)$, it follows that $a \circ b = b \circ a$ iff $\adj{F} = F$ and $a \circ b = 
-b \circ a$ iff $\adj{F} = -F$. 
\ep
Recall that every bilinear product has this form for some $F$ (Theorem \ref{prodhasF}).

Both types of operator are normal, so all the results of Section \ref{normal} apply to them.  (They're all
pretty trivial in these cases, I have to admit.)  Further, every 
linear operator is the sum of a symmetric and a skew symmetric operator, because
\begin{equation} F = \half\left(F + \adj{F}\right) + \half\left(F - \adj{F}\right). \end{equation}
Also, for any linear operator $F$, both $\adj{F}F$ and $F\adj{F}$ are symmetric.

Powers of symmetric and skew symmetric operators are themselves symmetric or skew symmetric.
\begin{thm}
Any power of a symmetric operator is symmetric.  Any even power of a skew symmetric operator is
symmetric, and any odd power is skew symmetric.
\end{thm}
\bp
Since $\adj{F^n} = (\adj{F})^n$, $\adj{F} = F$ implies $\adj{F^n} = F^n$, so $F^n$ is symmetric also, 
and $\adj{F} = -F$ implies $\adj{F^n} = (-1)^n F^n$, so $F^n$ is symmetric or skew as $n$ is even or odd. 
\ep

The spectral theorem says that every symmetric $F$ has a frame $\{a_i\}$ of eigenvectors with eigenvalues 
$\{\lambda_i\}$, which means $F(a)$ for any $a$ is given by
\begin{equation} F(a) = \sum_i \lambda_i (\scprod{a}{a^i}) a_i. \label{spectralthonvec} \end{equation}
Conversely, every $F$ of this form is symmetric.  Analogous results hold on infinite-dimensional spaces 
with various additional restrictions.

Skew symmetric operators also have a canonical form, which is expressed very nicely in geometric algebra.
As motivation, notice that if $F$ is skew, then $a \lin F(a) = 0$, so $F$ maps any vector to an orthogonal vector.  
Well, I know something else that does that: taking the dual by a bivector.  In fact, the function $F(a) = a \lin A_2$ 
for any bivector $A_2$ is skew, because the resulting bilinear product is antisymmetric:
\begin{align}
a \circ b & = a \lin (b \lin A_2) \nonumber \\
               & = (a \out b) \lin A_2 \nonumber \\
               & = -(b \out a) \lin A_2 \nonumber \\
               & = -b \lin (a \lin A_2) \nonumber \\
               & = -b \circ a.
\end{align}
It turns out all skew functions are of this form.
\begin{thm}
$F$ is skew iff $F(a) = a \lin A_2$ for a unique bivector $A_2$.
\end{thm}
\bp
I just finished showing that any $F$ of this form is skew.  Knowing $F$, I can reconstruct $A_2$ uniquely using 
any frame $\{a_i\}$ and Theorem \ref{expandArinvecs} 
when $r=2$:
\begin{align}
A_2 & = \half \sum_i a^i \out (a_i \lin A_2) \nonumber \\
        & = \half \sum_i a^i \out F(a_i).
\end{align}
Now assume $F$ is skew and let $A_2$ be defined as above.  I find that for any $a$,
\begin{align}
a \lin A_2 & = \half a \lin \left(\sum_i a^i \out F(a_i)\right) \nonumber \\
                 & = \half \sum_i (a \lin a^i) F(a_i) - \half \sum_i a^i (a \lin F(a_i)) \nonumber \\
                 & = \half \sum_i (a \lin a^i) F(a_i) + \half \sum_i a^i (F(a) \lin a_i) \nonumber \\
                 & = \half F\left( \sum_i (a \lin a^i) \, a_i \right) + \half \sum_i (F(a) \lin a_i) \, a^i \nonumber \\
                 & = \half F(a) + \half F(a) \nonumber \\
                 & = F(a).
\end{align}
\ep
Therefore every antisymmetric bilinear product is of the form $a \out b \lin A_2 = a \lin A_2 \rin b$ for some $A_2$.

\subsection{Isometries and orthogonal transformations}
\label{orth}

The final special linear operator is an \emph{isometry}, which preserves inner products: 
$F(u) \lin F(v) = u \lin v$.  (Equivalently, isometries preserve squares of vectors.)  Isometries 
are always one-to-one, because
\begin{align}
F(u) = 0 & \Longrightarrow F(u) \lin F(v) = 0 \quad \text{for all $v$} \nonumber \\
               & \Longrightarrow u \lin v = 0 \quad \text{for all $v$} \nonumber \\
               & \Longrightarrow u = 0.
\end{align}
So in finite dimensions, isometries are also onto and thus invertible.  An invertible
isometry is called an \emph{orthogonal transformation}.  The two are distinct only on 
infinite-dimensional spaces, but most of the results I'll show don't actually require invertibility,
so I'll continue to make the distinction.

Any power of an isometry is also an isometry, as is clear from the definition.  An isometry satisfies
\begin{equation} u \lin v = F(u) \lin F(v) = \adj{F}F(u) \lin v, \end{equation}
so if $F$ is an isometry then $\adj{F}F = \Id$.  If $F$ is also invertible, then its inverse has 
to be \adj{F}, so we also have $F\adj{F} = \Id$.  Therefore orthogonal transformations satisfy 
$\adj{F} = F^{-1}$ and are also normal, and as a bonus $F^{-1}$ is orthogonal too.  And as with
isometries, any power of an orthogonal transformation is also orthogonal.

\subsubsection{Isometries and versors}
\label{versors}

So far I've described three isometries: the parity operation (which was extended to the whole 
algebra as grade involution in Section \ref{gradeinv}), reflections, and rotations (both in Section 
\ref{projrefrot}).  Now a rotation is two reflections, and as I showed in Section \ref{refmultivector},
the parity operation is reflection in a volume element, which amounts to $n$ reflections in succession.
So every isometry I've shown so far is a composition of reflections.  That's no accident: the 
\emph{Cartan-Dieudonn\'{e} theorem} shows that every isometry in an $n$-dimensional 
space is the composition of at most $n$ reflections along axes.  That's fantastic news, because
reflections are easy to do in geometric algebra;  so now we have powerful tools to perform
and analyze any isometry at all.

So what does a general isometry look like?  Remembering Eq.~\eqref{refdefagain}, I find that 
the isometry $F$ that takes vector $u$ and reflects it along axes $a_1, a_2, \dotsc, a_r$ in succession is
\begin{align}
F(u) & = (-1)^r(a_r \dotsb a_2 \, a_1) u (a_1^{-1} a_2^{-1} \dotsb a_r^{-1}) \nonumber \\
        & =  (a_r \dotsb a_2 \, a_1) \grinv[r]{u} (a_r \dotsb a_2 \, a_1)^{-1} \nonumber \\
        & = A_r \grinv[r]{u} A_r^{-1}
\end{align}
where $A_r = a_r \dotsb a_2 \, a_1$ is an invertible $r$-versor.  Thus a general isometry in finite dimensions
is grade involution followed by conjugation with an invertible versor.  (This is why I defined versors in the 
first place, and it's also why I've been proving so many results not just for blades but for versors in
general.)  Now this looks a lot like Eq.~\eqref{refainAr} for reflecting a vector in a subspace; in 
fact, Eq.~\eqref{refainAr} is just a special case of this result, since a blade is a special type of versor.
Therefore this operation extends to the whole algebra the same way reflection in 
subspaces did in Section \ref{refmultivector}: a general isometry on multivectors takes the form
\begin{equation} F(B) = A_r \grinv[r]{B} A_r^{-1} \end{equation}
and it reduces to reflection in a subspace iff the versor $A_r$ is an $r$-blade.  This also makes it
clear that the isometries generated by $A_r$ and $A_r^{-1}$ are inverses of each other.

Even though every versor is associated with an isometry, the association isn't exactly 
one-to-one.  After all, $A_r$ and $\lambda A_r$ generate the same isometry for any $\lambda \neq 0$.
(The ultimate reason for this is that $a$ and $\lambda a$ represent the same axis, and thus the same
reflection.)  We can eliminate most of that ambiguity, however, by composing our versors out of 
unit vectors; in that case, $A_r$ is a unit versor.  That doesn't eliminate the sign ambiguity, but we 
can live with that.  Is there any further ambiguity?  Amazingly, no.  I'll show this in two steps.  First,
basically the same argument used to derive Eq.~\eqref{reflectI} shows that 
\begin{equation} A_r \grinv[r]{\I} A_r^{-1} = (-1)^r \I, \label{orthI} \end{equation}
so isometries divide into two classes: \emph{even} isometries, which leave \I\ alone and are 
represented by even versors, and \emph{odd} isometries, which change the sign of \I\ and
are represented by odd versors.  This also shows that an isometry is odd iff it's the composition of an 
even isometry and one reflection.  As I said I would back in Section \ref{axioms}, I'll now start
referring to any even invertible versor as a rotor, so rotors represent even isometries.

Now I'll prove the result.
\begin{thm}
Versors $A_r$ and $B_s$ represent the same isometry iff $A_r = \lambda B_s$ for some $\lambda \neq 0$.
\end{thm}
So if we consider only unit versors, the association of versors to isometries is exactly two-to-one.
\bp
If $A_r$ is a nonzero multiple of $B_s$, we know they represent the same isometry, so let's prove it the other way.
Suppose $A_r \grinv[r]{u} A_r^{-1}  = B_s \grinv[s]{u} B_s^{-1}$ for all $u$.  Since $r$ and $s$ are 
both even or both odd, I can drop the grade involutions and I'm left with $A_r u A_r^{-1}  = 
B_s u B_s^{-1}$, which can be sneakily rewritten
\begin{equation} \commute{B_s^{-1}A_r}{u} = 0 \quad \text{for all $u$.} \end{equation}
Then Theorem \ref{commutewithvecs} tells me that $B_s^{-1}A_r = \lambda + \mu \grade[-]{\I}$.  
But $B_s^{-1}A_r$ is even, so its odd part vanishes and I'm left with $B_s^{-1}A_r = \lambda$.
Now if $\lambda= 0$, by Theorem \ref{prodnonzeroiszero} both $A_r$ and $B_s$ would be null, 
which they aren't since I've been inverting both of them.  Therefore $\lambda \neq 0$ and 
$A_r = \lambda B_s$.
\ep
You might be tempted at this point to associate an isometry with a unique sequence of reflections,
but you can't.  That's because the factorization of a versor into vectors isn't unique.  For example, 
suppose $a$ and $b$ are orthogonal Euclidean unit vectors; then
\begin{equation} ab = \left(\frac{a-b}{\sqrt{2}}\right) \left(\frac{a+b}{\sqrt{2}}\right), \end{equation}
so in this case two different sequences of reflections give the same isometry.  In fact, reflections along
any orthonormal basis for a subspace will result in reflection in that subspace, so in that case infinitely 
many reflection sequences produce the same isometry.  But that's only because they all produce the 
same versor to within a sign.

\subsubsection{Rotors and biversors}
\label{rotors}

Nothing I've done in this section up to now has made any assumptions about the scalars, but for this last part
I assume the scalars are real.  Every rotor is a product of biversors, so I want to take a moment to 
examine them. Consider $ab$ where $a$ and $b$ are unit vectors; it represents the composition of reflections in 
the $b$ and $a$ directions in succession, so the resulting isometry acts in the $a \out b$ plane.  
My plan to analyze $ab$ is to expand it as $a \lin b + a \out b$ and figure out each piece separately.  I'll do that
by starting with Eq.~\eqref{prodsq} from back in Section \ref{motive},
\begin{equation} a^2b^2 = (a \lin b)^2 - (a \out b)^2. \label{abnormsq} \end{equation}
Since $a$ and $b$ are unit vectors, the left hand side of Eq.~\eqref{abnormsq} is $\pm1$.  In what follows, I will
set $\mu = a \lin b$, and I will set $a \out b = \lambda \bl{B}$ where \bl{B} is a $2$-blade.  In the cases
when $(a \out b)^2 \neq 0$, I'll choose $\lambda$ so \bl{B} is a unit blade; otherwise I'll come up with
some other way to choose $\lambda$.

Before I get into the general cases, I'll handle a special case that I'll need to refer back to later: let $a$ 
and $b$ be orthogonal.  Since $a \lin b = 0$, $(a \out b)^2 = \pm1$.  So $ab = a \out b$ is already a unit 
blade \bl{B}.  To see what isometry it generates, let $u$ lie in \bl{B}; then its product with \bl{B} is an 
inner product, so they anticommute, so
\begin{equation} \bl{B} u \bl{B}^{-1} = -u\bl{B}\bl{B}^{-1} = -u. \end{equation}
So versor \bl{B} generates a reflection in the plane it represents.  (Which we already knew from Section 
\ref{refvecinspace}.)

The result in the general case depends on the sign of $\bl{B}^2$.
\begin{enumerate}
\item Suppose first that $\bl{B}^2=-1$.  Then Eq.~\eqref{Isq} when $n=2$ tells me that the number of 
          negative-square vectors in a frame for \bl{B} is either $0$ or $2$, so the inner product on \bl{B} 
          is either positive definite or negative definite, which I call the \emph{Euclidean} or \emph{elliptic} 
          case.  Then $a^2$ and $b^2$ have the same sign, so $a^2b^2 = 1$.  Putting all this in 
          Eq.~\eqref{abnormsq}, I find
          \begin{equation} 1 = \mu^2 + \lambda^2. \end{equation}
          Therefore $\mu = \cos(\theta/2)$ and $\lambda = -\sin(\theta/2)$ for some $\theta$, so 
          \begin{align} 
          ab & = a \lin b + a \out b \nonumber \\
                & = \cos(\theta/2) - \bl{B} \sin(\theta/2) \nonumber \\
                & = \exp\left(-\bl{B} \theta/2\right)
          \label{definiterotor}
          \end{align}
          where the exponential is defined by its power series.  You may remember this from the end of Section 
          \ref{simpleapps}: it's a rotation through angle $\theta$ in the plane defined by \bl{B}.  When $\theta = \pi$, 
          I recover the special case I solved above: a rotation by $\pi$ in a Euclidean plane equals a reflection 
          in the plane.
\item Now suppose $\bl{B}^2 = 1$.  In this case the inner product is indefinite, which is called the \emph{hyperbolic} 
          case.  Now I have to give some thought to $a^2$ and $b^2$.  First let them have the same sign, 
          so Eq.~\eqref{abnormsq} becomes
          \begin{equation} 1 = \mu^2 - \lambda^2. \end{equation}
          Therefore $\mu = \pm\cosh(\phi/2)$ and $\lambda = \mp\sinh(\phi/2)$ for some $\phi$, so 
          \begin{align} 
          ab & = a \lin b + a \out b \nonumber \\
                & = \pm\left(\cosh(\phi/2) - \bl{B} \sinh(\phi/2)\right) \nonumber \\
                & = \pm\exp\left(-\bl{B} \phi/2\right)
          \end{align}
          where again the exponential is defined by its power series.  (This isometry, by the way, is a rotation 
          in the hyperbolic plane, and in special relativity it's a boost to velocity $c\tanh\phi$.)  This time I couldn't 
          absorb the sign of $a \lin b$ into a choice for the parameter, because $\cosh$ is always positive; but 
          that affects only the rotor, not the corresponding isometry.  So aside from that, the rotors for the last two 
          cases have the same polar form, and the difference in their expansions as scalar plus bivector is due to the 
          different behaviors of the area element.  
\item Sticking with $\bl{B}^2=1$, now I consider $a^2=-b^2$.  Eq.~\eqref{abnormsq} becomes
          \begin{equation} {-1} = \mu^2 - \lambda^2, \end{equation}
          so $\mu$ and $\lambda$ change roles: $\mu = \mp\sinh(\phi/2)$ and $\lambda = \pm\cosh(\phi/2)$ for 
          some $\phi$, so for the rotor I get 
          \begin{align} 
          ab & = a \lin b + a \out b \nonumber \\
                & = \mp\sinh(\phi/2) \pm \bl{B} \cosh(\phi/2) \nonumber \\
                & = \pm\bl{B} \left( \cosh(\phi/2) - \bl{B}\sinh(\phi/2)\right) \nonumber \\
                & = \pm\bl{B} \exp\left(-\bl{B} \phi/2\right).
          \end{align}
          This rotor is the product of the previous one and the area element.  As I showed in the special case 
          above, this extra factor generates a reflection in the plane.  Why is it showing up as a separate factor?  
          Because unlike the Euclidean case, there is no hyperbolic rotation that performs a reflection in the 
          plane, so it has to be included separately.  
\item Finally, suppose $\bl{B}^2 = 0$, which means by Theorem \ref{nulleqdegen} that the inner product is 
          degenerate.  (This also has a name: the \emph{parabolic} case.)  Now Eq.~\eqref{abnormsq} reduces
          to $\pm1 = \mu^2$.  This doesn't make sense if the left hand side can be $-1$, so let me show that it can't.
          The inner product may be degenerate, but it can't be identically zero, or every vector would be null and 
          there would be no axes in the plane to reflect along.  Therefore there's a non-null vector somewhere in 
          there which I'll call $v$, and the direction orthogonal to it is a null vector which together with $v$ spans 
          the plane.  Because of this, the length squared of any vector in the plane is just $v^2$ times the square 
          of its component along $v$, so they all have the same sign.  Thus $a^2 b^2 = 1$ and $\mu = \pm1$.  
          Let \bl{B} be any $2$-blade that's convenient to use to represent the plane; then $a \out b = \mp\lambda \bl{B}$ 
          for some $\lambda$, so the rotor becomes
          \begin{align} 
          ab & = a \lin b + a \out b \nonumber \\
                & = \pm\left(1 - \frac{\lambda}{2} \bl{B}\right) \nonumber \\
                & = \pm\exp\left(-\bl{B} \lambda/2\right) 
          \end{align}
          where once again the exponential is defined by its power series.

          You may wonder what this rotor does.  Its inverse is $\pm(1 + \frac{\lambda}{2}\bl{B})$, so for any vector 
          $u$
          \begin{align}
          (ab)u(ab)^{-1} & = \left(1 - \frac{\lambda}{2}\bl{B}\right)u\left(1 + \frac{\lambda}{2}\bl{B}\right) \nonumber \\
                                    & = u + \frac{\lambda}{2}(u\bl{B} - \bl{B}u) - \frac{\lambda^2}{4} \bl{B}u\bl{B} \nonumber \\
                                    & = u + \lambda u \lin \bl{B} - \frac{\lambda^2}{4} \bl{B}u\bl{B}.
          \end{align}
          Each term on the right hand side is a vector, and you can directly verify that the square of the whole thing 
          really is $u^2$.  The verification is an interesting exercise; you find that $u \lin \bl{B}$ is orthogonal to $u$, 
          $\bl{B}u\bl{B}$ is orthogonal to $u \lin \bl{B}$, $\bl{B}u\bl{B}$ is null, and the inner product of $u$ and 
          $\bl{B}u\bl{B}$ cancels the square of $u \lin \bl{B}$. 
\end{enumerate}

Putting all this together, I've shown that a general even isometry on a real vector space consists of any 
number of rotations in planes (elliptic, hyperbolic, or parabolic) and reflections in hyperbolic planes.  An 
odd isometry is the same thing plus one reflection along an axis.  All the rotations can be represented in 
the same polar form; the properties of the different area elements produce different types of rotations.  
If the whole space is Euclidean, things simplify further because there are no hyperbolic or parabolic planes: 
every isometry is a sequence of rotations in planes, preceded (or followed) by one reflection if it's odd.

The set of all isometries on a finite-dimensional real vector space forms a group called the \emph{orthogonal 
group} on that space.  All of this analysis tells us two main things about such groups:
\begin{enumerate}
\item The subset of even isometries forms a group of its own (it includes the identity and is closed
          under products), and the subset of odd isometries is a one-to-one copy of the even subgroup.  A 
          reflection along any axis provides the relation between the two subsets.
\item Aside from reflections in hyperbolic planes, all elements of the even subgroup are functions of 
          parameters that can be continuously varied to zero, which results in the identity transformation.
\end{enumerate}
These properties tell me that a space's orthogonal group is an example of a \emph{Lie group}.  This is a
group which can be divided into a finite number of isomorphic subsets, and within each subset the elements
can be labeled with a finite number of continuously-variable parameters.  Whichever subset is lucky enough
to contain the identity is a subgroup in its own right, so a Lie group is a continuously-parameterised group
together with some isomorphic copies.  (In our case, reflections along axes and in hyperbolic planes 
move us back and forth between the copies.)  After I learn how to use geometric algebra to study Lie groups
in general, I believe I'll be showing that they form even subalgebras which are generated by exponentiating 
bivectors.  But this is more than enough on isometries for now.

\subsection{Extending linear functions to the whole algebra}
\label{linearextensions}

So far, I have considered only linear functions defined on vectors; but functions on
vectors have an obvious extension to the whole algebra.  For example, consider the
$r$-dimensional space spanned by $\{a_i\}$; this is mapped by linear function $F$ to 
the space spanned by $\{F(a_i)\}$.  Since blades represent subspaces, it seems very 
natural to define $F$ not just on vectors but on blades too; I set 
\begin{equation} F(a_1\out \dotsb \out a_r) := F(a_1) \out \dotsb \out F(a_r). \end{equation}
If I then require this formal extension of $F$ to be linear over multivectors, I get  
$F(A \out B) = F(A) \out F(B)$ for any $A$ and $B$.  Well, almost any $A$ and $B$; my 
picture doesn't tell me what $F$ should do to scalars.  Assuming I figure that out, 
then I've found a way to naturally extend any linear function on vectors to a function on 
the whole geometric algebra that is not only linear but also respects outer products.  As a 
matter of fact, I've actually done this already no fewer than six times.   Four of the 
extensions were orthogonal projections, orthogonal rejections, reflections, and rotations 
in Section \ref{projrefrot}, and the fifth was general isometries in Section \ref{orth}.  
I defined all five on vectors to start with, and I extended them in exactly the way I just 
suggested: I had them respect outer products.  The sixth extension was grade involution 
in Section \ref{gradeinv}; I started with the parity operation $u \rightarrow -u$, but I extended it 
by making it respect not outer products but the full geometric product.  (It ended up 
respecting outer products too, as the third equality in Eqs.~\eqref{invprods} shows.)  
In retrospect, that looks a little daring; after all, a product of vectors contains terms of 
many different grades, and the rules I imposed on, say, three-fold and five-fold products
could have put conflicting requirements on trivectors.  So far, though, it looks like everything 
worked out.  Whew.  While my other five extensions look safer, because I didn't have 
different grades crossing over, they do raise a question: how \emph{do} they act
on products?  Do you operate on each factor separately and then multiply them back together?
That looks like it'd work for general isometries (and thus reflections and rotations) because 
the internal factors would cancel out, but I'm not too sure about projections and rejections.  
And then there's one property that all six extensions have in common: they leave scalars 
alone.  What's up with that?  Grade involution does it by definition, but the others were 
found to do so after their final definitions were stated.  How come?

\subsubsection{Outermorphisms}
\label{outermorphs}

To answer these questions, I want to lay some groundwork by describing something a little
more general.  A linear function on geometric algebras that preserves outer products is called
an \emph{outermorphism}.  It's not too hard to show that the composition of outermorphisms 
is also an outermorphism and that the inverse of an outermorphism, if it exists, is an outermorphism 
too.  However, a linear combination of outermorphisms is not an outermorphism.  To show this, let
$\mathcal{F} = \alpha \mathcal{F}_1 + \beta \mathcal{F}_2$ where $\mathcal{F}_1$ and 
$\mathcal{F}_2$ are outermorphisms, and try evaluating both $\mathcal{F}(A \out B)$ and 
$\mathcal{F}(A) \out \mathcal{F}(B)$.  You'll see the problem pretty quickly.  For this reason,
compositions and inverses of outermorphisms pop up frequently, but linear combinations don't.

Next, outermorphisms are very restricted in how they handle scalars.  First linearity has a 
say: an outermorphism $\mathcal{F}$ has to be linear over everything, not just vectors, which means 
$\mathcal{F}(\lambda) = \mathcal{F}(\lambda 1) = \lambda \mathcal{F}(1)$.  So $\mathcal{F}(1)$ 
determines $\mathcal{F}$ for all scalars.  But scalar multiplication is also an outer 
product, so $1 = 1 \out 1$.  Thus the outermorphism property requires $\mathcal{F}(1) = 
\mathcal{F}(1) \out \mathcal{F}(1)$.  So like $1$, $\mathcal{F}(1)$ must equal its outer product 
with itself.  Something that equals its own square is said to be \emph{idempotent}; since we 
have several different products, we have several different types of idempotency.  I've pointed out 
that $1$ is an \emph{outer idempotent}, and thanks to the outermorphism property, 
$\mathcal{F}(1)$ has to be an outer idempotent too.  It turns out there aren't too many of those.
\begin{thm}
If $A = A \out A$, then A = 0 or 1.
\end{thm}
\bp
Let $A = \sum_r A_r$; then $A = A \out A$ becomes
\begin{equation} A = \sum_{s, t} \grade[s+t]{A_s A_t}. \end{equation}
I'll look at this one grade at a time.  The grade-$r$ part of this expression is
\begin{equation} A_r = \sum_{s=0}^r \grade[r]{A_s A_{r-s}}. \end{equation}
When $r=0$, this becomes
\begin{equation} A_0 = A_0^2, \end{equation}
so $A_0$ is either $0$ or $1$.  When $r=1$, I find
\begin{equation} A_1 = 2 A_0 A_1. \end{equation}
Whether $A_0$ is $0$ or $1$, this equation requires $A_1=0$.

To show that all remaining $A_r$ vanish, I proceed by induction.  Suppose it's true for
$r-1$; then most of the terms in the sum for $A_r$ drop out, leaving
\begin{equation} A_r = 2 A_0 A_r. \end{equation}
Whether $A_0$ is $0$ or $1$, this gives me $A_r=0$, and that completes the proof.
\ep

So $\mathcal{F}(1) = 0$ or $1$.  That means outermorphisms can do only two
things to scalars.
\begin{thm}
If $\mathcal{F}$ is an outermorphism, then either $\mathcal{F} = 0$ or $\mathcal{F}(\lambda) 
= \lambda$.
\label{outerleavesscalars}
\end{thm}
This is why all six extensions left scalars alone; they had to.
\bp
Either $\mathcal{F}(1) = 0$ or $\mathcal{F}(1) = 1$.  In the former case, for any $A$ 
\begin{align}
\mathcal{F}(A) & = \mathcal{F}(1A) \nonumber \\
                           & = \mathcal{F}(1 \out A) \nonumber \\ 
                           & = \mathcal{F}(1) \out \mathcal{F}(A) \nonumber \\
                           & = 0, 
\end{align}
so $\mathcal{F} = 0$.  In the latter case, $\mathcal{F}(\lambda) = \lambda$ for all $\lambda$
by linearity.
\ep

Next, I'll take a passing look at adjoints.  Linear functions on geometric algebras 
have adjoints just as they do on any other vector spaces:  if $\mathcal{F}: \G^1 \rightarrow 
\G^2$ is linear, $\adj{\mathcal{F}}: \G^2 \rightarrow \G^1$ is a linear function given by
\begin{equation} 
\scprod{\adj{\mathcal{F}}(B)}{A} := \scprod{B}{\mathcal{F}(A)} \quad \text{for all $A \in \G^1$ and $B \in \G^2$.}  
\label{defineadj}
\end{equation}
Outermorphisms are linear, so they have adjoints which are linear.  I can't say more than that,
though, until I consider a special class of outermorphisms, which is what's next.

\subsubsection{Outermorphism extensions}
\label{outerextends}

Now, back to the reason we're here.  My goal is 
to start with a linear function from vectors to vectors, define it on scalars by having it leave 
them alone (as Theorem \ref{outerleavesscalars} says I have to do), and then extend it to 
the rest of the algebra by linearity and respecting outer products.  The result is an 
outermorphism that matches the original function on vectors.  Can I start with any linear 
function and do this?  Why sure; in fact, I can do it exactly one way.

\begin{thm}
Let $F$ be a nonzero linear function that maps vectors to vectors; then there exists
a unique outermorphism \Out{F} that reproduces $F$ when applied to vectors.
\end{thm}
\bp
Existence is obvious, because definition on scalars and vectors, plus the outermorphism
property, plus linearity is enough to define \Out{F} on any multivector.  Uniqueness follows 
for the same reason.
\ep

I had to specify $F \neq 0$ because technically if $F = 0$ then Theorem \ref{outerleavesscalars} 
allows two extensions: one is $\Out{F} = 0$, and the other is $\Out{F}(A) = \grade{A}$.  
Both extensions vanish on all vectors and all blades; they differ only in the scalar option 
from Theorem \ref{outerleavesscalars} they use.

It's clear that \Out{F} maps $r$-blades into $r$-blades, so
\begin{equation}
\grade[r]{\Out{F}(A)} = \Out{F}(\grade[r]{A}).
\end{equation}
So by requiring \Out{F} to preserve grade $1$, I find that it preserves all grades.  
It also follows that
\begin{align} 
\grinv{\Out{F}(A)} & = \Out{F}(\grinv{A}) \nonumber \\
\rev{\Out{F}(A)} & = \Out{F}(\rev{A}) \nonumber \\
\clifconj{\Out{F}(A)} & = \Out{F}(\clifconj{A}).
\end{align}
If \G[1] is the vector space of \G, then it's also pretty clear that
\begin{equation} \Out{\Id_{\G[1]}} = \Id_{\G}. \end{equation}

Now let's see how outermorphism extensions behave under composition, inverses, and adjoints; 
I'll show compositions and inverses first.
\begin{thm}
If $U$, $V$, and $W$ are the vector spaces of $\G^1$, $\G^2$, and $\G^3$ respectively, and 
$F: U \rightarrow V$ and $G: V \rightarrow W$ are linear, then
\begin{itemize}
\item[(a)] $\Out{GF} = \Out{G} \Out{F}$.
\item[(b)] if $F$ is invertible, $\Out{F}^{-1} = \Out{F^{-1}}$.
\end{itemize}
\label{outercompadjinv}
\end{thm}
\bp
For part (a), there's no problem with the action on scalars or vectors, so all I need to check is the 
outermorphism property.  I will check it for the product of vectors, which covers all the higher-grade 
cases too.
\begin{align}
\Out{G}\left[\Out{F}(u_1 \out \dotsb \out u_r)\right] & = \Out{G}\left[F(u_1) \out \dotsb \out  F(u_r)\right] 
                                                                                                                                                            \nonumber \\
                                        & = GF(u_1) \out \dotsb \out GF(u_r) \nonumber \\
                                        & = \Out{GF}(u_1 \out \dotsb \out u_r).
\end{align}
Part (b) follows from part (a):
\begin{align}
\Out{F^{-1}}\,\Out{F} & = \Out{F^{-1}F} \nonumber \\
                                    & = \Out{\Id_U} \nonumber \\
                                    & = \Id_{\G^1},
\end{align}
and similarly $\Out{F}\,\Out{F^{-1}} = \Id_{\G^2}$, so \Out{F^{-1}} is the inverse of \Out{F}.
\ep

As for adjoints, outermorphism extensions obey these relations, which (for the moment) use an 
amazing number of brackets.
\begin{thm}
\begin{align}
A \lin \left\{\Out{\adj{F}}(B)\right\} & = \Out{\adj{F}}\left(\left\{\Out{F}(A)\right\} \lin B\right) \nonumber \\
\left\{\Out{F}(A)\right\} \rin B & = \Out{F}\left(A \rin \left\{\Out{\adj{F}}(B)\right\} \right)
\end{align}
Therefore $\adj{\Out{F}} = \Out{\adj{F}}$.
\label{outeradj}
\end{thm}
Because this theorem is hard to read, I'm going to state it in words.  The purpose of the adjoint 
is to let you move $F$ from one side of the scalar product to the other, as long as you change $F$ 
to \adj{F} along the way.  This theorem says you can also move $\Out{F}$ from the ``high side" of
the inner product to the ``low side," as long as (a) you change \Out{F} to \Out{\adj{F}} and (b) you 
then act on the whole thing with \Out{F}.  

Take a look at two extreme cases.  If the multivector on the low side is a scalar, the inner products
become products, \Out{F} leaves the scalar alone, and it factors out.  That's why there's an extra \Out{F} 
(or \Out{\adj{F}}) acting on the whole thing.  On the other extreme, if both sides have the same grade
then the inner products are scalars and the extra \Out{F} or \Out{\adj{F}} drops out because it leaves 
scalars alone.  With a little tweaking, that gets us back to the definition of the adjoint in Eq.~\eqref{defineadj}.  
I'll do that tweaking in the proof.
\bp
I'll prove only the first relation; the second relation is the reverse of the first with a few substitutions.  
The result is true for general $A$ and $B$ if it's true for $A_r$ and $B_s$.  If $r > s$ both sides 
vanish identically, so let $r \leq s$.  If $r=0$, then $A_r=\lambda$ and both sides reduce to $\lambda 
\Out{\adj{F}}(B_s)$.  If $s=0$, then $r=0$ so we're back to the previous case.  For the remaining 
cases, I consider blades \bl{A_r} and \bl{B_s}.  Next I'll prove $r=1$ and any $s \geq 1$.  It's true 
for $s=1$, so assume it's true for $s-1$, let $\bl{B_s} = b \out \bl{B_{s-1}}$, and consider
\begin{align}
a \lin \Out{\adj{F}}(\bl{B_s}) & = a \lin \Out{\adj{F}}(b \out \bl{B_{s-1}}) \nonumber \\
                                       & = a \lin \left[ \adj{F}(b) \out \Out{\adj{F}}(\bl{B_{s-1}}) \right] \nonumber \\
                                       & = [a \lin \adj{F}(b)] \, \Out{\adj{F}}(\bl{B_{s-1}}) - \adj{F}(b) \out 
                                                                                                                   \left[a \lin \Out{\adj{F}}(\bl{B_{s-1}})\right] 
                                                                                                                                                               \nonumber \\
                                       & = [F(a) \lin b] \, \Out{\adj{F}}(\bl{B_{s-1}}) - \adj{F}(b) \out \Out{\adj{F}}\left( F(a) 
                                                                                                                   \lin \bl{B_{s-1}}\right)       \nonumber \\
                                       & = \Out{\adj{F}}\left[ (F(a) \lin b) \, \bl{B_{s-1}}\right] - \Out{\adj{F}}\left[b \out \left( F(a) 
                                                                                                          \lin \bl{B_{s-1}} \right) \right]    \nonumber \\
                                       & = \Out{\adj{F}}\left[ (F(a) \lin b) \, \bl{B_{s-1}} - b \out  \left( F(a) \lin \bl{B_{s-1}}\right)   
                                                                                                                                                 \right]    \nonumber \\
                                       & = \Out{\adj{F}}\left[ F(a) \lin (b \out \bl{B_{s-1}}) \right] \nonumber \\
                                       & = \Out{\adj{F}}(F(a) \lin \bl{B_s}).
\end{align}
Now for general $r \leq s$.  Fix $s$ and assume the result is true for $r-1$; then let $\bl{A_r} = 
\bl{A_{r-1}} \out a$ and consider
\begin{align}
\bl{A_r} \lin \Out{\adj{F}}(\bl{B_s}) & = (\bl{A_{r-1}} \out a) \lin \Out{\adj{F}}(\bl{B_s}) \nonumber \\
                                                  & = \bl{A_{r-1}} \lin (a \lin \Out{\adj{F}}(\bl{B_s})) \nonumber \\
                                                  & = \bl{A_{r-1}} \lin \Out{\adj{F}}(F(a) \lin \bl{B_s}) \nonumber \\
                                                  & = \Out{\adj{F}}\left[ \Out{F}(\bl{A_{r-1}}) \lin (F(a) \lin \bl{B_s}) \right] \nonumber \\
                                                  & = \Out{\adj{F}}\left[  (\Out{F}(\bl{A_{r-1}}) \out F(a)) \lin \bl{B_s} \right] \nonumber \\
                                                  & = \Out{\adj{F}}\left(  \Out{F}(\bl{A_{r-1}} \out a) \lin \bl{B_s} \right) \nonumber \\
                                                  & = \Out{\adj{F}}\left(  \Out{F}(\bl{A_r}) \lin \bl{B_s} \right).
\end{align}
And that takes care of all cases.

Finally, I'll show that $\adj{\Out{F}} = \Out{\adj{F}}$.
\begin{align}
\scprod{A}{\Out{\adj{F}}(B)} & = \grade{\rev{A} \lin \left\{\Out{\adj{F}}(B)\right\}}  \nonumber \\
                                                 & = \grade{\Out{\adj{F}}\left(\left\{\Out{F}(\rev{A})\right\} \lin B\right)} \nonumber \\
                                                 & = \Out{\adj{F}}\left(\grade{\left\{\Out{F}(\rev{A})\right\} \lin B}\right) \nonumber \\
                                                 & = \grade{\left\{\Out{F}(\rev{A})\right\} \lin B} \nonumber \\
                                                 & = \grade{\rev{\left\{\Out{F}(A)\right\}} \lin B} \nonumber \\
                                                 & = \scprod{\Out{F}(A)}{B}.
\end{align}
Thus $\Out{\adj{F}}$ satisfies Eq.~\eqref{defineadj} with $\mathcal{F}=\Out{F}$, so $\adj{\Out{F}} = 
\Out{\adj{F}}$.
\ep

Given the uniqueness of the outermorphism extension and its good behavior under composition, 
inverses, and adjoints, I will now happily drop the \Out{F} notation and let $F$ refer 
either to the linear function on vectors or the resulting outermorphism.  That certainly makes  
Theorem \ref{outeradj} easier to read: I'll take $A \lin F(B) = F\left(\adj{F}(A) \lin B\right)$ any day.

Recall that the restriction of $F$ to the subspace represented by \bl{A_r} is denoted $F_{\bl{A_r}}$.
This restriction and $F(\bl{A_r})$ are related in an important way.

\begin{thm}
$F(\bl{A_r}) = 0$ iff $\rank(F_{\bl{A_r}}) < r$, which is true iff $F_{\bl{A_r}}$ is not one-to-one. 
\label{Fbleq0}
\end{thm}
\bp
$F(\bl{A_r}) = 0$ iff $F$ maps \bl{A_r} to a subspace of dimension smaller than $r$, which means
$\rank(F_{\bl{A_r}}) < r$.  Since \bl{A_r} is finite-dimensional, by the rank-nullity theorem this is true
iff $\nullity(F_{\bl{A_r}}) > 0$, so $F$ is not one-to-one.
\ep
So $\bl{A_r} \in \Ker{F}$ iff $\Ker(F_{\bl{A_r}}) \neq \{0\}$.  Note the two different meanings of $F$
in these two statements.

Now only one mystery remains unsolved: why did grade involution turn out to be an 
outermorphism even though I made it preserve geometric products instead of outer products?  
Because the parity operation on which it's based has a special property.
\begin{thm}
The following conditions on $F$ are equivalent.
\begin{enumerate}
\item $F$ is an isometry.
\item $F(AB)=F(A)\,F(B)$ for all $A$ and $B$.
\item $F(A \lin B) = F(A) \lin F(B)$ and $F(A \rin B) = 
          F(A) \rin F(B)$ for all $A$ and $B$.
\end{enumerate}
\end{thm}
\bp
I first assume $F$ is an isometry.  That means $F(a) \lin F(b) = a \lin b$ for all vectors; 
but $a \lin b$ is a scalar, so $F(a \lin b) = a \lin b$, so $F(a \lin b) = F(a) \lin F(b)$.  
Since $F(a \out b) = F(a) \out F(b)$ by the outermorphism property, it follows 
that $F(ab) = F(a)F(b)$.  I can extend this to $F(a_1 a_2 \dotsb a_r)
=F(a_1)F(a_2) \dotsb F(a_r)$ by induction:  the result is true for $r=2$, so assume it's true 
for $r-1$ and consider
\begin{align}
F(a_1 a_2 \dotsb a_r) 
 & = F[a_1 \lin (a_2 \dotsb a_r)] + F[a_1 \out (a_2 \dotsb a_r)] \nonumber \\
 & = F\Biggl[\sum_{j=2}^r (-1)^{j-2} a_1 \lin a_j \, a_2 \dotsb \check{a}_j \dotsb a_r \Biggr] +
     F(a_1) \out F[a_2 \dotsb a_r] \nonumber \\
 & = \sum_{j=2}^r (-1)^{j-2} a_1 \lin a_j F(a_2) \dotsb \check{F}(a_j) \dotsb F(a_r) + 
     F(a_1) \out \{F(a_2) \dotsb F(a_r)\} \nonumber \\
 & = \sum_{j=2}^r (-1)^{j-2} F(a_1) \lin F(a_j) F(a_2) \dotsb \check{F}(a_j) \dotsb F(a_r) + 
     F(a_1) \out \{F(a_2) \dotsb F(a_r)\} \nonumber \\
 & = F(a_1) \lin \{F(a_2) \dotsb F(a_r)\} + F(a_1) \out \{F(a_2) \dotsb F(a_r)\} \nonumber \\
 & = F(a_1) F(a_2) \dotsb F(a_r).
\label{FprodeqprodF}
\end{align}
Now let's look at $F(AB)$; by linearity it's a sum of terms of the form 
$F(\bl{A_r}\bl{B_s})$.  If $r=0$ or $s=0$ then $F(\bl{A_r}\bl{B_s})=
F(\bl{A_r}) \, F(\bl{B_s})$ by linearity and Theorem \ref{outerleavesscalars}, 
so let $\bl{A_r}=a_1 a_2 \dotsb a_r$ and $\bl{B_s}=b_1 b_2 \dotsb b_s$; then 
Eq.~\eqref{FprodeqprodF} lets me show
\begin{align}
F(\bl{A_r}\bl{B_s}) & = F(a_1 a_2 \dotsb a_r b_1 b_2 \dotsb b_s) \nonumber \\
 & = F(a_1) F(a_2) \dotsb F(a_r) F(b_1) F(b_2) \dotsb F(b_s) \nonumber \\
 & = F(a_1 a_2 \dotsb a_r) F(b_1 b_2 \dotsb b_s) \nonumber \\
 & = F(\bl{A_r})\,F(\bl{B_s}).
\end{align}
Therefore $F(AB)=F(A)\,F(B)$.  Notice that the products among 
the $a_i$ equal outer products, and the same is true of products among the $b_j$; it is the 
product of $a_r$ and $b_1$ that does not equal an outer product, and this is the reason 
I needed $F(ab)=F(a)F(b)$, not just $F(a \out b)=F(a) \out F(b)$, for the proof.

Next, assume $F(AB)=F(A)\,F(B)$ for all $A$ and $B$.  Then, since $F$ 
commutes with all grade operators,
\begin{align}
F(A_r \lin B_s) & = F(\grade[s-r]{A_r B_s}) \nonumber \\
 & = \grade[s-r]{F(A_r B_s)} \nonumber \\
 & = \grade[s-r]{F(A_r) \, F(B_s)} \nonumber \\
 & = F(A_r) \lin F(B_s),
\end{align}
with the last line following because $F$ preserves grades.  Thus
\begin{equation} F(A \lin B) = F(A) \lin F(B) \end{equation}
 in general.  Replacing $s-r$ with $r-s$ in the proof yields the same 
result for the right inner product.

Finally, assume $F(A \lin B) = F(A) \lin F(B)$ for all $A$ and $B$; then $F(a) \lin F(b)
= F(a \lin b) = a \lin b$ for any two vectors $a$ and $b$, where the last equality follows
because $a \lin b$ is a scalar.  Therefore $F$ is an isometry, and all three results listed above 
are equivalent.

If I had assumed $F$ was not just an isometry but orthogonal, I could have used $\adj{F}=F^{-1}$
and Theorem \ref{outeradj} to prove the third part, but this way is better because it assumes less 
(on infinite-dimensional spaces at least).
\ep

Thus an isometry may be extended by having it respect either outer products or geometric
products, with the same result.  That won't work for anything else, though.  This is why grade involution
came out fine, and that's why I could indeed have extended reflections and rotations by respecting
products instead of outer products.  Projections and rejections had to be done the way I did them,
however.


\subsection{Eigenblades and invariant subspaces}
\label{eigen}

Vector $a$ is an \emph{eigenvector} of $F$ with eigenvalue $\lambda$ if $F(a) = \lambda a$,
or equivalently $a \in \Ker(F - \lambda\Id)$.  In this definition $a$ can't be zero, but $\lambda$ can.
Now suppose $a_1$ and $a_2$ are eigenvectors with eigenvalues $\lambda_1$ and $\lambda_2$
and let $\bl{A} = a_1 \out a_2$; then
\begin{align}
F(\bl{A}) & = F(a_1 \out a_2) \nonumber \\ 
                & = F(a_1) \out F(a_2) \nonumber \\
                & = \lambda_1 \lambda_2 \, a_1 \out a_2 \nonumber \\
                & =  \lambda_1 \lambda_2 \, \bl{A}.
\end{align}
So \bl{A} is an \emph{eigenblade} of $F$; it's mapped by $F$ to a multiple of itself.  

 An eigenblade is defined generally by $F(\bl{A}) = \lambda \bl{A}$, regardless of what the 
factors of \bl{A} do.  Rotation operators nicely illustrate the different ways eigenblades can 
arise, and how they are related (or not) to eigenvectors.  Let $F(A) = \exp(-\bl{B} \theta/2) A 
\exp(\bl{B} \theta/2)$, so $F$ is a rotation in plane \bl{B} through angle $\theta$.  Then
\begin{align}
F(\bl{B}) & = \exp(-\bl{B} \theta/2) \bl{B} \exp(\bl{B} \theta/2) \nonumber \\
                & = \exp(-\bl{B} \theta/2) \exp(\bl{B} \theta/2) \bl{B} \nonumber \\
                & = \bl{B},
\end{align}
so \bl{B} is an eigenplane of the rotation operator with eigenvalue $1$.  However, 
since every vector in \bl{B} gets rotated, in general none of them are eigenvectors.  (The 
exception is $\bl{B}^2 = -1$ and $\theta = \pi$, which is a reflection: every vector in the plane is an 
eigenvector with eigenvalue $-1$.)  You can check that \dual{\bl{B}} is also an eigenblade with 
eigenvalue $1$, but that's for a different reason: every vector in \dual{\bl{B}} is left alone by the 
rotation.  So $F(\bl{A}) = \bl{A}$ is consistent with $F(a) = a$ for every vector in \bl{A}, but it's 
consistent with many other things too. 

If $\lambda = 0$, then Theorem \ref{Fbleq0} tells me that $F$ is not one-to-one 
on \bl{A}, but that's all it tells me.  On the other hand, $\lambda \neq 0$ tells me all sorts of things.
First, Theorem \ref{Fbleq0} says $F$ is one-to-one on \bl{A}, so $F_{\bl{A}}$ is invertible.  But 
it also hints at what $\Range(F_{\bl{A}})$ is.  In fact, it's hard to see how $F$ could map \bl{A} 
to a multiple of itself unless it also mapped all members of \bl{A} back into \bl{A}.  That would 
make \bl{A} an \emph{invariant subspace} of $F$; that is, a subspace that is mapped to itself 
by $F$.  Put that together with Theorem \ref{Fbleq0} and you get this result.
\begin{thm}
Any eigenblade of $F$ with nonzero eigenvalue is an invariant subspace of $F$ on which $F$ 
is invertible.
\end{thm}
\bp
Suppose $F(\bl{A}) = \lambda\bl{A}$ and $\lambda \neq 0$.  We already know that $F$ is
invertible on \bl{A}, so I'll prove the first part.  If $a$ lies in \bl{A}, then $\bl{A} \out a = 0$, 
in which case
\begin{align}
\bl{A} \out F(a) & = \lambda^{-1} \lambda \bl{A} \out F(a) \nonumber \\
                           & = \lambda^{-1} F(\bl{A}) \out F(a) \nonumber \\
                           & = \lambda^{-1} F(\bl{A} \out a) \nonumber \\
                           & = 0,
\end{align}
so $F(a)$ lies in \bl{A} too. 
\ep
So $F$ maps \bl{A} invertibly onto \bl{A}.  Now this is true for any $\lambda \neq 0$; but what does the
actual value of $\lambda$ tell us?  Well, if $F(\bl{A}) = \lambda\bl{A}$ then $|F(\bl{A})|^2 = \lambda^2
|\bl{A}|^2$, so the value of $\lambda$ determines how much the norm squared of \bl{A} changes.  If 
the scalars are real, I can interpret this further.  Recalling Eq.~\eqref{weightdef} for the weight of a blade, I 
find that
\begin{equation} \weight(F(\bl{A})) = |\lambda| \weight(\bl{A}), \end{equation}
so $F$ multiplies the weight associated with \bl{A} by $|\lambda|$; and $F$ changes the orientation
of \bl{A} if $\lambda$ is negative.  This suggests to me that $\lambda$ is actually the determinant of 
$F_{\bl{A}}$, since it seems to be the factor by which the volume of \bl{A} changes.  The idea that the determinant 
of a linear transformation is actually an eigenvalue, not of the original transformation but of its 
outermorphism extension, is worth following up on and has general validity, even if the scalars aren't
real.  So I'll do that next.

\subsection{The determinant}
\label{determinant}

The \emph{determinant} of a linear transformation is the factor by which it multiplies the volume
element of the space it acts on.  Finding it in geometric algebra is easy; we consider $F(\I)$.  This 
is an $n$-blade, so it has to be a multiple of \I.  (Put another way, \I\ is an eigenblade of all linear 
transformations.)  That multiple is the determinant, or
\begin{equation} F(\I) =: \det(F) \I. \end{equation}
An equivalent definition is
\begin{equation} \det(F) = \dual{F(\I)}. \end{equation}
This way of defining the determinant is very intuitive and also very easy to use in calculations, as I'll show.

First, it's obvious that $\det(\Id) = 1$.  Now let $F$ and $G$ be linear transformations; since
\begin{align}
\det(FG)\I & = FG(\I) \nonumber \\
                  & = F\left( \det(G) \I \right) \nonumber \\
                  & = \det(G) F(\I) \nonumber \\
                  & = \det(G) \det(F) \I, 
\end{align}
I find with minimum fuss that
\begin{equation} \det(FG) = \det(F) \det(G). \end{equation}
Therefore if $F$ is invertible,
\begin{align} 
\det(F^{-1}) \det(F) & = \det(F^{-1}F) \nonumber \\
                                  & = \det(\Id) \nonumber \\
                                  & = 1 
\end{align}
so
\begin{equation} \det(F^{-1}) = \det(F)^{-1}. \end{equation}
That tells me that $\det(F) \neq 0$ if $F$ is invertible.  I'll use that later.

Now for adjoints.  From the definition in Eq.~\eqref{defineadj},
\begin{align}
\scprod{\adj{F}(\I)}{\I} & = \scprod{\I}{F(\I)} \nonumber \\
\det(\adj{F}) \scprod{\I}{\I} & = \det(F) \scprod{\I}{\I} \nonumber \\
\det(\adj{F}) & = \det(F).
\end{align}
It's easy when you know how.

Next I'll calculate the determinants of some specific operators.  
\begin{itemize}
\item From Eq.~\eqref{projrejI} I get the determinants of orthogonal projections and rejections:
          \begin{align}
          \det(P_{\bl{A_r}}) & = \delta_{rn} \nonumber \\
          \det(R_{\bl{A_r}}) & = \delta_{r0}.
          \end{align}
\item Let $F$ be a symmetric operator with a frame $\{a_i\}$ of eigenvectors with eigenvalues 
         $\{\lambda_i\}$.  Then $a_1 \out \dotsb \out a_n$ is a volume element, so
         \begin{align}
         \det(F) \, a_1 \out \dotsb \out a_n & = F(a_1 \out \dotsb \out a_n) \nonumber \\
                                                                    & = F(a_1) \out \dotsb \out F(a_n) \nonumber \\
                                                                    & = \lambda_1 \dotsb \lambda_n \, a_1 \out \dotsb \out a_n
         \end{align}
         so
         \begin{equation} \det(F) = \lambda_1 \dotsb \lambda_n. \end{equation}
\item If $F$ is orthogonal, then $F(v) = A_r \grinv[r]{v} A_r^{-1}$ for some rotor $A_r$, so 
         Eq.~\eqref{orthI} shows that
         \begin{equation} \det(F) = (-1)^r. \end{equation}
\end{itemize}

In real matrix algebra, there's a well-known relationship between determinants, invertibility, and adjoints:
a matrix is invertible iff its determinant is nonzero, in which case its inverse is its adjugate (adjoint of
the cofactor matrix) divided by its determinant.  The geometric algebra equivalent, as you would expect,
applies to the linear operator itself, not its matrix representation on some basis.  On top of that, it's valid 
for any multivector. 
\begin{thm}
$F$ is invertible iff $\det(F) \neq 0$, and for any multivector $A$
\begin{equation} F^{-1}(A) = \frac{\dual{\adj{F}(\invdual{A})}}{\det(F)}. \end{equation}
\end{thm}
Recall that \dual{} is the duality transform and \invdual{} is its inverse.
\bp
I've already proven that if $F$ is invertible, $\det(F) \neq 0$, so let's go the other way.  Suppose
$\det(F) \neq 0$ and let $G$ be defined by $G(A) = \dual{\adj{F}(\invdual{A})}$.  Then I use the
result $A \lin \adj{F}(B) = \adj{F}(F(A) \lin B)$ from Theorem \ref{outeradj} to get
\begin{align}
GF(A) & = G[F(A)] \nonumber \\
            & = \dual{\adj{F}[\invdual{F(A)}]} \nonumber \\
            & = \adj{F}[F(A) \lin \I] \I^{-1} \nonumber \\
            & = [A \lin \adj{F}(\I)] \I^{-1} \nonumber \\
            & = \det(\adj{F})  (A \lin \I) \I^{-1} \nonumber \\
            & = \det(F) A \I \I^{-1} \nonumber \\
            & = \det(F) A.
\end{align}
A similar argument shows $FG(A) = \det(F) A$, so $F^{-1} = G/\det(F)$.
\ep
You can see that you need the full apparatus of geometric algebra to do this: I take the dual, which is a 
geometric product with a volume element, and I need to use the outermorphism extension of $F$, 
because calculating $F^{-1}(a)$ involves calculating \adj{F}(\invdual{a}), and \invdual{a} is not 
a vector.  (Unless we're in a two-dimensional space, I suppose.)

\section{Applications}
\label{apps}

\subsection{Classical particle mechanics}
\label{classicalmech}

The most obvious place to apply geometric algebra is classical mechanics, since 
it relies heavily on vector algebra already.  In this section only I'll adopt the notational 
conventions of classical mechanics, so vectors are denoted $\bm{a}$, $\bm{b}$, 
and so on, the magnitude of vector $\bm{a}$ is denoted $a$, unit vectors are 
indicated with an overhat $\hat{\  }$, and the derivative of any quantity with respect 
to time is indicated by an overdot $\dot{\ }$.   Since boldface means something else, I
will not use boldface for blades in this section.  The material in this section is largely drawn
from \cite{NewFounds} and \cite{GAforphys}.

\subsubsection{Angular momentum as a bivector}
\label{Lbivec}

As a particle moves over time, its position vector $\bm{r}$ sweeps out area
at a rate that a picture easily shows to be
\begin{equation} \dot{A} = \half \bm{r} \out \bm{v} \end{equation}
where $\bm{v} = \dot{\bm{r}}$ is the particle's velocity vector.  
Unsurprisingly, the rate at which area is swept out is a $2$-blade.  
This blade is proportional to the dynamical quantity
\begin{equation} 
L := \bm{r} \out \bm{p} = m \bm{r} \out \bm{v} = 2 m \dot{A}, 
\end{equation}
called the \emph{angular momentum}.  In standard vector algebra, angular 
momentum is defined to be the vector $\bm{L} = \bm{r} \times \bm{p}$, the
cross product of $\bm{r}$ and $\bm{p}$; the definition given here is the
dual of that vector (see Eq.~\eqref{crossprod}), which is more natural given
the association with areas.  Nonetheless, since the algebraic properties of the
outer and cross products are so similar, much of what one knows from the 
standard treatment holds without change; for example,
\begin{align}
\dot{L} & = m \bm{v} \out \bm{v} + m \bm{r} \out \dot{\bm{v}} \nonumber \\
        & = m \bm{r} \out \dot{\bm{v}} \nonumber \\
        & = \bm{r} \out \bm{F},
\end{align}
so $L$ is conserved iff the force $\bm{F}$ is central (parallel or antiparallel 
to $\bm{r}$).  Since $L$ is conserved iff $\dot{A}=0$, I have Kepler's Second 
Law: the position vector of a particle subject to central forces sweeps out 
equal areas in equal times.  Further, the plane in which central force motion 
takes place is $L$ itself.  Writing $\bm{r} = r \bm{\hat{r}}$, which implies $\bm{v} 
= \dot{r} \bm{\hat{r}} + r \dot{\bm{\hat{r}}}$, I find
\begin{align}
L & = \bm{r} \out \bm{p} = m \bm{r} \out \bm{v} \nonumber \\
  & = m r \bm{\hat{r}} \out ( \dot{r} \bm{\hat{r}} + 
                                r \dot{\bm{\hat{r}}} ) \nonumber \\
  & = m r^2 \bm{\hat{r}} \out \dot{\bm{\hat{r}}}
\end{align}
since $\bm{\hat{r}} \out \bm{\hat{r}} = 0$.  But I know a bit more than that; 
$\bm{\hat{r}}$ is a unit vector, or $\bm{\hat{r}} \lin \bm{\hat{r}} = 1$, the time 
derivative of which is $\bm{\hat{r}} \lin \dot{\bm{\hat{r}}} = 0$. This is just the 
familiar fact that a constant-length vector and its time derivative must always 
be perpendicular.  (Incidentally, this shows that $\bm{v}= \dot{r} \bm{\hat{r}} + 
r \dot{\bm{\hat{r}}}$ is a decomposition into radial and tangential components.)  
If $\bm{\hat{r}} \lin \dot{\bm{\hat{r}}} = 0$, then $\bm{\hat{r}} \out \dot{\bm{\hat{r}}} 
= \bm{\hat{r}} \dot{\bm{\hat{r}}}$, so
\begin{equation} 
L = m r^2 \bm{\hat{r}} \dot{\bm{\hat{r}}}. 
\label{Ldef}
\end{equation}
Now this is nice because the geometric product has better properties than the 
outer product, and this is the first algebraic feature of this treatment that 
is genuinely new.  Since $L$ is a bivector,
\begin{equation}
L = -\rev{L} = -mr^2 \dot{\bm{\hat{r}}} \bm{\hat{r}}
\end{equation}
so the scalar $l$, the magnitude of $L$, is given by
\begin{equation}
l^2 := |L|^2 = \rev{L}L = -L^2 = m^2 r^4 \dot{\bm{\hat{r}}}^2.
\end{equation}
Notice that $l$ equals the magnitude of the angular momentum vector from standard
treatments.  It should, of course, since bivector $L$ and the angular momentum
vector are duals.

\subsubsection{The Kepler problem}

The Kepler problem is to determine the motion of a point particle of mass $m$ 
moving in a potential of the form $V=-k/r$, where $r$ is the particle's distance from
some fixed origin.  The particle experiences a force
\begin{equation}
\bm{F} = -\frac{k}{r^2}\bm{\hat{r}}
\end{equation}
where the constant $k$ is positive for an attractive force and negative for a 
repulsive force, so the particle's acceleration is given by
\begin{equation}
\dot{\bm{v}} = -\frac{k}{mr^2}\bm{\hat{r}}.
\label{Keplera}
\end{equation}
Now take a look at Eqs.~\eqref{Ldef} and \eqref{Keplera}.  One is proportional 
to $r^2$ while the other is inversely proportional to $r^2$, so their product 
is independent of $r$.  In fact, let me calculate the product:
\begin{align}
L \dot{\bm{v}} & = \left(-mr^2 \dot{\bm{\hat{r}}} \bm{\hat{r}}\right) 
                   \left(-\frac{k}{mr^2}\bm{\hat{r}}\right) \nonumber \\
               & = k \dot{\bm{\hat{r}}}, \nonumber
\end{align}
and since $L$ is conserved and $k$ is a constant, this implies
\begin{equation} 
\frac{d}{dt}\left(L\bm{v} - k\bm{\hat{r}}\right) = 0. 
\end{equation}
Well, look at that: another constant of motion.  The second term in the constant, 
$k\bm{\hat{r}}$, is clearly a vector, and the first term can be written
\begin{align}
L \bm{v} & = L \rin \bm{v} + L \out \bm{v} \nonumber \\
         & = L \rin \bm{v} + m \bm{r} \out \bm{v} \out \bm{v} \nonumber \\
         & = L \rin \bm{v}.
\end{align}
This is hardly a surprise; $\bm{v}$ is a vector in the plane defined by $L$, so by 
Theorems \ref{innerprodmean} and \ref{AsubspaceofB}, $L \bm{v} = L \rin \bm{v}$ 
is a nonzero 
vector in the plane of $L$ perpendicular to $\bm{v}$.  Thus the conserved quantity 
is a vector in the plane of motion; it is often called the ``Laplace-Runge-Lenz 
vector,'' and in traditional vector algebra treatments of the Kepler problem it typically 
appears at the end as the result of a great deal of work.  Here it was the first thing I found.

I would actually prefer to define a dimensionless conserved vector, and this 
quantity clearly has dimensions of $k$, so I define a conserved vector $\bm{e}$
by
\begin{equation} \bm{e} := \frac{L\bm{v}}{k} - \bm{\hat{r}}. \label{ecvec} \end{equation}
I'd like to use this equation to get further expressions describing the motion
of the particle; first is the polar equation, $r$ as a function of direction.  Since 
the expression for $\bm{e}$ has $\bm{\hat{r}}$ in it and $\bm{\hat{r}} \bm{r} = r$, 
it follows that I can get an equation for $r$ by multiplying Eq.~\eqref{ecvec} by 
$k\bm{r}$, with the result
\begin{equation} 
L\bm{v}\bm{r} = k(\bm{\hat{r}}\bm{r} + \bm{e}\bm{r}). \end{equation}
The left hand side equals
\begin{align}
L\bm{v}\bm{r} & = L (\bm{v} \lin \bm{r} + \bm{v} \out \bm{r} ) \nonumber \\
 & = (\bm{r} \lin \bm{v})L - \frac{L^2}{m} \nonumber \\
 & = \frac{l^2}{m} + (\bm{r} \lin \bm{v})L 
\label{Lvr}
\end{align}
while $\bm{\hat{r}}\bm{r} = r$ and $\bm{e}\bm{r} = e r\cos\theta + \bm{e} 
\out \bm{r}$, so putting it all together
\begin{equation}
\frac{l^2}{m} + (\bm{r} \lin \bm{v})L = k(r + e r\cos\theta + \bm{e} 
                                           \out \bm{r}), 
\end{equation}
or on separating the scalar and bivector parts,
\begin{align}
\frac{l^2}{m} & =  k(r + e r\cos\theta) \nonumber \\ 
(\bm{r} \lin \bm{v})L & = \bm{e} \out \bm{r}. 
\end{align}
The scalar equation can be solved for $r$ with the result
\begin{equation} r = \frac{l^2/mk}{1 + e\cos\theta}, \end{equation}
which is the equation for a conic section with eccentricity $e$ and one 
focus at the origin.  Since the length of $\bm{e}$ is the eccentricity 
of the orbit, $\bm{e}$ is naturally called the \emph{eccentricity vector}, 
which is the name I'll use for it henceforth.

The direction of $\bm{e}$ also has a geometrical meaning, but it's
different in the attractive and repulsive cases, so I'll do one at a time.  First
I assume $k > 0$ and I note that $r$ equals its minimum and maximum values 
when $\theta = 0$ and $\pi$ respectively, which means that $\bm{e}$ points 
toward the particle's point of closest approach, called its \emph{periapsis}, 
and away from its point of farthest retreat, called the \emph{apoapsis}.  
(Fun fact: these two points are called the perigee and apogee if you're orbiting
the earth, the perihelion and aphelion if you're orbiting the sun, and the 
pericynthion and apocynthion if you're orbiting the moon.  So now you know.)  

Now the repulsive case.  If $k < 0$, we run into a problem:  $r$ has to be 
non-negative, so we have to have $1 + e\cos\theta \leq 0$ for at least some values 
of $\theta$, and the orbit may include only those values.  This is possible iff 
$e > 1$, with the result that the orbit is a hyperbola.  In this case, $r$ takes on its 
smallest value when $\theta = \pi$, so in the repulsive case the eccentricity vector 
points away from the periapsis.

The motion in the Kepler problem is completely determined by two vectors,
the initial position and velocity, which are themselves determined by six parameters.  
The conserved angular momentum supplies three parameters because it's a bivector, 
and the conserved eccentricity vector supplies two more (only two because the angular
momentum fixes the plane of the motion), so the motion is completely determined 
by these two constants plus one further parameter, which may be taken to be the 
initial value of $\theta$.  If that's the case, then anything that doesn't depend on the 
starting point, such as any other constants of motion, should be a function of only $L$ 
and $\bm{e}$.  I'll now show that this is the case for the energy by finding the magnitude 
of the eccentricity vector.
\begin{align}
(L\bm{v} - k\bm{\hat{r}})^2 & = k^2 e^2 \nonumber \\
(L\bm{v})^2 - 2 k (L\bm{v}) \lin \bm{\hat{r}} + k^2 & = k^2 e^2
\label{esq}
\end{align}
Using the fact that a vector equals its own reverse, the first term on the left 
hand side can be calculated as
\begin{align}
(L\bm{v})^2 & = L\bm{v} L\bm{v} \nonumber \\
            & = \rev{(L\bm{v})}L\bm{v} \nonumber \\
            & = \bm{v}\rev{L}L\bm{v} \nonumber \\
            & = l^2 v^2.
\end{align}
The second term on the left is $-2k$ times
\begin{align}
(L\bm{v}) \lin \bm{\hat{r}} & = \frac{(L\bm{v}) \lin \bm{r}}{r} \nonumber \\
                           & = \frac{\grade{L\bm{v}\bm{r}}}{r} \nonumber \\
                           & = \frac{l^2}{mr},
\end{align}
where in the last line I used Eq.~\eqref{Lvr}, so now Eq.~\eqref{esq} becomes
\begin{align}
l^2 v^2 - \frac{2kl^2}{mr} & = k^2(e^2-1) \nonumber \\
\frac{2l^2}{m}\left(\half mv^2 - \frac{k}{r}\right) & = k^2(e^2-1)
\end{align}
or
\begin{equation}
E = \frac{mk^2}{2l^2}(e^2-1).
\end{equation}
This gives the energy in terms of $l$ and $e$.

I have derived all the main results of the Kepler problem (except for the time evolution) 
a whole lot more easily than standard treatments do.  In fact, many textbooks don't 
even get to the eccentricity vector.  Here geometric algebra is clearly superior to standard 
vector algebra both for solving the equations and for understanding the results.

\appendix

\section{Summary of definitions and formulas}
\label{app:summary}

\setcounter{axiom}{0}
\setcounter{equation}{0}

\subsection{Notation}
\label{app:notation}

\begin{tabbing}
\G \hspace{.85in} \= Geometric algebra \\
\G[r] \> Grade-$r$ subspace of \G\ (space of $r$-vectors) \\
$\G^n$ \> Geometric algebra of an $n$-dimensional vector space \\ 
\\
$A$, $B$, etc. \> General multivector \\
$\lambda$, $\mu$, etc. \> Scalar \\
$a$, $b$, $u$, $v$, etc. \> Vector \\
$A_r$ \> $r$-vector (sometimes grade-$r$ part of $A$) \\
\bl{A_r} \> $r$-blade \\
$A_+$ \> Even-grade multivector \\
$A_-$ \> Odd-grade multivector \\
\I \> Volume element \\ 
\\
\grade[r]{A} \> Grade-$r$ part of $A$ \\
\grade{A} \> Scalar (grade-$0$) part of $A$ \\
\grade[+]{A} \> Even-grade part of $A$ \\
\grade[-]{A} \> Odd-grade part of $A$ \\
$A^{-1}$ \> Inverse of $A$ \\
\grinv{A} \> Grade involution of $A$ \\
\grinv[r]{A} \> $r$ times grade involuted $A$ \\
\rev{A} \> Reverse of $A$ \\
\clifconj{A} \> Clifford conjugate of $A$ \\
$|A|^2$ \> Squared norm of $A$ \\
\dual{A} \> Dual of $A$ \\
\invdual{A} \> Inverse dual of $A$ \\
\\
$AB$ \> Geometric product of $A$ and $B$ \\
$A \lin B$ \> Left inner product of $A$ into $B$ \\
$A \rin B$ \> Right inner product of $A$ by $B$ \\
$A \out B$ \> Outer product of $A$ and $B$ \\
\scprod{A}{B} \> Scalar product of $A$ and $B$ \\
\commute{A}{B} \> Commutator of $A$ and $B$ \\
$P_{\bl{A_r}}(B)$ \> Orthogonal projection of $B$ into \bl{A_r} \\
$R_{\bl{A_r}}(B)$ \> Orthogonal rejection of $B$ from \bl{A_r} \\
\\
$U$, $V$, $W$, etc. \> Vector space \\
$F$, $G$, etc. \> Linear function of vectors (or its outermorphism extension) \\
$F_U$ \> Restriction of $F$ to subspace $U$ \\
$F_{\bl{A}}$ \> Restriction of $F$ to subspace represented by \bl{A} \\
$\Id$ \> Identity function \\
$\Ker(F)$ \> Kernel of $F$ \\
$\Range(F)$ \> Range of $F$ \\
$\nullity(F)$ \> Nullity of $F$ \\
$\rank(F)$ \> Rank of $F$ \\ 
\adj{F} \> Adjoint of $F$ \\
$\det(F)$ \> Determinant of $F$ \\
$\mathcal{F}$ \> Outermorphism 
\end{tabbing}

\subsection{Axioms}
\label{app:axioms}

A geometric algebra \G\ is a set with two composition laws, addition and multiplication, 
that satisfy these axioms.

\begin{axiom}
\G\ is a ring with unit.  The additive identity is called $0$ and 
the multiplicative identity is called $1$.
\end{axiom}
\begin{axiom}
\G\ contains a field \G[0] of characteristic zero which includes $0$ 
and $1$.
\end{axiom}
\begin{axiom}
\G\ contains a subset $\G[1]$ closed under addition, and 
 $\lambda \in \G[0], v \in \G[1]$ implies $\lambda v = v \lambda \in \G[1]$.
\end{axiom}
\begin{axiom}
The square of every vector is a scalar.
\end{axiom}
\begin{axiom}
The inner product is nondegenerate.
\end{axiom}
\begin{axiom}
If $\G[0] = \G[1]$, then $\G = \G[0]$.  Otherwise, \G\ is the direct sum of all the \G[r].
\end{axiom}

\subsection{Contents of a geometric algebra}
\label{app:contents}

An $r$-blade \bl{A_r} is the outer product of $r$ vectors, $a_1 \out \dotsb \out a_r$.
It represents the subspace spanned by $\{a_j\}_{j=1,\dotsc,r}$, with a weight and orientation
if the scalars are real. \\

\noindent $\bl{A_r} = 0$ iff the $a_j$ are linearly dependent. \\

\noindent \bl{A_r} and \bl{B_r} define the same subspace iff $\bl{A_r} = \lambda \bl{B_r}$. \\ 

\noindent If \bl{A_r} is a proper subspace of \bl{A_s}, then \bl{A_r} can be factored out of 
\bl{A_s} from either the left ($\bl{A_s} = \bl{A_r} \out \bl{A_{s-r}}$) or the right ($\bl{A_s} = 
\bl{A_{s-r}} \out \bl{A_r}$). The grade-$s-r$ factors in each case may be chosen to be the 
same except for at most a sign. \\

\noindent $a \out \bl{A_r} = 0$ iff $a$ lies in \bl{A_r}. \\

\noindent $a \lin \bl{A_r} = 0$ iff $a$ is orthogonal to \bl{A_r}. \\

\noindent The reflection of multivector $B$ in subspace \bl{A_r} is $\bl{A_r} \grinv[r]{B} \bl{A_r}^{-1}$. \\

\noindent \bl{A_r}, \grinv{\bl{A_r}}, \rev{\bl{A_r}}, \clifconj{\bl{A_r}}, and $\bl{A_r}^{-1}$ 
(if it exists) represent the same subspace.

\subsection{The inner, outer, and geometric products}
\label{app:inoutgeom}

\begin{equation} 
A_r B_s = \sum_{j=0}^{\min\{r,s\}} \grade[|r-s|+2j]{A_r B_s}
\end{equation}

\begin{equation} 
\grade[r+s-2j]{A_r B_s} = (-1)^{rs-j} \grade[r+s-2j]{B_s A_r}
\end{equation}

\begin{align} 
\grade{A B} & = \grade{B A} \\
 & = \grade{\grinv{A} \grinv{B}} \\
 & = \grade{\rev{A} \rev{B}} \\
 & = \grade{\clifconj{A} \clifconj{B}}
\end{align}

\begin{align}
A \lin B & = \sum_{r,s} \grade[s-r]{A_r B_s} \\
A \rin B & = \sum_{r,s} \grade[r-s]{A_r B_s}  \\
A \out B & = \sum_{r,s} \grade[r+s]{A_r B_s}
\end{align}

\begin{align} 
A_r \lin B_s & = (-1)^{r(s-1)} B_s \rin A_r \\
A_r \out B_s & = (-1)^{rs} B_s \out A_r
\end{align}

\begin{equation}
a_1 \out a_2 \out \dotsb \out a_r = \grade[r]{a_1 a_2 \dotsb a_r}
\end{equation}

\begin{align}
a \lin A & = \half(a A - \grinv{A} a) \\
a \out A & = \half(a A + \grinv{A} a)
\end{align}

\begin{align}
A \rin  a & = -a \lin \grinv{A} \\
A \out a & = \ \, \, a \out \grinv{A}
\end{align}

\begin{equation}
a \out A \out b = -b \out A \out a
\end{equation}

\begin{align}
a \lin (A B) & = (a \lin A) B + \grinv{A} (a \lin B) \\
                   & = (a \out A) B - \grinv{A} (a \out B) 
\end{align}
\begin{align}
a \out (A B) & = (a \out A) B - \grinv{A} (a \lin B) \\
                   & = (a \lin A) B + \grinv{A} (a \out B)
\end{align}
\begin{align}
a \lin (A \out B) & = (a \lin A) \out B + \grinv{A} \out (a \lin B) \\
a \out (A \rin B) & = (a \out A) \rin B - \grinv{A} \rin (a \lin B) \\
a \out (A \lin B) & = (a \lin A) \lin B + \grinv{A} \lin (a \out B)
\end{align}

\begin{align}
a \lin (a_1 \out a_2 \out \dotsb \out a_r) & = \sum_{j=1}^r (-1)^{j-1} 
    (a \lin a_j) \, a_1 \out a_2 \out \dotsb \out \check{a}_j \out 
    \dotsb \out a_r  \\
a_1 \out (a_2 \out \dotsb \out a_r) & = a_1 \out a_2 \out \dotsb \out a_r
\end{align}

\noindent If $r \leq s$ then
\begin{equation}
B_r \lin (a_1 \out a_2 \out \dotsb \out a_s) = \sum
    (-1)^{\sum_{j=1}^r (i_j - j)} (B_r \lin a_{i_1} \out a_{i_2} \out 
    \dotsb \out a_{i_r}) \, a_{i_{r+1}} \out \dotsb \out 
    a_{i_s}
\end{equation}
where the sum is performed over all possible choices of $\{a_{i_j}\}_{j=1, \dotsc,r}$ 
out of $\{a_i\}_{i=1,\dotsc,s}$, and in each term $i_1$ through 
$i_r$ and $i_{r+1}$ through $i_s$ separately are in ascending order.

\begin{align}
A \out (B \out C) & = (A \out B) \out C \\
A \lin (B \rin C) & = (A \lin B) \rin C  \\
A \lin (B \lin C) & = (A \out B) \lin C  \\
A \rin (B \out C) & = (A \rin B) \rin C
\end{align}

\noindent If $A = a_1 a_2 \dotsb a_r$, then 
\begin{equation} A B_s \rev{A} = \grade[s]{A B_s \rev{A}} \end{equation}
and
\begin{equation} (A B \rev{A}) \out (A C \rev{A}) = |A|^2 A (B \out C) \rev{A} \end{equation}

\subsection{The geometric meaning of the inner and outer products}
\label{app:geomeaninout}

$\bl{A_r} \out \bl{B_s} = 0$ iff \bl{A_r} and \bl{B_s} share nonzero vectors. \\

\noindent $\bl{A_r} \out \bl{B_s}$, if nonzero, represents the direct sum of \bl{A_r} and \bl{B_s}. \\

\noindent $\bl{A_r} \lin \bl{B_s} = 0$ iff \bl{A_r} contains a nonzero vector orthogonal to \bl{B_s}. \\
 
\noindent $\bl{A_r} \lin \bl{B_s}$, if nonzero, represents the orthogonal 
complement of \bl{A_r} in \bl{B_s}. \\

\noindent If \bl{A_r} and \bl{B_s} are orthogonal, then $\bl{A_r} \bl{B_s} = \bl{A_r} \out 
\bl{B_s}$. \\

\noindent If \bl{A_r} is a subspace of \bl{B_s}, then $\bl{A_r} \bl{B_s} = \bl{A_r} \lin \bl{B_s}$. \\

\noindent The converses of the previous two statements are true if (1) $r=1$ or $s=1$ or 
(2) \bl{A_r} or \bl{B_s} is invertible.

\begin{align}
P_{\bl{A_r}}(B) & = B \lin \bl{A_r} \bl{A_r}^{-1} = (B \lin \bl{A_r}) \lin \bl{A_r}^{-1} \\
R_{\bl{A_r}}(B) & = B \out \bl{A_r} \bl{A_r}^{-1} = B \out \bl{A_r} \rin \bl{A_r}^{-1}
\end{align}

\subsection{Grade involution}
\label{app:grinv}

\begin{align}
\grinv{\lambda} & = \ \ \lambda \\
\grinv{a} & = -a \\
\grinv{(AB)} & = \grinv{A} \grinv{B} \\
\grinv{(A + B)} & = \grinv{A} + \grinv{B}
\end{align}

\begin{equation} \grinv{A_r} = (-1)^r A_r  \end{equation}

\begin{equation} 
\grinv{A} = \grade[+]{A} - \grade[-]{A} 
\end{equation}

\begin{equation} 
\grinv{A} = \I \grinv[n]{A} \I^{-1}
\end{equation}

\begin{equation}
\grade[\pm]{A} = \half (A \pm \grinv{A})
\end{equation}

\begin{equation} \doublegrinv{A} = A \end{equation} 

\begin{equation} \grinv{(A^{-1})} = (\grinv{A})^{-1} \end{equation}

\begin{align}
\grinv{(A \lin B)} & = \grinv{A} \lin \grinv{B} \\
\grinv{(A \rin B)} & = \grinv{A} \rin \grinv{B} \\
\grinv{(A \out B)} & = \grinv{A} \out \grinv{B}
\end{align}

\begin{equation} A \I = \I \grinv[(n-1)]{A} \end{equation}

\subsection{Reversion}
\label{app:reverse}

\begin{align}
\rev{\lambda} & = \lambda  \\
\rev{a} & = a  \\
\rev{(AB)} & = \rev{B} \rev{A} \\
\rev{(A + B)} & = \rev{A} + \rev{B}
\end{align}

\begin{equation} \rev{A_r} = (-1)^{r(r-1)/2} A_r \end{equation}

\begin{equation} \doublerev{A} = A \end{equation} 

\begin{equation} \rev{(A^{-1})} = (\rev{A})^{-1} \end{equation}

\begin{align}
\rev{(A \lin B)} & = \rev{B} \rin \rev{A} \\
\rev{(A \rin B)} & = \rev{B} \lin \rev{A} \\
\rev{(A \out B)} & = \rev{B} \out \rev{A}
\end{align}

\subsection{Clifford conjugation}
\label{app:clifconj}

\begin{align}
\clifconj{\lambda} & = \ \ \lambda \\
\clifconj{a} & = -a \\
\clifconj{(AB)} & = \clifconj{B} \clifconj{A} \\
\clifconj{(A + B)} & = \clifconj{A} + \clifconj{B}
\end{align}

\begin{equation} \clifconj{A} = \grinvrev{A} = \revgrinv{A} \end{equation}

\begin{equation} \clifconj{A_r} = (-1)^{r(r+1)/2} A_r \end{equation}

\begin{equation} \doubleclifconj{A} = A \end{equation}

\begin{equation} \clifconj{(A^{-1})} = (\clifconj{A})^{-1} \end{equation}

\begin{align}
\clifconj{(A \lin B)} & = \clifconj{B} \rin \clifconj{A} \\
\clifconj{(A \rin B)} & = \clifconj{B} \lin \clifconj{A} \\
\clifconj{(A \out B)} & = \clifconj{B} \out \clifconj{A}
\end{align}

\subsection{The scalar product and norm}
\label{app:scprod}

\begin{align} 
\scprod{A}{B} & = \grade{\rev{A} B} \\
 & = \grade{\rev{A} \lin B} \\
 & = \grade{\rev{A} \rin B}
\end{align}

\begin{align}
\scprod{A}{B} & = \sum_r \scprod{A_r}{B_r} \\
& = \sum_r \rev{A_r} \lin B_r \\
& = \sum_r \rev{A_r} \rin B_r
\end{align}  

\begin{align}
\scprod{A}{B} & = \scprod{B}{A} \\ 
 & = \scprod{\grinv{A}}{\grinv{B}} \\
 & = \scprod{\rev{A}}{\rev{B}} \\
 & = \scprod{\clifconj{A}}{\clifconj{B}}
\end{align}

\begin{align}
\scprod{A}{(BC)} & = \scprod{(\rev{B} A)}{C} \\
\scprod{A}{(B \rin C)} & = \scprod{(\rev{B} \rin A)}{C}  \\
\scprod{A}{(B \lin C)} & = \scprod{(\rev{B} \out A)}{C}  \\
\scprod{A}{(B \out C)} & = \scprod{(\rev{B} \lin A)}{C}
\end{align}

\noindent Multivector $A$ is uniquely determined by either of the following:
\begin{enumerate}
\item \scprod{A}{B} for every multivector $B$.
\item \grade{A} and $a \lin A$ for every vector $a$.
\end{enumerate}

\begin{align}
|A|^2 & = \scprod{A}{A} \\
 & = |\grinv{A}|^2 \\
 & = |\rev{A}|^2 \\
 & = |\clifconj{A}|^2
\end{align}

\noindent If $A = a_1 a_2 \dotsb a_r$, then 
\begin{description}
\item[(a)] $|A|^2 = \rev{A} A = a_1^2 a_2^2 \dotsb a_r^2$.
\item[(b)] $A^{-1}$ exists iff $|A|^2 \neq 0$, in which case $A^{-1} = \rev{A}/|A|^2$
          and $|A^{-1}|^2 = |A|^{-2}$.
\item[(c)] $\scprod{(AB)}{(AC)} = \scprod{(BA)}{(CA)} = |A|^2 \, \scprod{B}{C}$.
\end{description}

\noindent For any blade \bl{A_r}, 
\begin{description}
\item[(a)] $|\bl{A_r}|^2=0$ iff the inner product is degenerate on \bl{A_r}.
\item[(b)] $\bl{A_r}^{-1}$ exists iff $|\bl{A_r}|^2 \neq 0$, in which case
          $\bl{A_r}^{-1} = (-1)^{r(r-1)/2} \bl{A_r}/|\bl{A_r}|^2 = \bl{A_r}/\bl{A_r}^2$.
\end{description}

\subsection{The dual}
\label{app:dual}

A volume element \I\ is a unit $n$-blade.

\begin{align} 
\dual{A} & = A \lin \I^{-1} \\ 
               & = A \I^{-1} 
\end{align}

\begin{align}
\invdual{A} & = A \lin \I \\
                    & = A \I \\
                    & = \I^2 \dual{A} 
\end{align}

\noindent \dual{\bl{A}} is the orthogonal complement of \bl{A}.

\begin{align} 
\dual{(AB)} & = A\dual{B} \\
\dual{(A \out B)} & = A \lin \dual{B} \\
\dual{(A \lin B)} & = A \out \dual{B}
\end{align} 

\begin{equation}
(\dual{A})^{-1} = \I A^{-1}
\end{equation}

\begin{equation}
\dual{(\grinv{A})} = (-1)^n \grinv{(\dual{A})} 
\end{equation}

\begin{equation}
\dual{(\rev{A})} = \dual{\left[\rev{\I} \rev{(\dual{A})}\right]}
\end{equation}

\begin{equation}
\dual{(\clifconj{A})} = \dual{\left[\clifconj{\I} \clifconj{(\dual{A})}\right]}
\end{equation}

\begin{equation}
\scprod{\dual{A}}{\dual{B}} = |\I|^{-2} \, \scprod{A}{B}
\end{equation}

\begin{equation} 
\dual{\bl{A_r}} \grinv[(n-r)]{B} (\dual{\bl{A_r}})^{-1} = \grinv{(\bl{A_r} \grinv[r]{B} \bl{A_r}^{-1})}
\end{equation}

\begin{equation} R_{\bl{A}}(a) = P_{\dual{\bl{A}}}(a) \end{equation} 

\subsection{The commutator}
\label{app:commute}

\begin{equation} \commute{A}{B} = \half(AB - BA) \end{equation}

\begin{equation}
\commute{A}{(BC)} = (\commute{A}{B})C + B(\commute{A}{C})
\end{equation}

\begin{equation}
\commute{A}{(\commute{B}{C})} + \commute{B}{(\commute{C}{A})} + \commute{C}{(\commute{A}{B})} = 0.
\end{equation}

\begin{align}
\grinv{(\commute{A}{B})} & = \commute{\grinv{A}}{\grinv{B}} \\
\rev{(\commute{A}{B})} & = \commute{\rev{B}}{\rev{A}} \\
\clifconj{(\commute{A}{B})} & = \commute{\clifconj{B}}{\clifconj{A}}
\end{align} 

\begin{equation} \commute{\lambda}{A} = 0 \end{equation}

\begin{align}
\commute{a}{A} & = a \lin \grade[+]{A} + a \out \grade[-]{A} \\
\commute{A}{a} & = \grade[+]{A} \rin a + \grade[-]{A} \out a
\end{align}

\begin{equation}
\commute{A_2}{A_r} = \grade[r]{A_2 A_r}
\end{equation}

\begin{align}
\commute{A_2}{(B \lin C)} & = (\commute{A_2}{B}) \lin C + B \lin (\commute{A_2}{C}) \\
\commute{A_2}{(B \rin C)} & = (\commute{A_2}{B}) \rin C + B \rin (\commute{A_2}{C})  \\ 
\commute{A_2}{(B \out C)} & = (\commute{A_2}{B}) \out C + B \out (\commute{A_2}{C})
\end{align}

\begin{equation}
\commute{A_2}{(a_1 \out a_2 \out \dotsb \out a_r)} = \sum_{j=1}^r a_1 
    \out a_2 \out \dotsb \out (A_2 \rin a_j) \out \dotsb \out a_r
\end{equation}

\noindent $A$ commutes with all multivectors iff $A$ commutes with all vectors iff $A = \lambda + \mu\grade[-]{\bl{I}}$.

\subsection{Frames and bases}
\label{app:frames}

If $\{a_i\}_{i=1,\dotsc,n}$ is a frame with volume element $a_N = a_1 \out \dotsb 
\out a_n$, the reciprocal frame is given by
\begin{equation}
a^i = (-1)^{i-1} (a_1 \out a_2 \out \dotsb \out \check{a}_i \out 
           \dotsb \out a_n) a_N^{-1}.
\end{equation}
It satisfies
\begin{equation} a^i \lin a_j = \delta^i_j.  \end{equation}
Let $I$ be an increasing string of indices $i_1, i_2, \dotsc, i_r$; then $a_I$ and $a^I$ are
\begin{align} 
a_I & = a_{i_1} \out a_{i_2} \out \dotsb \out a_{i_r} \\
a^I & = a^{i_1} \out a^{i_2} \out \dotsb \out a^{i_r}. 
\end{align}
They satisfy
\begin{equation} \scprod{a_I}{a^J} = \delta_I^J, \end{equation}
and for any multivector $A$,
\begin{align}
A & = \sum_I A^I a_I  \quad \text{where}  \quad A^I = \scprod{A}{a^I} \\
   & = \sum_I A_I a^I   \quad \text{where}  \quad A_I = \scprod{A}{a_I}.
\end{align}
If $I$ is increasing and $I^c$ is the increasing string of indices complementary to $I$, then
\begin{equation} a^I = (-1)^{\sum_{j=1}^r(i_j-1)} a_{I^c} \, a_N^{-1}. \end{equation}
A frame and its reciprocal satisfy these identities:
\begin{equation} 
\sum_i a^i \, a_i \lin A_r = \sum_i a^i \out (a_i \lin A_r)  = r A_r  \quad \text{for any $A_r$.} 
\end{equation}
\begin{equation} 
\sum_i a_i \, a^i \lin A_r = \sum_i a_i \out (a^i \lin A_r)  = r A_r  \quad \text{for any $A_r$.} 
\end{equation}
\begin{equation} \sum_i a_i\,a^i = \sum_i a^i\,a_i = n. \end{equation}
The volume element of the reciprocal frame, $a^N = a^1 \out \dotsb 
\out a^n$, is also given by
\begin{equation} a^N = \frac{a_N}{|a_N|^2}. \end{equation}
\begin{align}
P_{a_1 \out \dotsb \out a_r}(B) & = \sum_I (\scprod{B}{a_I}) a^I \\
                                                       & = \sum_I (\scprod{B}{a^I}) a_I 
\end{align}

\subsection{The adjoint of a linear operator}
\label{app:adj}

\begin{equation} \scprod{\adj{F}(B)}{A} = \scprod{B}{F(A)} \end{equation}

\begin{align}
A \lin \adj{F}(B) & = \adj{F}\left(F(A) \lin B\right) \nonumber \\
F(A) \rin B & = F\left(A \rin \adj{F}(B) \right)
\end{align}
 
\begin{equation} \doubleadj{F} = F \end{equation}

\begin{equation} \adj{GF} = \adj{F}\,\adj{G} \end{equation}  

\begin{equation} \adj{F}^{-1} = \adj{F^{-1}} \end{equation}

\begin{equation} \det(\adj{F}) = \det(F) \end{equation}

\subsection{Symmetric and skew symmetric operators}
\label{app:symmskew}

$F$ is symmetric if $\adj{F} = F$ and skew symmetric (or skew) if $\adj{F} = -F$. \\

\noindent $F$ is (skew) symmetric iff $a \lin F(b)$ is (anti)symmetric. \\

\noindent $F$ is symmetric iff $F(a) = \sum_i \lambda_i (\scprod{a}{a^i}) a_i$ for some
frame $\{a_i\}$ of eigenvectors with eigenvalues $\{\lambda_i\}$. \\


\noindent $F$ is skew iff $F(a) = a \lin A_2$ for some bivector $A_2$.

\subsection{Isometries and orthogonal transformations}
\label{app:isoorth}

$F$ is an isometry if $F(u) \lin F(v) = u \lin v$ for all $u$ and $v$. \\

\noindent $F$ is an isometry iff $\adj{F}F = \Id$. \\

\noindent $F$ is an orthogonal transformation if $F$ is an invertible isometry. \\

\noindent $F$ is orthogonal iff $\adj{F} = F^{-1}$. \\

\noindent $F$ is an isometry on a finite-dimensional space iff $F(a) = A_r \grinv[r]{a} A_r^{-1}$ for some invertible $r$-versor $A_r$. \\

\noindent The extension of orthogonal $F$ on a finite-dimensional space to all multivectors is $F(B) = A_r \grinv[r]{B} A_r^{-1}$.

\subsection{Eigenvalues, invariant subspaces, and determinants}
\label{app:eigen}

$\bl{A} \neq 0$ is an eigenblade of $F$ if $F(\bl{A}) = \lambda \bl{A}$. \\

\noindent An eigenblade of $F$ is an invariant subspace on which $F$ is invertible. The eigenvalue is $\det(F_{\bl{A}})$.

\begin{equation} \det F = \dual{F(\I)} \end{equation}

\begin{equation} F^{-1}(A) = \frac{\dual{\adj{F}(\invdual{A})}}{\det(F)} \end{equation}

\section{Topics for future versions}
\label{app:future}

These are the subjects I plan to add to the notes next, in no particular order.  The items with
asterisks are most interesting to me at the moment.

\begin{itemize}
\item Linear algebra
         \begin{itemize}
         \item More on invariant subspaces and determinants$^*$
         \item Representing a general linear operator as a sequence of multivector multiplications$^*$
         \end{itemize}
\item Differential and integral calculus
          \begin{itemize}
          \item The directed integral of a multivector
          \item The derivative of a multivector-valued function defined in terms of the directed integral$^*$
          \item Recovering traditional vector calculus
          \item The fundamental theorem of calculus and its corollaries (Gauss' theorem, Stokes' theorem, Green's theorem, etc.)$^*$
          \item Taylor series$^*$
          \item Generalizations of Cauchy's integral formula$^*$
          \item The invertibility of the derivative (cf.\ the exterior derivative)
          \item Solutions to standard ODEs and PDEs (simple harmonic oscillator, wave equation, etc.)$^*$
          \item Fourier analysis
          \item Manifold theory
          \item Lie groups and Lie algebras
          \item Curvature$^*$
          \end{itemize}
\item Geometry
          \begin{itemize}
          \item Meet and join of subspaces
          \item Projective splits (e.g.\ Minkowski spacetime into any observer's space + time)$^*$
          \item Different models of space (Euclidean, projective, conformal)
          \item Geometric algebra on a vector space without an inner product
          \end{itemize}
\item Physics
          \begin{itemize}
          \item Rotational dynamics and the inertia tensor$^*$
          \item Relativistic particle mechanics$^*$
          \item Electricity and magnetism in 3D and 4D$^*$
          \item Lagrangian and Hamiltonian mechanics$^*$
          \item Continuum mechanics and elasticity theory$^*$
          \item The Dirac equation
          \item General relativity
          \item The Galilei and Lorentz groups and their Lie algebras
          \end{itemize}
\end{itemize}



\begin{thebibliography}{99}
\addcontentsline{toc}{section}{References}
\bibitem{STA} David Hestenes, \textit{Space-Time Algebra} (New York: Gordon and Breach, 
               1966).
\bibitem{CAtoGC} David Hestenes and Garret Sobczyk, \textit{Clifford Algebra to
               Geometric Calculus} (Dordrecht: D.\ Reidel Publishing Company,    
               1984).
\bibitem{NewFounds} David Hestenes, \textit{New Foundations for Classical
               Mechanics}, 2nd ed. (Dordrecht: Kluwer Academic Publishers, 1999).
\bibitem{GAforphys} Chris Doran and Anthony Lasenby, \textit{Geometric Algebra for
                Physicists} (Cambridge: Cambridge University Press, 2003).
\bibitem{GAforCS} Leo Dorst, Daniel Fontijne, and Stephen Mann, \textit{Geometric 
                Algebra for Computer Science: An Object-Oriented Approach to Geometry}, 
                rev. ed. (Amsterdam: Morgan Kaufmann, 2007).
\bibitem{LandGA} Alan Macdonald, \textit{Linear and Geometric Algebra} (Charleston: 
                CreateSpace, 2011).
\bibitem{CAS} Pertti Lounesto, \textit{Clifford Algebras and Spinors}, London Mathematical
               Society Lecture Note Series 286, 2nd ed. (Cambridge: Cambridge University 
               Press, 2001).
\bibitem{elementary} Alan Macdonald, ``An Elementary Construction of the Geometric 
                Algebra," Adv.\ Appl.\ Cliff.\ Alg.\ \textbf{12}, 1-6 (2002).  An improved version is 
                available at \url{http://faculty.luther.edu/~macdonal/}.
\end{thebibliography}
\end{document}